\documentclass[onecolumn,amsmath,amssymb,nofootinbib,12pt]{article}
\pdfoutput=1 
\usepackage{jheppub} 

\usepackage{amsfonts}
\usepackage{amsmath}
\allowdisplaybreaks[4]        
\usepackage{amssymb}
\usepackage{euscript}     
\usepackage{color}         
\usepackage{tensor}        
\usepackage{amsthm}
\usepackage{graphicx}
\usepackage{subfigure}
\usepackage{bbm}
\usepackage[header,title,page,titletoc]{appendix}  

\usepackage[numbers]{natbib}  
\usepackage{float}

\newcommand{\labell}[1]{\label{#1}} 

\def\a{\mathfrak{a}}
\def\ha{\tilde{\a}}

\def\w{\mathfrak{w}}

\renewcommand{\(}{\left(}
\renewcommand{\)}{\right)}
\renewcommand{\[}{\left[}
\renewcommand{\]}{\right]}

\newcommand{\cO}{{\cal O}} 
\newcommand{\bra}[1]{{\left\langle{#1}\right\vert}}
\newcommand{\ket}[1]{{\left\vert{#1}\right\rangle}}

\newcommand{\be}{\begin{equation}}
\newcommand{\ee}{\end{equation}}
\newcommand{\bea}{\begin{eqnarray}}
\newcommand{\eea}{\end{eqnarray}}
\newcommand{\beq}{\begin{equation}}
\newcommand{\eeq}{\end{equation}}
\newcommand{\beqa}{\begin{eqnarray}}
\newcommand{\eeqa}{\end{eqnarray}}
\newcommand{\beqar}{\begin{eqnarray*}}
\newcommand{\eeqar}{\end{eqnarray*}}

\newcommand{\eg}{{\it e.g.,}\ }
\newcommand{\ie}{{\it i.e.,}\ }
\newcommand{\reef}[1]{(\ref{#1})}

\newcommand{\mt}[1]{\textrm{\tiny #1}}

\newcommand{\mC}{\mathcal{C}}
\newcommand{\mO}{\mathcal{O}}
\newcommand{\mR}{\mathbb{R}}
\newcommand{\mH}{\mathbb{H}}

\newcommand{\mL}{\mathcal{L}}

\newcommand{\mr}{\mt{R}}

\newcommand{\veps}{\varepsilon}

\newcommand{\cev}[1]{\reflectbox{\ensuremath{\vec{ \reflectbox{\ensuremath{#1}}}}}}
\newcommand{\s}{\sigma}
\newcommand{\wrr}{\omega_{\mt R}}
\newcommand{\id}{\mathbbm{1}}

\preprint{arXiv:1807.07677 [hep-th]}
\title{\boldmath Circuit Complexity for Coherent States}

\author[a,b]{Minyong Guo,}
\author[a,c]{Juan Hernandez,}
\author[a]{Robert C. Myers}
\author[a,c]{and Shan-Ming Ruan}

 \affiliation[a]{Perimeter Institute for Theoretical Physics,  Waterloo, Ontario N2L 2Y5, Canada}
\affiliation[b]{Department of Physics, Beijing Normal University,
	Beijing 100875, P.R. China}
\affiliation[c]{Department of Physics and Astronomy,
	University of Waterloo, Waterloo, Ontario\\ N2L 3G1, Canada}
	
\emailAdd{guominyong@gmail.com}
\emailAdd{jhernandez@perimeterinstitute.ca}
\emailAdd{rmyers@perimeterinstitute.ca}
\emailAdd{sruan@perimeterinstitute.ca}

\date{\today}

\abstract{We examine the circuit complexity of coherent states in a free scalar field theory, applying Nielsen's geometric approach as in \cite{Jeff}. The complexity of the coherent states have the same UV divergences as the vacuum state complexity and so we consider the finite increase of the complexity of these states over the vacuum state. One observation is that generally, the optimal circuits introduce entanglement between the normal modes at intermediate stages even though our reference state and target states are not entangled in this basis. We also compare our results from Nielsen's approach with those found using the Fubini-Study method of \cite{Chapman:2017rqy}. For general coherent states, we find that the complexities, as well as the optimal circuits, derived from these two approaches, are different. }

\begin{document} 

	\maketitle
	\flushbottom

\section {Introduction} \label{intro}

In recent years, a new bridge has begun to develop connecting  quantum information theory to quantum gravity and quantum field theory. In particular, understanding the relation between quantum entanglement and the emergence of semi-classical spacetime geometry \cite{VanRaamsdonk:2010pw,Lashkari:2013koa,Faulkner:2013ica} has become an active field of research. Gauge/gravity duality \cite{Maldacena:1997re,revue,joh} has been the central arena for the exploration of these connections and much of the understanding of the connection between entanglement and geometry has come from investigations of holographic entanglement entropy \cite{Ryu:2006bv,Ryu:2006ef,Rangamani:2016dms}. However, it has become clear that holographic entanglement entropy is not able to probe the bulk spacetime far behind the event horizon of black holes  \cite{Hartman:2013qma,Susskind:2014moa}. Inspired by this problem, Susskind \cite{Susskind:2014jwa,Susskind:2014rva,Susskind:2014moa} proposed the study of new bulk observables, which he conjectured should be the gravitational dual of the circuit complexity in the boundary theory. In particular, there are two proposals for `holographic complexity': complexity=volume (CV) \cite{Stanford:2014jda,Susskind:2014rva} and complexity=action (CA)  \cite{Brown:2015bva,Brown:2015lvg}. The CV conjecture states that the complexity of the boundary state is proportional to the volume of an extremal codimension-one surface extending the boundary time slice into the bulk. The CA conjecture identifies the complexity of the boundary state with the gravitational action evaluated at special bulk region called the Wheeler-DeWitt patch, \ie the causal development of the bulk surface identified in the previous approach. Both conjectures bring to our attention new gravitational observables which contain information about the spacetime region deep behind the black hole horizon and they have been vigorously investigated in the recent literature, \eg \cite{Roberts:2014isa, Alishahiha:2015rta, Carmi,Cai:2016xho, Lehner:2016vdi, Chapman:2016hwi, Brown:2016wib, Couch:2016exn, Brown:2017jil, Reynolds:2017lwq,  Carmi:2017jqz, Zhao:2017isy, Moosa:2017yvt, Swingle:2017zcd, Fu:2018kcp, An:2018xhv,Alishahiha:2018tep, Chen:2018mcc, Chapman:2018dem, Agon:2018zso,  Chapman:2018lsv,  Couch:2018phr,Kim:2017qrq}.
	
An obstacle in this program is that precise comparisons with the boundary theory are not yet possible because we still do not have a precise definition of circuit complexity for states in quantum field theory. However, some preliminary steps towards developing such a definition have been taken in the past year, \eg \cite{Jeff,Chapman:2017rqy, Hackl:2018ptj, Khan:2018rzm, Reynolds:2017jfs, Alves:2018qfv, prep2,Hashimoto:2017fga, Yang:2017nfn,  Kim:2017qrq,   therm0, Yang:2018nda, Abt:2017pmf,Abt:2018ywl,Molina-Vilaplana:2018sfn}. In particular, refs.~\cite{Jeff} adapted the approach of Nielsen and his collaborators \cite{nielsen2006quantum,nielsen2008,Nielsen:2006} to translate the task of finding the complexity of the ground state of a free scalar field theory into a geometric problem of finding optimal geodesics in an associated geometry. As similar geometric approach was developed for this question in \cite{Chapman:2017rqy} based on the Fubini-Study metric.\footnote{We must add that a complementary approach to understand complexity in quantum field theory using path integral techniques is being developed by \cite{Caputa:2017yrh,Czech:2017ryf, Caputa:2017urj, Bhattacharyya:2018wym, Caputa:2018kdj}.} In fact, both approaches produced the same simple circuit to prepare the vacuum state for a simple (unentangled) reference state and assigned the same complexity to the vacuum. In these calculations the field theory must be regulated since the complexity is dominated by contributions from the high energy modes and the result is UV divergent. However, an interesting agreement was found in comparing the structure of these divergences with those appearing in holographic complexity. In particular, the leading divergence found for holographic complexity \cite{Carmi} takes the form
\beq\label{leaderH}
\mC_\mt{A}\sim \frac{V}{\delta^{d-1}}\,\log(\ell/\delta)\,, \qquad\qquad
\mC_\mt{V}\sim \frac{V}{\delta^{d-1}}\,,
\eeq
where $V$ is the spatial volume, $\delta$ is the short-distance cutoff, and $d$ is the spacetime dimension of the boundary theory. The scale $\ell$ is undetermined and arises because of ambiguities in defining the gravitational action on regions with null boundary segments \cite{Lehner:2016vdi, Chapman:2018lsv}. An analogous ambiguity appears in evaluating the complexity for the scalar field theory because an undetermined scale must be introduced to define the reference state, and the leading divergence of the vacuum has precisely the same form as shown above for $\mC_\mt{A}$ \cite{Jeff,Chapman:2017rqy}. Of course, in either calculation, the interesting logarithmic factor can be removed by choosing $\ell \sim \delta$ and so this does not rule out the CV conjecture.

In this paper, we are extending the investigations of complexity in refs.~\cite{Jeff,Chapman:2017rqy} by examining the complexity of excited states in the free scalar field theory. In particular, we develop the additional techniques needed to evaluate the complexity of coherent states in which the scalar field acquires a nonvanishing expectation value. An exploratory investigation of the complexity of coherent states already appears in \cite{Yang:2017nfn}, however, the analysis there differs in many essential ways from our approach and there is no substantive overlap between the previous work and the present paper, as we will describe in more detail in section \ref{discuss}. Here, we might add that the complexity of excited fermionic states was considered in \cite{Hackl:2018ptj}. But this was a special case where the excited states were still Gaussian states and so no new ingredients were needed beyond those needed to evaluate the complexity of the vacuum. Further, refs.~\cite{Alves:2018qfv,prep2} examined the complexity of excitations for the free scalar produced by a quench of the mass term. However, again these excited states could be assessed using the same techniques used to evaluated the vacuum complexity. To prepare the coherent states, we must introduce a new class of gates in our circuits and in particular, this requires introducing a new  (undetermined) scale into our model for the complexity. We develop the extended geometry associated with this larger gate set for both the Nielsen and Fubini-Study approaches and the resulting optimal circuits and complexities exhibit a number of interesting features. For example, we find that the optimal circuits introduce entanglement between the normal modes at intermediate stages even though our reference state and target states are not entangled in this basis. Further for general coherent states, we show that the complexities, as well as the optimal circuits, derived by Nielsen and Fubini-Study approaches are different.

	\subsection{Nielsen, geometry and complexity}
	\label{sec:intro2}
	
\begin{figure}[htbp]
		\centering
		\subfigure{\includegraphics[width=5.5in]{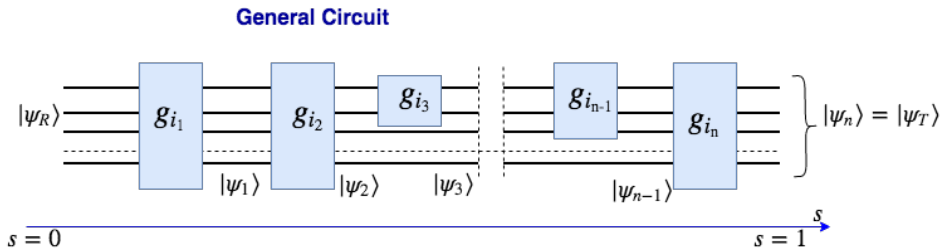}}
		\caption{A general quantum circuit where $\ket{\psi_\mt{T}}$ is prepared beginning with $\ket{\psi_\mt{R}}$ and applying a sequence of elementary unitaries $g_{i}$. We also indicate all of the intermediate states $\ket{\psi_i}$ that are produced after every step.}\label{general_circuit}
	\end{figure}

	In this section, we briefly review the salient ideas required to apply  Nielsen's geometric approach to circuit complexity \cite{nielsen2006quantum,nielsen2008,Nielsen:2006}  to evaluate the complexity of state in a quantum field theory, as developed in \cite{Jeff}. In this setting, complexity is a measure of the difficulty or cost to prepare the particular target state  $\ket{\psi_\mt{T}}$ starting with a certain simple reference state $| \psi_\mt{R} \rangle$. We are using a quantum circuit model where the preparation is accomplished by applying a series of elementary unitaries, chosen from a particular set of gates $\{g_1,\cdots,g_\mt{N}\}$. That is,\footnote{When working with discrete gates as discussed here, we will typically only  prepare $\ket{\psi_\mt{T}}$ within some  tolerance $\varepsilon$, \eg $\Vert \,\ket{\psi_\mt{T}}-U_\mt{T}\ket{\psi_\mt{R}}\Vert^2 \le \varepsilon $. However, with the continous construction of unitaries introduced in eq.~\reef{unitaries}, we are always able to exactly prepare the target states with a finite cost, and so we will not need to introduce a tolerance.}
	\begin{equation}\label{circuit_def}
	\ket{\psi_\mt{T}}= U_\mt{T}\, \ket{\psi_\mt{R}}= g_{i_\mt{n}}\cdots g_{i_\mt{2}}\,g_{i_\mt{1}}\ket{\psi_\mt{R}}\,,  
	\end{equation} 
	whose circuit is shown in Figure \ref{general_circuit}.
Now in general, we must expect that there are a large (\eg infinite) number of circuits or sequences of elementary gates which will accomplish the above transformation. The complexity of the target state $\ket{\psi_\mt{T}}$ is then defined as the minimum number of gates needed to construct a unitary $U_\mt{T}$ satisfying eq.~\reef{circuit_def}. We stress that this optimal number will depend on the choices for the reference state $\ket{\psi_\mt{R}}$ and for the gate set $\{g_1,\cdots,g_\mt{N}\}$, however, one can still obtain interesting physical insights by comparing the complexities for families of target states. Nonetheless, given a particular set of choices, the main challenge is to identify the optimal circuit from amongst the infinite range of possibilities to prepare a certain target state.

To overcome this challenge, Nielsen and collaborators \cite{nielsen2006quantum,nielsen2008,Nielsen:2006}	developed a geometric method. Adapting this approach to evaluate the complexity of QFT states \cite{Jeff}, one begins with a continuum construction of the unitary transformations acting on the states
	\begin{equation}\label{unitaries}
	U(\s) = \cev{\mathcal{P}} \exp \[ -i \int^\s_0\!\!\! d s\, H( s)\], \quad \text{where} \quad H(s)= \sum_I Y^I(s)\,\mathcal{O}_I
	\end{equation}
where $s$ parametrizes the circuit and $\cev{\mathcal{P}} $ indicates right-to-left path ordering. The (path-dependent) Hamiltonian $H(s)$ is  expanded in terms of a basis of Hermitian operators $\mathcal{O}_I$, which we think of as generators for elementary gates $g_I\sim\exp[-i\varepsilon \mO_I]$ (where $\varepsilon$ would be an infinitesimal parameter). The control functions $Y^I(s)$ then specify which gates (and how many times they) are applied at a particular point $s$ in the circuit. Note that eq.~\reef{unitaries} specifies not only the full transformation $U_\mt{T}$ in eq.~\reef{circuit_def} but also a trajectory $U(\s)$ through the space of unitaries, or through the space of states with $\ket{\psi(\s)}=U(\s)\ket{\psi_\mt{R}}$, where $0\le\s\le1$. The circuits of interest are then the trajectories satisfying the boundary conditions 
	\begin{equation}
	U(\s=0)= \mathbbm{1}\,, \qquad  U(\s=1)= U_\mt{T}\,,
	\end{equation}
\ie we start from the identity and end with the desired unitary $U_\mt{T}$ producing the desired transformation in eq.~\reef{circuit_def}. From this perspective, the $Y^I(s)$ can be understood as the tangent vectors to the trajectories with
	\begin{equation}
	Y^I(s)\,\mathcal{O}_I = i \,\partial_s U(s)\,U^{-1}(s)\,,
	\label{tangent}
	\end{equation}
which will play a important role later.  

Then Nielsen's approach for identifying the optimal circuit is to minimize the cost defined as
	\begin{equation}\label{cost_D}
	\mathcal{D}(U(\s))= \int^1_0 ds ~ F \( U(s), Y^I(s)  \),  
	\end{equation}
where $F$ is a local cost function depending on the position $U(s)$ and the tangent vector $Y^I(s)$. This question is then similar to the physical problem of identifying a particle trajectory by minimizing the action with Lagrangian $F ( U(s), Y^I(s) )$.  While the precise form of the cost function $F$ is not fixed, there are a number of desirable features for  reasonable cost functions \cite{Nielsen:2006}: 1) Smoothness, 2) Positivity, 3) Triangle inequality and 4) Positive homogeneity -- see also \cite{Jeff,comparison-paper}. Some simple examples of cost functions that satisfy the above constraints are 
	\begin{equation}\label{function_F}
	F_1(U,Y)=\sum_I \left|Y^I\right|~,\qquad\qquad
	F_2(U,Y)=\sqrt{\sum_I  \(Y^I\)^2}~.
	\end{equation}
Given the role of the $Y^I(s)$ as control functions, the $F_1$ measure comes the closest to the original concept of counting the number of gates. The $F_2$ measure can be recognized as the proper distance in a Riemannian geometry, and this choice reduces the problem of identifying the optimal circuit to finding the shortest geodesic connecting the reference and target states in this geometry. 

Another class of cost functions introduced in \cite{Jeff} take the form 
		\begin{equation}\label{function_Fkappa}
	\begin{split}
	F_\kappa(U,Y)=\sum_I \left|Y^I\right|^\kappa~.
	\end{split}
	\end{equation}
These $\kappa$ cost functions  can be thought of as a generalization of the $F_1$ cost function. The corresponding vacuum complexity compares well with the results from holographic complexity but these cost functions do not satisfy the `homogeneity' property above, \ie the cost \reef{cost_D} is not invariant under reparametrizations of $s$. We also note that the $\kappa=2$ cost function will yield exactly the same extremal trajectories or optimal circuits as the $F_2$ cost function. An interesting suggestion in \cite{Hackl:2018ptj} was to construct a family of new cost functions using the Schatten norm (\eg see \cite{bhatia2013matrix,watrous2018theory,gil2003operator}) 
\beq
F_p(U,Y)=\Vert V \Vert_p =\Big[
{\rm Tr}\!\(\( V^\dagger\,V\)^{p/2}\)\Big]^{1/p}\,,
\eeq 
where $V=Y^I(s)\mO_I$ is the tangent vector defined as an operator which transforms the states (see details in section \ref{Schat}). These cost functions satisfy all of the desired properties and further are independent of the particular choice of basis for the $\mO_I$ -- a issue for the $F_1$ measure and the general $\kappa$ cost functions (for $\kappa\ne2$) \cite{Jeff}.

Before closing this short review, we must mention the group theoretic structure that naturally appears in applying this approach to evaluate the complexity of QFT states. For the problem to be tractable, one only considers a limited basis of operators $\mO_I$ to constructing the unitaries \reef{unitaries}. A practical restriction is that this basis should then form a closed  algebra, and hence in many examples, the $\mO_I$ provide a representation of a Lie algebra $\mathfrak{g}$, \ie $[\mO_I,\mO_J]=i f_{IJ}{}^K \mO_K$. For example, in examining the complexity of fermionic Gaussian states, an $\mathrm{O}(2N)$ group structure emerges \cite{Hackl:2018ptj}. In \cite{Jeff}, a $\mathrm{GL}(N,\mathbb{R})$ algebra appeared in evaluating the complexity of the  ground state of a free scalar field, and the latter was extended to an $\mathrm{Sp}(2N,\mathbb{R})$ algebra examining the corresponding thermofield double state in \cite{therm0} -- see also \cite{Hackl:2018ptj}. In the following, we will find that an $\mR^N\rtimes \mathrm{GL}(N,\mathbb{R})$ algebra plays a central role in evaluating the complexity of coherent states. The utility of this group theoretic perspective is that the physical details of the basis operators $\mO_I$ can be pushed to the background. Instead, the generators in eq.~\reef{unitaries} are simply elements of the Lie algebra $\mathfrak{g}$, and we can choose the most convenient  representation for the particular calculations of interest.\\

The rest of the paper is organized as follows: In section~\ref{sec:2ho}, we study the complexity for coherent Gaussian states in a system of two coupled oscillators. For general states, we must resort to numerical methods to evaluate the complexity in section \ref{sec:numerics}, however, we also produce some analytic results for simple cases in which only one mode is excited in section \ref{sec:simple} or in which the excitations have small amplitudes in section \ref{sec:small}. This initial analysis is based on the $F_2$ and $\kappa=2$ cost functions, and so
we discuss analogous results for the $F_1$ cost function in section \ref{costf1} and the $p=1$ Schatten norm in section \ref{Schat}. We extend our results to a free scalar field theory by a lattice regularization in section~\ref{sec:QFT}. In section \ref{app:info-metric}, we apply the geometric approach based on the Fubini-Study metric \cite{Chapman:2017rqy} to reinvestigate the complexity of our coherent states for two coupled harmonic oscillators. The results for this simple system are also compared with our results in section \ref{sec:2ho} using Nielsen's approach with the $F_2$ cost function. We conclude with a discussion of our results and possible future directions in section \ref{discuss}.

\section{Complexity of coupled harmonic oscillators} 
    \label{sec:2ho}
    
Our goal is to evaluate the complexity of coherent states in a free scalar field theory, applying the techniques of \cite{Jeff}. However, as a warm-up exercise, we begin here by considering coherent states in the simpler system of two coupled harmonic oscillators. In this section, our focus will be on the $F_2$ cost function \reef{function_F}, and also on the $\kappa=2$ cost function \reef{function_Fkappa} which are extremized by the same trajectories. We will turn to consider other cost functions in section \ref{sec:F1}. Our approach here closely follows that in \cite{Jeff} and we refer the reader there for a more detailed discussion.
    
\subsection{Gate set and group structure} \label{sec:gates}

Let us consider two coupled harmonic oscillators with the Hamiltonian 
\beqa
H&=&\frac{1}{2m}\left[ p_1^2+p_2^2+m^2\omega^2\( x_1^2+x_2^2\)+m^2 \Omega^2 \( x_1-x_2\)^2 \right]
\nonumber\\
&=& \frac{1}{2m}\[ p_+^2+ m^2 \omega_+^2 x_+^2\]+ \frac{1}{2m} \[p_-^2+ m^2 \omega_-^2 x_-^2\]~,\labell{qm1}
\eeqa
where in the second line, the two oscillators were decoupled by introducing the normal modes,
      \begin{equation}\label{Phybasis}
      x_{\pm}=\frac{1}{\sqrt{2}} (x_1\pm x_2)\,, \qquad  \omega_+=\omega\,, \qquad \omega_-=\sqrt{\omega^2+2\Omega^2}\,.
      \end{equation}
Given the decoupled Hamiltonian, the ground state wave function is easily written as
\beqa
\psi_0(x_1,x_2)&=&\frac{\(m^2\,\omega_+\omega_-\)^{1/4}}{\sqrt{\pi}}\ \mathrm{exp}\!\left[-\frac{m}{2}\(\omega_+\, x_+^2+\omega_-\,  x_-^2\)\right]
\labell{targetPhys}\\
&=&
\frac{\(m^2\,\omega_+\omega_-\)^{1/4}}{\sqrt{\pi}}\ \mathrm{exp}\left[-\frac{m(\omega_++\omega_-)}4 \(x_1^2+x_2^2\) + \frac{m}{2}\(\omega_--\omega_+\)x_1 x_2\right]\,.
\nonumber
\eeqa
While the normal modes are completely unentangled here, from the perspective of the physical coordinates $\lbrace x_1,x_2\rbrace$, the ground state is an entangled state. Ref.~\cite{Jeff} developed the techniques needed to evaluate the complexity of this state relative to a factorized Gaussian state as the reference state,
\begin{equation}\label{eq:refPhys}
     \psi_\mt{R}(x_1,x_2)=\frac{\wrr}{\sqrt{\pi}}\ \mathrm{exp}\!\left[-\frac{\wrr^2}{2}( x_1^2+x_2^2 )\right]=\frac{\wrr}{\sqrt{\pi}}\ \mathrm{exp}\!\left[-\frac{\wrr^2}{2}( x_+^2+x_-^2 )\right]\,.
     \end{equation}
where the reference frequency $\wrr$, which characterizes the width of this state, is left as a free parameter.\footnote{Note that our notation is slightly different from that in \cite{Jeff}. In particular, the latter had $m\omega_0$ in place of $\wrr^2$ in eq.~\reef{eq:refPhys}.} We note that this reference state is unentangled in both the physical and the normal-mode bases.

Now we would like to extend the calculations in \cite{Jeff} to evaluate the complexity of coherent states of the form 
     \begin{equation}
     \label{target1}
     \psi_{\mt T}( x_+, x_-) = \frac{\(m^2\,\omega_+\omega_-\)^{1/4}}{\sqrt{\pi}}\ \mathrm{exp}\left[-\frac{m}{2}\(\omega_+ (x_+ -a_+)^2+\omega_- (x_- - a_-)^2\)\right]\,.
     \end{equation}
These coherent states are characterized by the expectation
values
\beq
\langle \psi_{\mt T}|x_\pm|\psi_{\mt T}\rangle = a_\pm\,,
\label{expect}
\eeq
which vanish in the ground state \reef{targetPhys}. Alternatively, in terms of the physical coordinates, we have 
\beq
\langle \psi_{\mt T}|x_1|\psi_{\mt T}\rangle = \frac{a_++a_-}{\sqrt2}
\qquad{\rm and}\qquad 
\langle \psi_{\mt T}|x_2|\psi_{\mt T}\rangle = \frac{a_+-a_-}{\sqrt2}\,.
\label{expect2}
\eeq
The coherent states in eq.~\reef{target1} are written in terms of the normal modes since this simplifies the calculations below, as shown in~\cite{Jeff}, and this will be our working basis throughout the rest of this paper.\footnote{With the states chosen in eq.~\reef{target1}, we are focusing on real wavefunctions with $\langle \psi_{\mt T}|x_i|\psi_{\mt T}\rangle \ne 0$ but $\langle \psi_{\mt T}|p_i|\psi_{\mt T}\rangle = 0$. In principle, by considering complex wavefunctions, one could examine more general states which also have $\langle \psi_{\mt T}|p_i|\psi_{\mt T}\rangle \ne 0$, as would naturally arise from the time evolution of the wavefunctions in eq.~\reef{target1}. We note that this would require the extending the $GL(2,\mathbb{R})$ algebra appearing below to $Sp(4,\mathbb{R})$, \eg see \cite{Hackl:2018ptj,therm0}. We thank Lucas Hackl for a discussion on this point.}
  
The next step is to identify the set of elementary unitary gates with which we will construct the desired unitary $U$, which implements
\beq
|\psi_\mt{T}\rangle =U\,|\psi_\mt{R}\rangle\,.
\eeq
With the new shift parameters $a_\pm$, we need additional gates than those described by the $GL(2,\mathbb{R})$ algebra in~\cite{Jeff}. However, the full complement of gates required to construct an arbitrary Gaussian state were discussed in~\cite{Jeff} and for the coherent states of the form~\eqref{target1}, we only need three types of elementary gates:
\beqa
 {\rm scaling\ gates:} &&     \qquad Q_{ii}=e^{\frac{i\veps}{2} (x_i p_i+p_ix_i)}=e^{\varepsilon/2}\,e^{i\veps x_i p_i} ~, 
\nonumber \\
 {\rm entangling\ gates:}&&\qquad
     Q_{ij}=e^{i\veps x_ip_j}\ \ ({\rm with}\ i\ne j)~,
 \labell{4gates}    \\
 {\rm shift\ gates:}&&\qquad Q_{0i}=e^{i\veps x_0p_i}\,,
\nonumber
\eeqa
where the $i,j$ can be either $\{1,2\}$ or $\{+,-\}$, but as mentioned above, we will work in the normal mode basis, \ie
$i,j\in\{+,-\}$. Further we recall that $\veps$ is a small (dimensionless) parameter which ensures that these gates only make small changes to the states on which they act. The dimensionful parameter $x_0$ appearing in the shift gates is another free parameter (a c-number) which characterizes our complexity model. As we discuss below, we might simplify the model by setting $x_0\sim 1/\wrr$ (or $x_0\sim \delta$ in the QFT calculations). 
The action of these gates is illustrated with the following examples:
     \begin{equation}
     \begin{split}
     \label{eq:trans-6-gates}
     Q_{++}\,\psi(x_+,x_-)&=e^{\varepsilon /2}\psi\( e^\varepsilon x_+,x_-\) \qquad  \mathrm{scale~} x_+ \to e^\varepsilon  x_+ \,,\\
     Q_{-+}\,\psi(x_+,x_-)&=\psi(x_++\varepsilon  x_-,x_-) \qquad  \mathrm{shift~} x_+ \mathrm{~by~} \varepsilon  x_- \,,\\
          Q_{0+}\,\psi(x_+,x_-)&=\psi(x_++\varepsilon x_0,x_-) \qquad  \mathrm{shift~} x_+ \mathrm{~by~}\varepsilon x_0 \,.
     \end{split}
     \end{equation}

Note that our set of elementary gates~\eqref{4gates} can be summarized by
   \begin{equation}
   \label{eq:6gates}
   Q_{ai} =\exp\[i\veps \mO_{ai}\]= e^{\frac{i \veps}{2} (x_a p_i + p_i x_a)}\,,
   \end{equation}
where $a \in \{+,-,0\}$ and $i \in \{+,-\}$.  We have also introduced the notation $\mO_{ai}$ to denote the Hermitian generators of these elementary gates. 

Now following \cite{Jeff} to make further progress, next, we construct a matrix representation of these gates. In general, we are interested in coherent states of the form
   \begin{equation}\label{waveA}
   \psi(x_+,x_-) = \mathcal{N}\, \exp\left[ -\frac{1}{2}\( x_a\,A^{ab}\,x_b -  c\, x_0^2\) \right]\,.
   \end{equation}
where again the sums over $a, b$ run over $\{+,-,0\}$, and $A$ is a symmetric 3$\times$3 matrix with $A^{00}=c$. We introduced the term $c x_0^2$ above to eliminate this c-number contribution from the exponent and hence $\mathcal{N}$ is the normalization constant. It will be convenient to keep $A^{00}$ in the following calculations, but we stress that its value will be unimportant since the wavefunction \reef{waveA} is independent of this coefficient.\footnote{Since the elementary gates~\eqref{4gates}  are unitary, they preserve the normalization of the wavefunctions. However, the normalization is an inessential feature which can be restored given an $A$ and so we will lose track of it when working with the matrix representation below.}

Of course, the matrix $A$ completely determines the wave function, and so instead of working with these wavefunctions directly, we focus our attention on the five-dimensional space of $A$'s.  Again, the full space of symmetric 3$\times$3 matrices would be six-dimensional but since as explained above, the wavefunctions are independent of $A^{00}$, we have a five-dimensional space of distinct wavefunctions. With this matrix form,  we can represent the reference state \eqref{eq:refPhys} as 
   \begin{equation}
   \label{eq:A-ref}
   \psi_\mt{R}(x_+,x_-)\rightarrow A_\mt{R}=\left(
   \begin{array}{ccc}
   \wrr^2 & 0 & 0 \\
   0 & \wrr^2 & 0 \\
   0 & 0 & c_\mt{R} \\
   \end{array}
   \right)\,,
   \end{equation}
and  the  target state \eqref{target1} is represented by 
   \begin{equation}
   \label{eq:A-target1}
   \psi_\mt{T}(x_+,x_-)\rightarrow A_\mt{T}= m\,
   \begin{pmatrix}
   \omega_+ & 0 & - \a_+ \, \omega_+ \\ 0 &  \omega_- & -\a_- \, \omega_- \\ -\a_+ \, \omega_+ & - \a_-\,\omega_- & c_\mt{T}
   \end{pmatrix}\,, 
   \end{equation}
where $\a_\pm \equiv a_\pm/x_0$. We emphasize once more that the values of $c_\mt{R}$ and $c_\mt{T}$ are completely undetermined since they do not affect the wavefunctions.
   
By considering the action of the elementary gates \reef{eq:trans-6-gates} on the general wavefunctions \reef{waveA}, we produce a 3$\times$3 matrix representation of the gates which transforms the A as follows 
   \begin{equation}
A \to   A' = Q_{ai}\, A\,\, Q_{ai}^T\,,
   \label{matrix0}
   \end{equation}
   where
   \begin{equation}
   Q_{ai} = \exp\!\left[\varepsilon\,M_{ai}\right]
   \qquad{\rm with}\quad
   \left[M_{ai}\right]{}_{cd} = \delta_{ac}\,\delta_{id}\,.
   \label{matrix}
   \end{equation}
An easy way to verify this result is to consider the action of the matrices $Q_{ai}$ on the vector $\mathbf{\tilde{x}}^T=(x_+,x_-,x_0)$ and confirm that the result agrees with the transformation by the original gates~\eqref{4gates}, \eg we can compare
   \begin{equation}
   \begin{split}
   \mathbf{\tilde{x}}^T  Q_{++} &= (e^{\varepsilon}x_+,x_-,x_0)\, ,\\
   \mathbf{\tilde{x}}^T Q_{-+} &= (x_++\varepsilon x_-, x_-,x_0)\,, \\
   \mathbf{\tilde{x}}^T Q_{0+} &= (x_+ + \varepsilon x_0 ,x_-,x_0)\,,
   \end{split}
   \end{equation}
with the transformations in eq.~\reef{eq:trans-6-gates}.
   Explicitly, the six generators $M_{ai}$ are
   \begin{equation}\label{longM}
   \begin{split}
   M_{++}=\begin{pmatrix}
   1&  0        & 0\\
   0 &  0  & 0 \\
   0  &  0  &   0
   \end{pmatrix} ,\ \ 
 &M_{--}=  \begin{pmatrix}
   0&  0        & 0\\
   0 &  1  & 0 \\
   0  &  0  &   0
   \end{pmatrix} , \ \
   M_{-+}=\begin{pmatrix}
   0&  0        & 0\\
   1 &  0  & 0 \\
   0  &  0  &   0
   \end{pmatrix} ,\ \ 
   M_{+-}=\begin{pmatrix}
   0&  1       & 0\\
   0 &  0  & 0 \\
   0  &  0  &   0
   \end{pmatrix} ,\\
  & M_{0+}=\begin{pmatrix}
   0&  0        & 0\\
   0 &  0  & 0 \\
   1 &  0  &   0
   \end{pmatrix} ,\quad
   M_{0-}=\begin{pmatrix}
   0&  0        & 0\\
   0 &  0  & 0 \\
   0  &  1  &   0
   \end{pmatrix}. 
   \end{split}
   \end{equation}
The convenience of this representation is that we can define a simple inner product of these matrix generators \eqref{matrix},
   \begin{equation}\label{inner}
   {\rm tr}\( M_I M_J^T\) = \delta_{IJ}\,,   
   \end{equation}
where $I,J\in \lbrace++,--,-+,+-,0+,0- \rbrace$.
We will use this inner product in a moment in constructing a metric on the six-dimensional
space of unitary transformations generated by our elementary gates \reef{4gates}.
    
Now we recall from \cite{Jeff} that the four generators $M_{ij}$ for the scaling and entangling gates (appearing in the first line of eq.~\reef{longM}) form a $GL(2,\mathbb{R})$ algebra. More generally if we consider the action of a string of the elementary gates on  $x_+$ and $x_-$, we find that they are transformed as $x_i \rightarrow G_i{}^j\, x_j + v_i$ (where $G$ is a $GL(2,\mathbb{R})$ matrix). That is, our gates produce affine transformations of the coordinates. Hence the full group generated the six gates  $Q_{ai}$ has a structure similar to that of the Poincar\'e group. The $GL(2,\mR)$ of the scaling and entangling gatess plays the role of the Lorentz group $O(1,3)$ and the  $\mR^2$ of translations generated by the $Q_{0i}$ is analogous to the translations in Minkowski space. Hence, the structure of our algebra here\footnote{For $N$ harmonic oscillators, it is straightforward to generalize this discussion to show that the algebra of the generators of $N (N+1)$ elementary gates acting on the coordinates of the harmonic oscillators form a fundamental representation of $\mR^N \rtimes GL(N,\mR)$. \label{foot66}}  is the semidirect product of $\mR^2$  by general linear transformations $GL(2,\mathbb{R})$,
\ie
     \begin{equation}\label{groupX}
     \mR^{2} \rtimes GL(2,\mathbb{R})\,.
     \end{equation}

\subsection{Six-dimensional geometry and its geodesics}
  \label{subsec:6-geometry}

The group structure \reef{groupX} is manifest by any transformation $U$ generated by the $M_{ai}$ taking the form
   \begin{equation}
   U=
   \begin{pmatrix}
   U_2  & \mathbf{0} \\
   \mathbf{u}^T                     &   1 
   \end{pmatrix} \label{oblique}
   \end{equation}
where $\mathbf{u}^T = (u_+,u_-)\in \mR^2$ and $U_2 \in GL(2,\mR)$. It will be convenient to parametrize the latter with the following polar decomposition
   \begin{equation}\label{GL2_matrix}
   U_{2} = e^y R(-x)\, S(\rho)\, R(z)=
   e^y \begin{pmatrix}
   \cos x & - \sin x\\
   \sin x & \ \cos x
   \end{pmatrix}
   \begin{pmatrix}
   e^{\rho} & 0 \\
   0 & e^{-\rho} 
   \end{pmatrix}
   \begin{pmatrix}
   \ \cos z & \sin z\\
   -\sin z & \cos z
   \end{pmatrix}\,,
   \end{equation}
where $R$ denotes a rotation matrix and $S$ is a `squeezing' matrix.
This construction then introduces the coordinates $\mathbf{y}^T = (y,\rho,x,z,u_+,u_-)$ on the group of affine transformations \reef{groupX}. 

There is also a surjective, but not injective, map that associates a wavefunction of the form~\eqref{waveA} to every group element, given by
   \begin{equation}
 \psi_\mathbf{y}(x_+,x_-) = U(\mathbf{y})\,\psi_\mt{R}(x_+,x_-)\,,
   \end{equation}
with the reference state given in eq.~\reef{eq:refPhys}. The corresponding transformation using the matrix representation \reef{oblique} becomes
\small
\begin{equation}\label{eq:A-target}
\begin{split}
   A(\mathbf{y})&= U(\mathbf{y})\,A_\mt{R}\, U^T(\mathbf{y}) \\
   &=\wrr^2 \left(
   \begin{array}{ccc}
   e^{2 y} (\cosh (2 \rho )+\cos (2x) \sinh (2 \rho )) & e^{2 y} \sin (2x) \sinh (2 \rho ) & \Lambda_+ \\
   e^{2 y} \sin (2x ) \sinh (2 \rho ) & e^{2 y} (\cosh (2 \rho )-\cos (2x) \sinh (2 \rho )) & \Lambda_- \\
   \Lambda_+ & \Lambda_- & u_+^2+u_-^2+c_\mt{R} \\
   \end{array}
   \right)\,, 
   \end{split}
\end{equation}
\normalsize
where $A_\mt{R}$ is given in eq.~\reef{eq:A-ref} 
and
\begin{equation}\label{split}
\begin{split}
   \Lambda_+ &=e^{y+\rho} \cos(x) (u_+\cos (z) + u_- \sin(z))-e^{y-\rho}\sin(x)(u_- \cos(z)-u_+ \sin (z))\,,\\
   \Lambda_- &=e^{y-\rho} \cos(x) (u_-\cos (z) - u_+\sin(z))+e^{y+\rho}\sin(x)(u_+\cos(z)+u_-\sin (z))\,.
\end{split}
\end{equation}
In fact, one can see that $\mathbf{\Lambda}=U_2\cdot\mathbf{u}$ where we have assembled the vector $\mathbf{\Lambda}^T=(\Lambda_+,\Lambda_-)$. This observation is useful because it allows us to see that the final matrix $A(\mathbf{y})$ is independent of $z$ in the following sense: First, it is obvious that the upper-left 2$\times$2 block in eq.~\reef{eq:A-target} is invariant under arbitrary shifts $z \to z'= z+ \delta z$, \ie this block is completely independent of $z$. Now given the form of $U_2$ in eq.~\reef{GL2_matrix}, it is also evident that $\mathbf{\Lambda}$ is invariant as long as we accompany the shift of $z$ with a rotation $\mathbf{u}\to \mathbf{u}'= R(-\delta z)\cdot\mathbf{u}$.  Finally, this rotation also leaves invariant the component $[A(\mathbf{y})]^{00}$, as can be seen by writing this term as $\mathbf{u}^T\cdot\mathbf{u}$. This result reflects the fact that the map from the space of unitary transformations to Gaussian states is surjective but not injective, \ie the space of unitaries which we are considering is six-dimensional while our space of Gaussian states is only five-dimensional.\footnote{Further, the fact that this mismatch appears as $A(\mathbf{y})$ being independent of $z$ is a reflection of the rotation invariance of the reference state~\eqref{eq:A-ref}. This symmetry can be made more explicit by reparametrizing the group elements as
$$ U = \begin{pmatrix}
   	\id_2 & \mathbf{0}\\
   	\mathbf{v}^T & 1
   	\end{pmatrix}
   	\begin{pmatrix}
   	U_2(y,\rho,x,z) & \mathbf{0}\\
   	\mathbf{0}^T & 1
   	\end{pmatrix}\,,$$
   with $\mathbf{v}^T=(v_+,v_-)$. We then find 
$$ \mathbf{\tilde{x}}^T{\cdot}A{\cdot}\mathbf{\tilde{x}} = (\mathbf{x}+x_0\mathbf{v})^T{\cdot} A_2{\cdot}(\mathbf{x}+x_0\mathbf{v})\,, $$ where $\mathbf{x}^T = (x_+,x_-)$ and $A_2=A_2(y,\rho,x)$ is the $2\times 2$ upper-left matrix in eq.~\reef{eq:A-target}. The wavefunction is then clearly independent of $z$. We chose not to use this parametrization because the metric in these coordinates is much more complicated. 
\label{footy44}} 

Now following \cite{Jeff}, we replace the unitaries \reef{unitaries} by their matrix counterparts
	\begin{equation}\label{Munitaries}
	U(\s) = \cev{\mathcal{P}} \exp \[  \int^\s_0\!\! ds\, H(s)\] \quad \text{where} \quad H(s)= \sum_I Y^I(s)\,M_I\,,
	\end{equation}
with the generators $M_I$ given in eq.~\reef{longM}. Now using eq.~\reef{inner}, we can solve for the coefficients $Y_I(\s)$ as
   \begin{equation} \label{velo1}
   Y^I(\s)={\rm tr}\! \left(\partial_\s U(\s)\, U^{-1}(\s)\, M_I^T\right)\,.
   \end{equation}
Further, for the parametrization of the group elements in $R^2 \rtimes GL(2,R)$ in eq.~\eqref{oblique}, we can define a metric on the space of unitary transformations as\footnote{More generally, one could replace $\delta_{IJ}\to G_{IJ}$ in constructing this metric. However, the present choice assigns the same cost to all of elementary gates and further it corresponds to the $F_2$ cost function introduced in eq.~\reef{function_F}. Following a construction analogous to that in \cite{Hackl} (see also \cite{Hackl:2018ptj,therm0}), we can also construct the metric by defining
$$ds^2=\text{tr}(dUU^{-1}\,A_\mt{R}\, (dUU^{-1})^T\,a_\mt{R})$$
where $a_\mt{R}$ is the inverse of $A_\mt{R}$, \ie $ [A_\mt{R}]^{ac} \,[a_\mt{R}]_{cb}=\delta^a{}_b$. In this case, the metric takes the form
\beqa 
ds^2&=& 2\, dy^2 + 2\, d\rho^2 + 2\, dx^2 - 4\, {\rm cosh}(2\rho) dx\, dz + 2\, {\rm cosh}(4\rho) dz^2
\nonumber\\
& &\qquad+\, \kappa\,e^{-2y} \big[  {\rm cosh}(2\rho) (du_+^2 + du_-^2) - {\rm cos}(2z) \,{\rm sinh}(2\rho) (du_+^2-du_-^2) 
\nonumber\\
&&\qquad\qquad\qquad\qquad\qquad- 2\, {\rm sin}(2z) \,{\rm sinh}(2\rho)\, du_+\, du_- \big]\,,
\nonumber
\eeqa
with $\kappa=c_\mt{R}/\wrr^2$. Of course, this metric agrees with eq.~\eqref{metric-ds} when we choose $c_\mt{R}=\wrr^2$, \ie with $A_\mt{R}\propto\id$. Recall that up to this point, $c_\mt{R}$ was a spurious parameter but with the above construction, it plays an essential role in defining the geometry. In particular, if we were to adopt this approach, we would have to restrict our attention to $c_\mt{R}\ge0$ to produce a sensible geometry.}
\begin{eqnarray}
\labell{metric-ds}
ds^2&=& \delta_{IJ}\,\text{tr}(dUU^{-1}M^T_I)\,\text{tr}(dUU^{-1}M^T_J)\\
&=& 2\, dy^2 + 2\, d\rho^2 + 2\, dx^2 - 4\, {\rm cosh}(2\rho) dx\, dz + 2\, {\rm cosh}(4\rho) dz^2
\nonumber\\
  & &\qquad+\, e^{-2y} \big[  {\rm cosh}(2\rho) (du_+^2 + du_-^2) - {\rm cos}(2z) \,{\rm sinh}(2\rho) (du_+^2-du_-^2) 
\nonumber\\
&&\qquad\qquad\qquad\qquad\qquad- 2\, {\rm sin}(2z) \,{\rm sinh}(2\rho)\, du_+\, du_- \big]\,.
\nonumber
\end{eqnarray}

An intuitive cost function in this context is the $F_2$ norm \reef{function_F}, which becomes
  \begin{equation}
  \label{eq:distance}
  {\cal D}_{2} \left(U\right) = \int_0^1 \!\!\!ds\ \sqrt{g_{ab} \,\dot{x}^a\, \dot{x}^b}\,,
  \end{equation} 
\ie this simply corresponds to evaluating the geodesic distance in geometry defined by eq.~\reef{metric-ds}. However, as was argued in~\cite{Jeff} (see also \cite{Chapman:2017rqy}), this cost function does not reproduce the expected UV divergences found in holographic complexity \cite{Carmi}.  However, this situation can be remedied by using the $\kappa=2$ cost function \reef{function_Fkappa}, which corresponds to
  \begin{equation}
  \label{eq:distance2}
  {\cal D}_{\kappa=2} \left(U\right) = \int_0^1 \!\!\!ds\ g_{ab} \,\dot{x}^a\, \dot{x}^b\,.
  \end{equation} 
Of course, from a physicist's perspective, this can be seen as the action of a test particle moving in the same geometry and so it yields the same extremal trajectories. We will also consider two alternative cost functions in section \ref{sec:F1}, the $F_1$ and Schatten measures, but in the following we will focus on finding the circuits that minimize the distance~\eqref{cost_D} using the cost functions~\eqref{eq:distance} and~\eqref{eq:distance2}.

Now the complexity is  given by the shortest unitary connecting the reference and target state, \ie
${\cal C}\left(A_\mt{T}\right) = {\rm min}_U {\cal D}\left(U\right)$ with
  \begin{equation}
  \label{bound4}
 A_\mt{T}= U(\s=1)\,A_\mt{R}\,U^T(\s=1)  \qquad
  {\rm and}\qquad U(\s=0)= \id\,.
  \end{equation}
With the cost functions in eqs.~\reef{eq:distance} and \reef{eq:distance2}, this corresponds to finding a geodesic from the origin in the geometry \reef{metric-ds} to the point corresponding to the desired transformation $U(\s=1)$.  However, as we described for the transformation in eq.~\reef{eq:A-target}, the target state is independent of one of the coordinates in $U$ or alternatively, the reference state is invariant under a family of transformations (known as the stabilizer group, \eg see \cite{Hackl:2018ptj,therm0}). Hence for any target state $A_\mt{T}$, there exists a one-parameter family of transformations satisfying the boundary conditions in eq.~\reef{bound4}. Thus, there is a one-parameter family of geodesics connecting the reference state to the target state and the complexity will be determined by the length of the shortest geodesic in this family.
  
For simplicity, we describe the determination of the geodesics in terms of extremizing eq.~\reef{eq:distance2}, which takes the form of a particle action with Lagrangian
\beqa
  {\cal L}_0 & =& 2\, \dot{y}^2 + 2\, \dot{\rho}^2 + 2\,( \dot{x} -  {\rm cosh}(2\rho)\, \dot{z})^2 + 2\, {\rm sinh}^2(2\rho)\, \dot{z}^2 \labell{eq:L2}\\
  &+& e^{-2y} \left(  {\rm cosh}(2\rho) (\dot{u}_+^2 + \dot{u}_-^2) - {\rm cos}(2z) \,{\rm sinh}(2\rho) (\dot{u}_+^2-\dot{u}_-^2) - 2\, {\rm sin}(2z) \,{\rm sinh}(2\rho)\, \dot{u}_+\, \dot{u}_- \right)\,.\nonumber
\eeqa
We solve the resulting `equations of motion' analytically for simpler target states in section~\ref{sec:simple} and provide numerical solutions for general target states of the form~\eqref{eq:A-target1} in section~\ref{sec:numerics}.

\subsection{Solving for simple geodesics}
\label{sec:simple}

While we were not able to find analytic solutions for the geodesics to arbitrary target states~\eqref{eq:A-target1}, for some simpler set of target states, the optimal path between the reference and target states remains in a $\mathbb{H}^2 \times \mR$ slice of the full geometry \eqref{metric-ds}. We begin by describing these simple geodesics which have an analytic solution. In section~\ref{sec:numerics}, we confirm numerically that these are indeed globally the shortest geodesics for the particular target states of interest.

First, we can use the freedom to reparametrize $s$ in the cost function \eqref{eq:distance} to fix 
\beq
k=\sqrt{g_{ab} \,\dot{x}^a\, \dot{x}^b}
\label{kk4}
\eeq
where $k$ is a (positive) constant. This means that when evaluated for the optimal trajectory, the complexity with this cost function is simply given by $\mC_2=k$. Similarly, with the $\kappa=2$ cost function \eqref{eq:distance2}, the complexity is given by $\mC_{\kappa=2}=k^2$. 

Now to identify simple geodesics, we begin by looking at the equations of motions for $x(s)$ and $z(s)$:
\beqa
		0 &=& \partial_s\left( \dot{x} - {\rm cosh}(2\rho) \dot{z} \right)\,,
\label{eom-x}\\
0 &=& \partial_s \left(2{\rm cosh}(4\rho) \dot{z} - 2 {\rm cosh}(2\rho) \dot{x} \right)
\label{eom-z}\\
&&\qquad\qquad + e^{-2y}\, {\rm sinh}(2\rho)\left(2 \cos(2z) \dot{u}_+\dot{u}_--\sin(2z)(\dot{u}_+^2 - \dot{u}_-^2) \right)\,.
\nonumber
\eeqa
Now if $\dot{u}_\pm=0$ (\eg if we were simply preparing the vacuum state as in \cite{Jeff}), these equations would be solved by $x(s)=\bar{x}$ and $z(s)=\bar{z}$, \ie setting both of these coordinates to be constant along the trajectory. These are the geodesics that we will focus on below.

To pick appropriate values for $\bar{x}$ and $\bar{z}$, we must examine the boundary conditions. 
The initial boundary condition $U(s=0)=\id$ and comparing to eqs.~\reef{oblique} and \reef{GL2_matrix} then gives 
\beq
x_0=z_0\,,\qquad \rho_0=0\,, \qquad y_0=0\,, \qquad u_{\pm 0}=0\,,
\label{init}
\eeq
where the subscript notation indicates  $y_{a0}=y_a(s=0)$.\footnote{We will use a similar notation for the final boundary conditions at $s=1$.} The first restriction implies that we must choose $\bar{x}=\bar{z}$ for our simple geodesics. Similarly for the final boundary conditions, comparing \eqref{eq:A-target1} and \eqref{eq:A-target}, we see that $\sin(2x_1)=0$ is required to produce a  target state which is unentangled in normal mode basis. Hence this end point condition requires $\sin(2\bar{x}) = 0$, \ie $\bar{x}=n\pi/2$. Combining these conditions for the final state from eq.~\eqref{eq:A-target} (with $c_\mt{R}=0$) gives at $s=1$,
\beqa
A(s=1)&=&U(s=1)\,A_\mt{R}\, U^T(s=1)
\label{eq:A-target-special}\\
& =&\wrr^2\,\left(
\begin{array}{ccc}
e^{2 (y_{1}+\rho_1)} & 0 & e^{y_{1}+\rho_1} u_{+1} \\
0 & e^{2  (y_{1}-\rho_1)}  & e^{y_{1}-\rho_1}u_{-1} \\
e^{y_{1}+\rho_1} u_{+1}\ \  & e^{y_{1}-\rho_1}u_{-1} &\ \  u_{+1}^2+u_{-1}^2 \\
\end{array}
\right)\,,
\nonumber
\eeqa
where implicitly we have assumed $\cos(2\bar{x}) = + 1$ (\ie $\bar{x}=0$ or $\pi$). For the case $\cos(2\bar{x}) = - 1$ (\ie $\bar{x}=\frac\pi2$ or $\frac{3\pi}2$), we simply interchange $y_{1}+\rho_1
\leftrightarrow y_{1}-\rho_1$ in the above state. To simplify the following, we will proceed with the analysis assuming that $\cos(2\bar{x}) = 1$. 

With these choices, the $z$ equation \reef{eom-z} reduces to
\begin{equation}
e^{-2y} {\rm sinh}\left(2\rho\right) \dot{u}_+\dot{u}_- = 0\,.
\end{equation}
Together with the initial boundary conditions \reef{init}, this is satisfied for $\rho=0$ or $u_+=0$ or $u_-=0$. The first of these possibilities is inconsistent with the final boundary condition if the normal mode frequencies given in eq.~\reef{Phybasis} are not equal, \ie $\Omega\ne0$. Hence we must choose one of the latter two possibilities. That is, the consistency of our simple geodesics (with constant $x$ and $z$) demands that we only shift {\it one} of the normal modes to produce either a nonvanishing expectation value $\langle x_+\rangle$ or $\langle x_-\rangle$, but not both!\footnote{Of course, with more general geodesics, we are able to prepare states where both 
$\langle x_+\rangle$ and $\langle x_-\rangle$ are nonvanishing, as we will examine in section \ref{sec:numerics}.} We continue our discussion here with the choice $u_-(s)=0$, \ie we consider states with $\langle x_+\rangle\ne0$ and $\langle x_-\rangle=0$.

To ensure that the choice $\bar{x} = \bar{z}$, ${\rm sin}(2\bar{z})= 0$ and $u_-=0$ is still a geodesic of the full geometry~\eqref{metric-ds}, we verify that the equation of motion for $u_-$ is satisfied, \ie
\begin{equation}
\partial_s \left(e^{-2(y-\rho)} \dot{u}_- \right)=0\,,
\end{equation}
which is indeed the case. The induced geometry on the slice given by these choices becomes
\begin{equation}
\label{eq:metric3d2}
ds^2 = dy^2_+  + e^{-2y_+} du_+^2 + dy^2_-\,,
\end{equation}
where we have introduced $y_\pm=y\pm\rho$.  
Our analysis has guaranteed that finding geodesics $(y_+(s), y_-(s), u_+(s))$ in the induced metric~\eqref{eq:metric3d2} will give us geodesics $(y_+(s),$ $y_-(s)$, $x= n\pi$, $z = n \pi$, $u_+=0=u_-)$ in the full six-dimensional geometry described by eq.~\eqref{metric-ds}. It is straightforward to see that the three-dimensional geometry~\eqref{eq:metric3d2} is the direct product of two-dimensional hyperbolic space (covered by $y_+$ and $u_+$) with the real line (covered by $y_-$). Since two components of this geometry are maximally symmetric, it would be straightforward to evaluate the distance between any two points in this geometry using standard formulae. However, it is useful for us to have explicit expressions describing the geodesics and so we proceed by evaluating the equations of motion in the $\mathbb{H}^2 \times \mathbb{R}$ geometry.

Of course, from eq.~\reef{init}, the initial conditions for the geodesics are simply: $y_{+0}=0=y_{-0}=u_{+0}$. To determine the final boundary conditions, we begin with eq.~\reef{eq:A-target-special} for the final state, which simplifies to
\begin{equation}\label{eq:A-target-special-1}
\begin{split}
A(s=1)& =\wrr^2\,\left(
\begin{array}{ccc}
e^{2 y_{+1}} & 0 & e^{y_{+1}} u_{+1} \\
0 & e^{2 y_{-1}}  & 0 \\
e^{y_{+1}} u_{+1} & 0 & u_{+1}^2 \\
\end{array}
\right)\,.
\end{split}
\end{equation}
Requiring that this state matches the target state $A_\mt{T}$ with $a_-=0$ in eq.~\eqref{eq:A-target1} yields the following boundary conditions at $s=1$:
\begin{equation}
\label{eq:boundary-cond}
y_{+1} = \frac{1}{2} \log  \w_+\,,\quad y_{-1} = \frac{1}{2} \log  \w_-\,, \quad u_{+1} =- \sqrt{\w_+}\,\a_+\,,
\end{equation}
where for convenience, we are using the following dimensionless ratios:\footnote{Of course, $\a_\pm$ are the same quantities which already appeared in eq.~\reef{eq:A-target1}.}
\beq
\w_\pm = \frac{m \omega_\pm}{\wrr^2}\qquad
{\rm and}\qquad
\a_\pm = \frac{a_\pm}{x_0}\,.
\label{dimless}
\eeq

Now to find the path which these geodesics follow, we extremize the cost function (either eq.~\reef{eq:distance} or \reef{eq:distance2}) subject to the restriction that the motion is only in the three-dimensional subspace described by eq.~\eqref{eq:metric3d2}. Each of the three equations of motion can be integrated to yield the following first order equations
\begin{equation}\label{HR_solution}
\begin{aligned}
\dot{y}_+ =A-B\,u_+\,,\quad \dot{u}_+=B\,e^{2y_+}\,,\quad
\dot{y}_-=C\,,
\end{aligned}
\end{equation}	
where the three integration constants correspond to the velocities at $s=0$, \eg $\dot{y}_+|_{s=0} =A$.
These equations can be integrated and after imposing the initial conditions, the solution becomes
\beq
\label{eq:solution}
y_+(s)=\frac{1}{2}\log\Big(\frac{\Delta^2}{B^2}\text{sech}^{2}(\alpha(s))\Big)\,,\quad
u_+(s)=\frac{\Delta}{B}\tanh(\alpha(s))+\frac{A}{B}\,,\quad
y_-(s)= C s\,,
\eeq
where $\Delta = \sqrt{A^2+B^2}$ and $\alpha(s)=\Delta s-\text{arctanh}(\frac{A}{\Delta})$. Furthermore, the final conditions~\eqref{eq:boundary-cond} fixes the integration constants as
 \begin{equation}\label{solution2}
 \begin{split}
 A &= \frac{\a_+^2\w_+ +\w_+ -1}{\sqrt{(\a_+^2\w_+ +\w_+ -1)^2+4\a_+^2 \w_+}}\, \text{arccosh}\!\(\frac{\a_+^2\w_+ +\w_++1}{2 \sqrt{\w_+}}\)  \,,\\
  B &= \pm 2 \sqrt{\frac{\a_+^2\w_+}{(\a_+^2\w_+ +\w_+ -1)^2+4\a_+^2 \w_+}}\, \text{arccosh}\(\frac{\a_+^2 \w_+ +\w_++1}{2 \sqrt{\w_+}}\)  \,,\\
  C &= \frac 12 \log \w_- \,,
 \end{split}
 \end{equation}
where the sign of $B$  is chosen to match  the sign of $u_{+1}$ (\ie the opposite sign as $\a_+$).\footnote{Here, we assume the definition of `arccosh' is such that it always yields a positive result.} Further, it follows that
\beqa 
\Delta 
&=&\text{arccosh}\(\frac{\a_+^2 \w_+ +\w_++1}{2 \sqrt{\w_+}}\) \labell{DELTA}\\
&=& \log\!\[\frac{(\a_+^2\w_++\w_++1)+ \sqrt{(\a_+^2\w_++\w_++1)^2-4\w_+}}{2\sqrt{\w_+}} \]\,.
\nonumber
\eeqa
For these simple geodesics, the constraint \reef{kk4} reduces to  
\begin{equation}\label{kk5}
k_\mt{sim}^2=\dot{ y}_+^2 + e^{-2y_+} \dot{u}_+^2+ \dot{y}_-^2  = \Delta^2+C^2 \,.
\end{equation}

As will be shown in section~\ref{sec:numerics}, these geodesics are indeed the shortest ones connecting the reference state \reef{eq:A-ref}
to the target states \eqref{eq:A-target1} with $a_-=0$ in the full geometry. Therefore eq.~\reef{kk5} yields the complexity of the coherent states with the $F_2$ and $\kappa=2$ cost functions, \ie $\mC_2=k_\mt{sim}$ and $\mC_{\kappa=2}=k_\mt{sim}^2$. As a check, one can easily verify that in the limit $\a_+ \to 0$, this result \reef{kk5} yields the complexity of the ground state found in~\cite{Jeff}, \ie
\begin{equation}
\label{eq:d-a0}
\mC_{\kappa=2,\mt{vac}}=k^2_\mt{sim}\big|_{\a_+ \to 0}= \frac{1}{4} \[\left(\log  \w_+ \right)^2 + \left(\log  \w_- \right)^2\]\,.
\end{equation}

As expected, the difference in complexity between the coherent states and the ground state comes from the normal mode which has been translated ($x_+$ in this case). It is interesting to examine the difference in various limiting cases.\footnote{In the following, we focus on the complexity for the ${\kappa=2}$ cost function rather that the $F_2$ measure. There are two motivations for doing so: First, the $\kappa=2$ complexity reproduces the expected UV divergences of holographic complexity as was found in~\cite{Jeff}. Second, as we will see in section~\ref{sec:complexity}, the change in $F_2$ complexity $\Delta{\cal C}_2$ vanishes when generalizing our results to free scalar field theory. In contrast, the change in $\kappa=2$ complexity $\Delta {\cal C}_{\kappa=2}$ remains finite when generalizing to field theory.}  That is, let us consider the asymptotic behavior of
\beqa
\Delta\mC_{\kappa=2}&=& \mC_{\kappa=2}-\mC_{\kappa=2,\mt{vac}}=\Delta^2-\frac14\left(\log  \w_+ \right)^2
\labell{complex8}\\
&=&\(\log\!\[\frac{(1+\a_+^2\w_++\w_+)+ \sqrt{(1+\a_+^2\w_++\w_+)^2-4\w_+}}{2\sqrt{\w_+}} \]\)^2-\frac14\left(\log  \w_+ \right)^2\,.
\nonumber
\eeqa
Expanding for small $|\a_+|$, eq.~\reef{complex8} yields
\begin{equation}\label{smalla}
\Delta\mC_{\kappa=2} =  \log (\w_+) \,\frac{\w_+\, \a_+^2}{\w_+-1}  + \[1+\frac{\log (\w_+)\,(\w_+ +1)}{2\(\w_+-1\)}   \] \(\frac{\w_+\, \a_+^2}{\w_+-1}\)^2+ {\cal O}(\a_+^6)\,,
\end{equation}
while for large $|\a_+|$, we find
\begin{equation}\label{biga}
\Delta\mC_{\kappa=2} = (\log  \a_+^2)^2 + \[ \log  (\w_+) + 2\frac{\w_++1}{\w_+\, \a_+^2}\]\log  \a_+^2   +
\log  (\w_+) \frac{\w_++1}{\w_+\, \a_+^2}
+ {\cal O}\!\(\frac{\log  \a_+^2}{ \a_+^{4}}\)\,.
\end{equation}
There is no divergent term for $\w_+=1$ where the above expansion for small $\a_+$ doesn't apply and the change of complexity  is simplified as 
\begin{equation}
\Delta\mC_{\kappa=2}\(\w_+=1\) = \( \text{arccosh}\(\frac{\a_+^2  +2}{2}\) \)^2\,.
\end{equation}

\subsection{Numerical results in full geometry}
\label{sec:numerics}

In this section, we describe our results for numerical solutions of the geodesic equations. Our approach was to derive the second order differential equations from the variation of eq.~\reef{eq:L2} and then use the pseudo-spectral method where Chebyshev polynomials were used in the $s$ direction. Combining the Dirichlet boundary conditions at $s=0$ and $s=1$, the solutions can be uniquely determined. One subtlety is that in the initial conditions \reef{init}, the value of $x_0=z_0$ is not fixed. However, with our method, this parameter is easily fixed by scanning through a range of values for $x_0$ and demanding that the solution is smooth in the vicinity of $s=0$. 

 \begin{figure}[htbp]
 	\centering
 	\hspace{-13pt}
     \vspace{-20pt}
 	\subfigure{\includegraphics[width=3in]{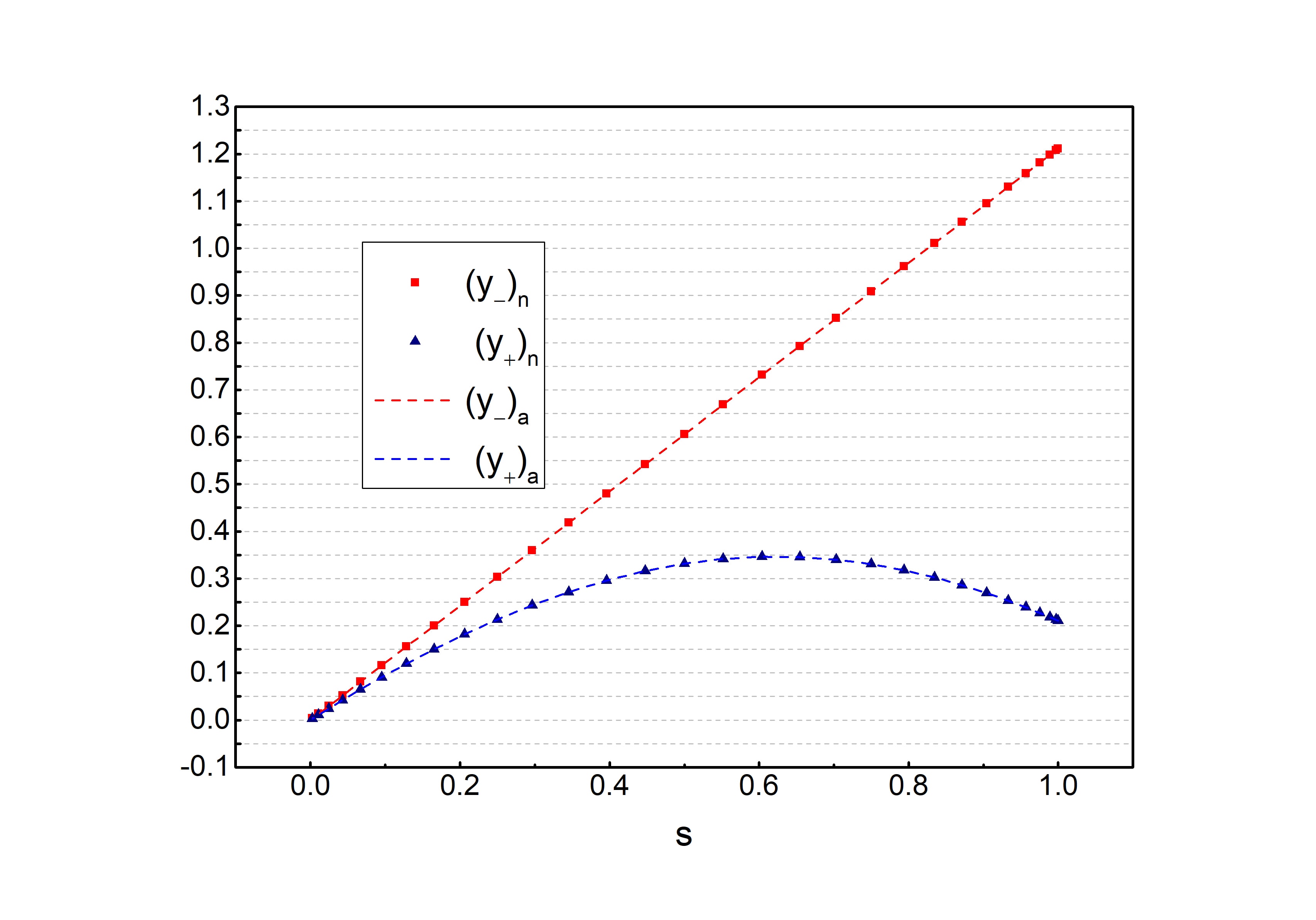}
 	\hspace{-20pt}
 	\includegraphics[width=3in]{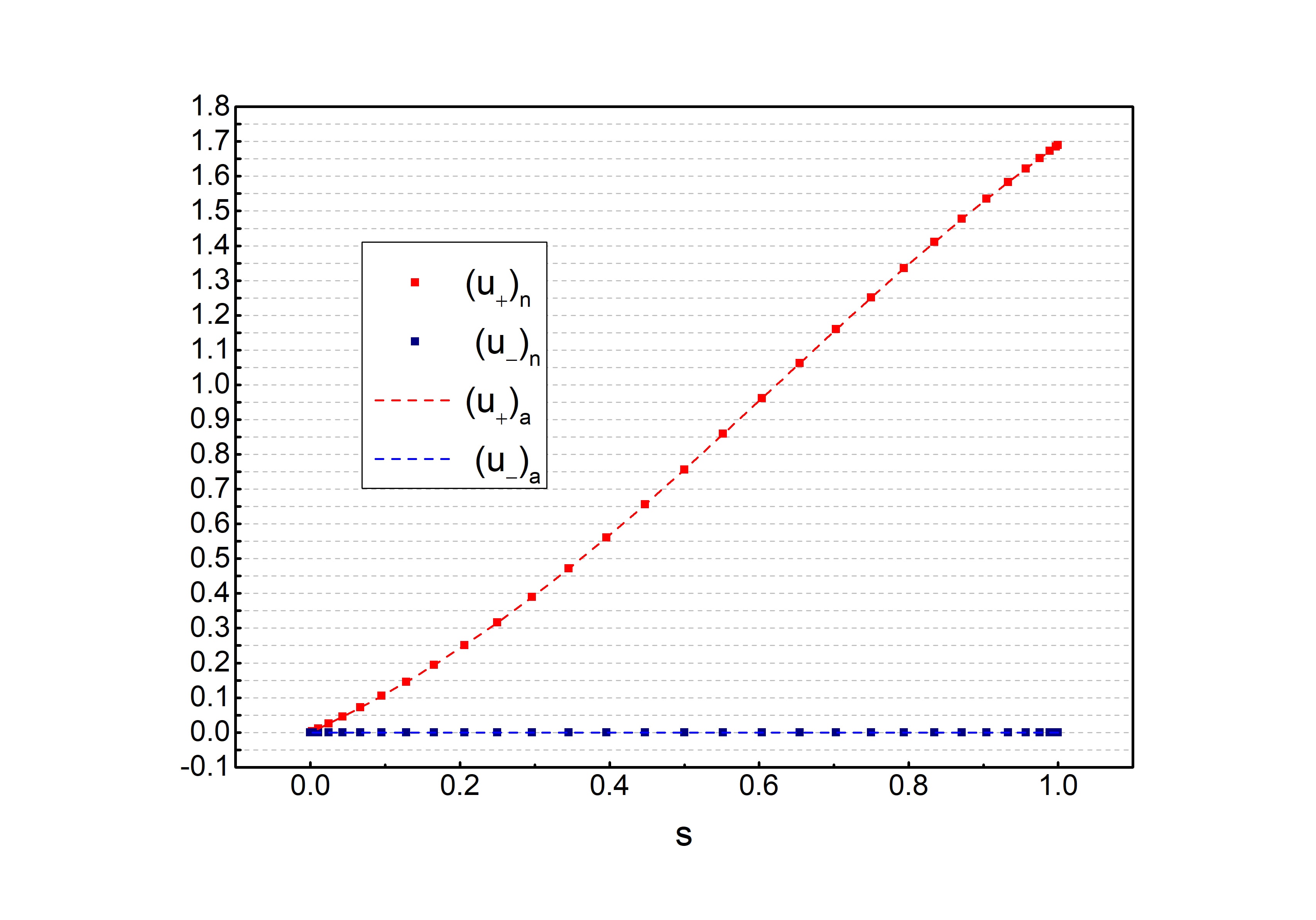}}
 	\hspace{-0pt}
     \vspace{-0pt}
 	\subfigure{\includegraphics[width=3in]{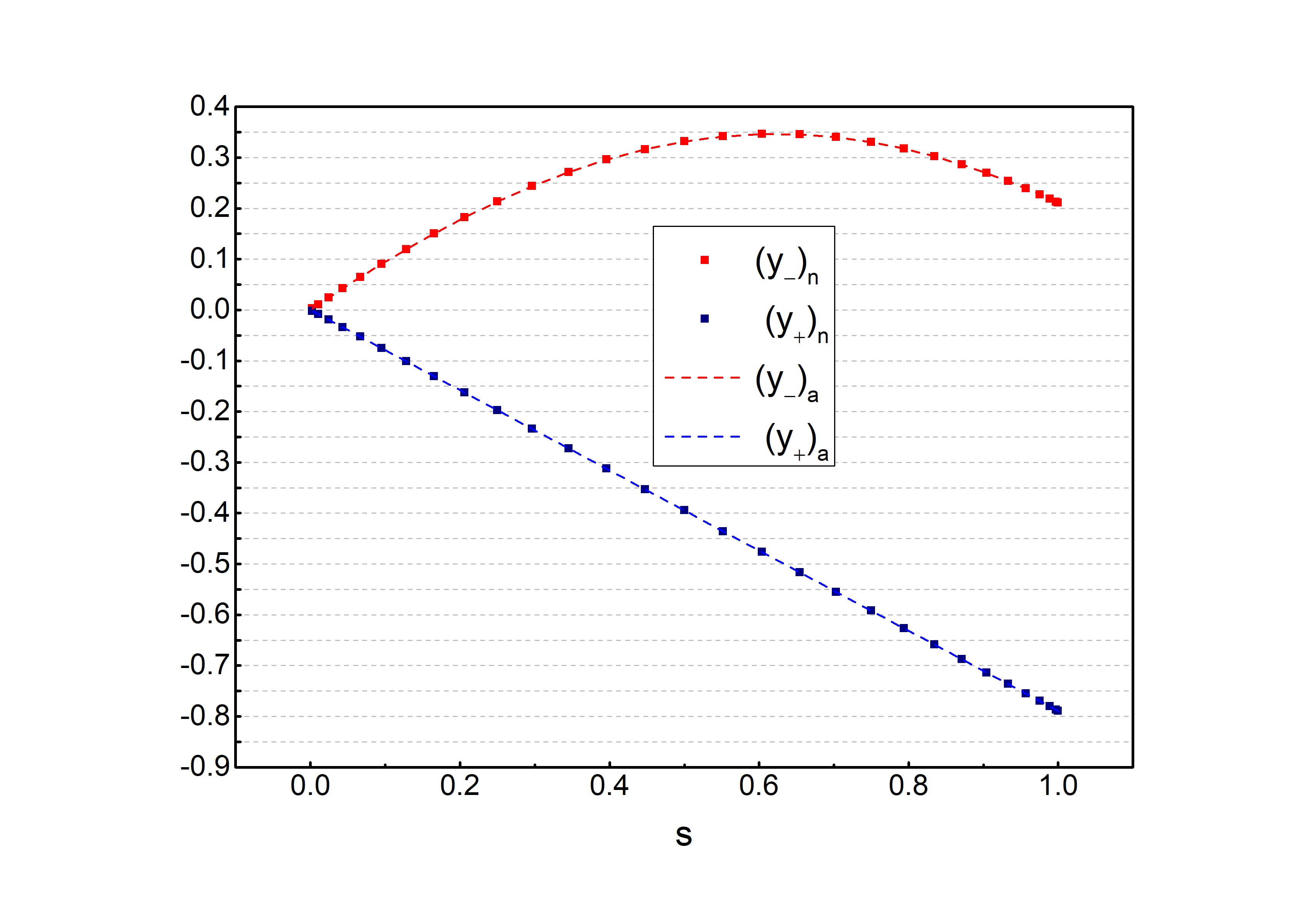}
 	\hspace{-20pt}
 	\includegraphics[width=3in]{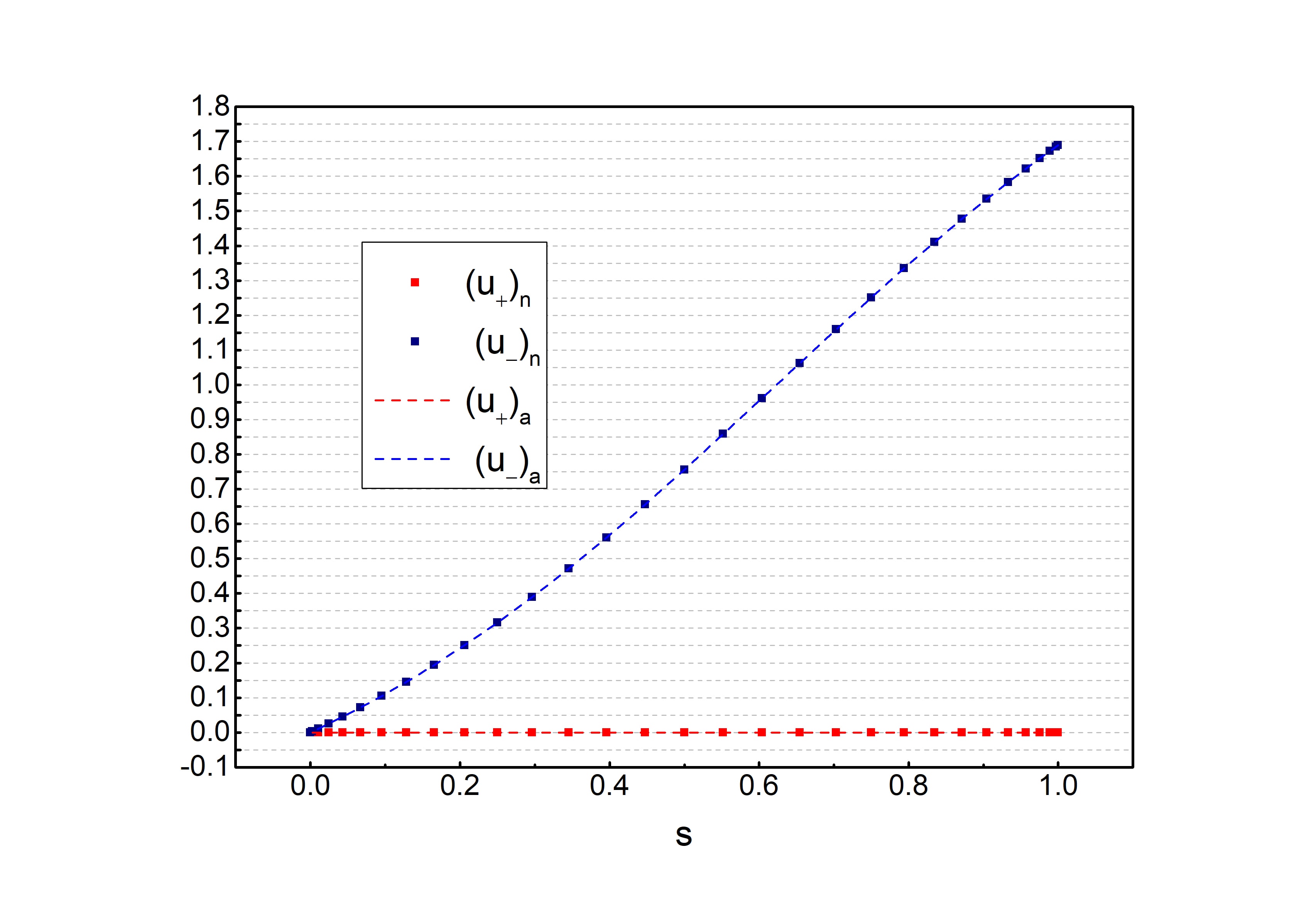}}
 	\setlength{\abovecaptionskip}{-15pt}
    \setlength{\belowcaptionskip}{0pt}
 	\caption{Comparing the numerical solutions to analytic solutions for the simple geodesics \reef{eq:solution}. The top two graphs show the geodesic ending at $y_{+1}=0.211,\ y_{-1}=1.211,\ u_{+1}=1.690,\ u_{-1}=0 $, while the bottom two graphs show the geodesic ending at $y_{+1}=-0.790,\ y_{-1}=0.211,\ u_{+1}=0,\ u_{-1}=1.690 $. These values were chosen to produce simple values for $\Lambda_+$ and $\Lambda_-$, \ie $\Lambda_+=\Lambda_-=1$. The subscripts \lq\lq{}n\rq\rq{} and \lq\lq{}a\rq\rq{} are used to indicate the numerical and analytical solutions, respectively. }\label{cu}
 \end{figure}
Our first application was to verify our numerical results by comparing them with the analytic solutions for the simple geodesics found in the previous section. An example is shown in figure \ref{cu}. As expected, if $u_+$ ($u_-$) ends at zero, it remains zero along the entire trajectory, and the scale coordinate $y_-$ ($y_+$) follows a straight line. The other two coordinates follow curved paths, as expected from eqs.~\eqref{eq:solution} and \reef{solution2}. In every case, we found excellent agreement between the numerical and the analytic solutions.

\begin{figure}[t]
	\centering
	\subfigure{\includegraphics[width=3in]{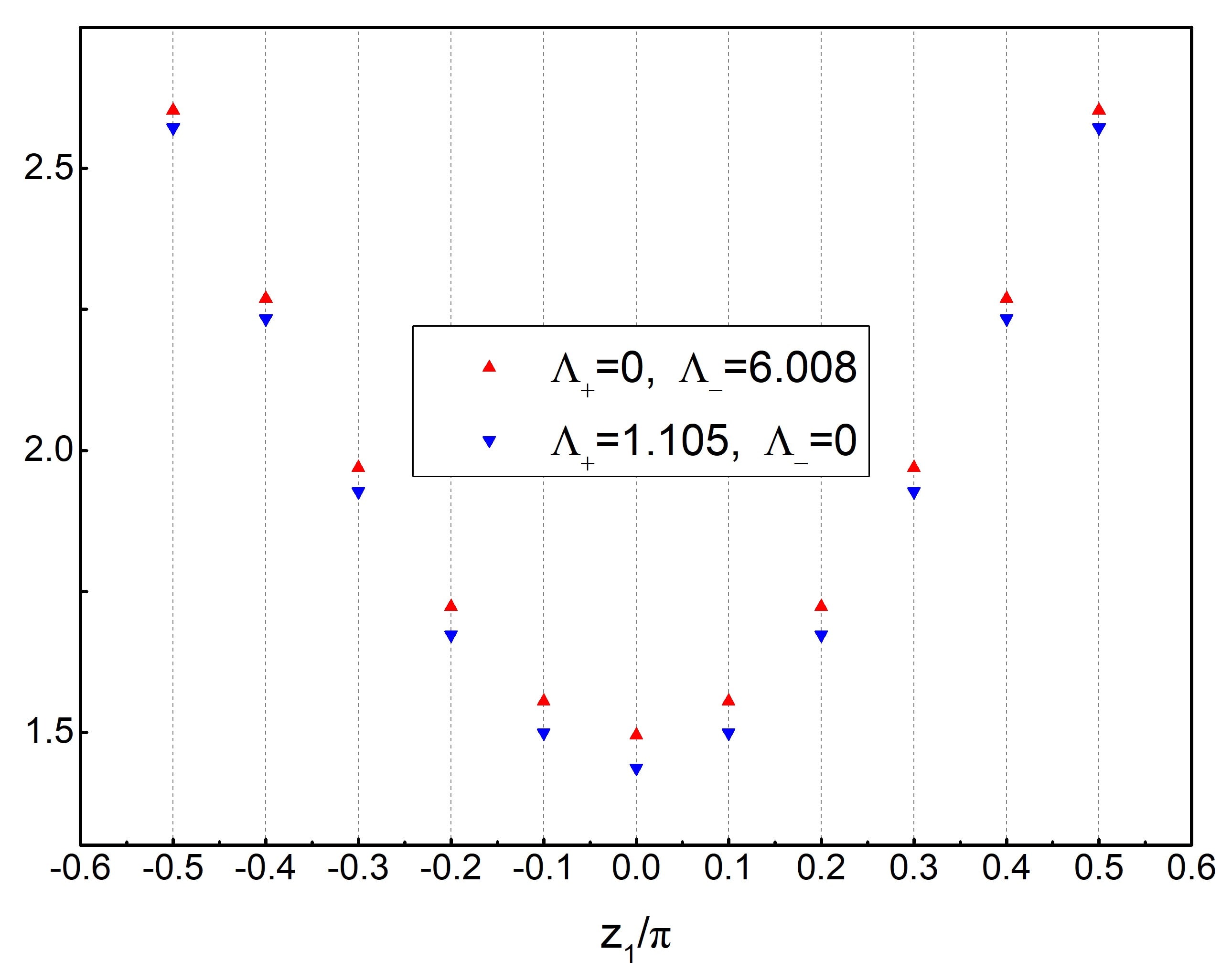}}
	\setlength{\abovecaptionskip}{-4pt}
    \setlength{\belowcaptionskip}{0pt}
	\caption{Lengths of a family of geodesics ($k$) connecting to two target states, in which a single mode is excited, for different final values of the $z$ angle. The red upper triangles represent geodesics reaching the state with $ y_{+1}=0.1,\ y_{-1}=1.1,\ \Lambda_{+}= 0,\  \Lambda_{-}=6.008$. The blue lower triangles are for $y_{+1}=0.1,\  y_{-1}=1.1,\  \Lambda_{+}=1.105,\ \Lambda_{-}=0$. In both cases, the minimum value arises at $z=0$, \ie the optimal geodesic corresponds to one of the simple geodesics found in the previous section.
}\label{ol}
\end{figure}

Next we considered the family of geodesics connecting the reference state to a specific target state with $a_{-1}=0$ (or $a_{+1}=0$), as shown in figure \ref{ol}. Recall that as described below eqs.~\reef{eq:A-target} and \reef{split}, the final state was independent of $z_1$ (as long as the final values $u_{\pm1}$ were rotated appropriately). In the figure, we see that the shortest geodesic is that for which $z_1 = 0$. That is, for all the examples that we examined, our numerics confirm that the optimal geodesics correspond to the simple geodesics derived in the previous section. Hence these numerical studies provide strong evidence for the claim that the simple geodesics are indeed the shortest ones for the target states in which  only one of normal modes is shifted.
 \begin{figure}[h]
 	\centering
 	\hspace{-10pt}
 	\subfigure{\includegraphics[width=2.3in]{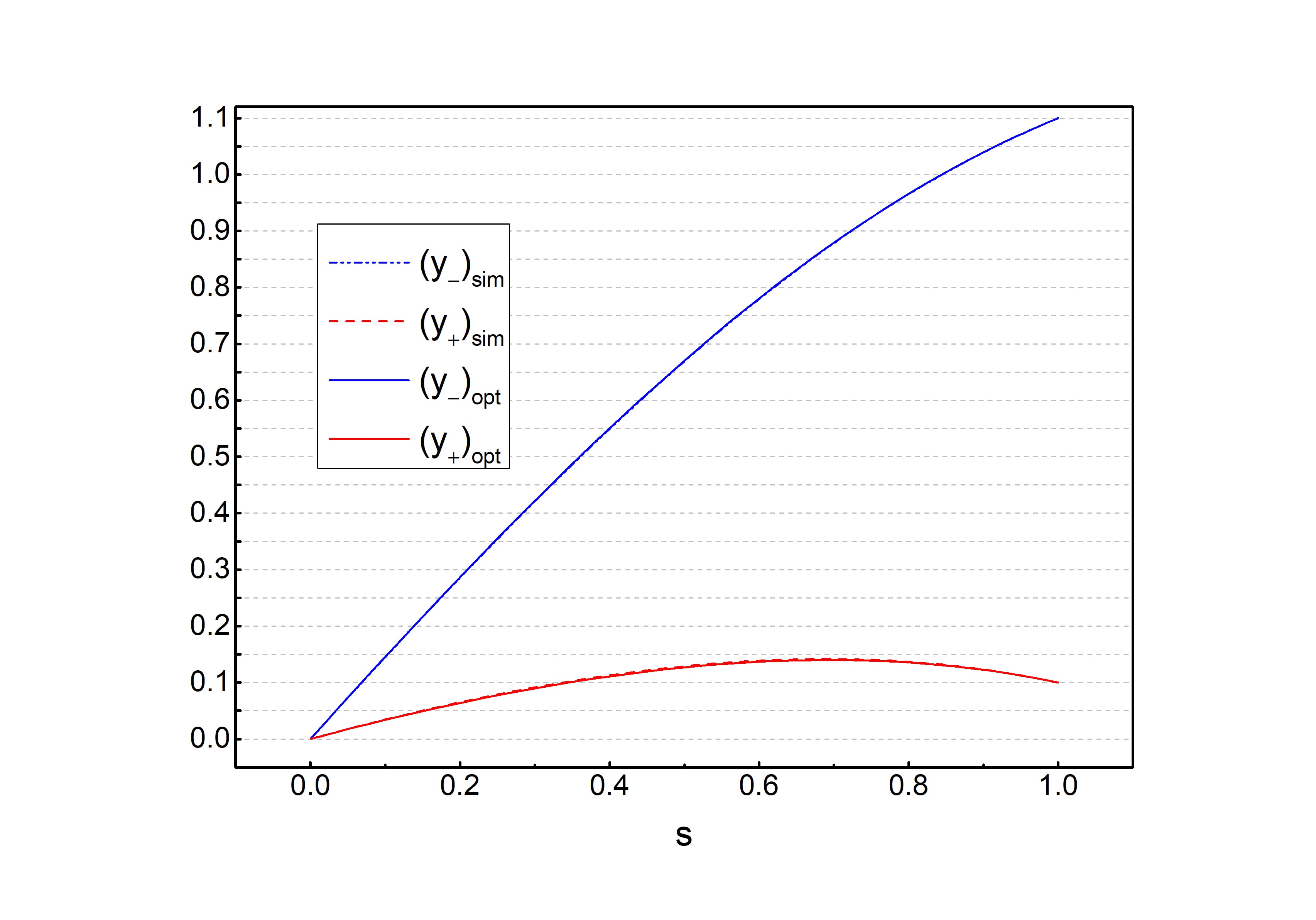}
 	 \hspace{-25pt}
 	\includegraphics[width=2.3in]{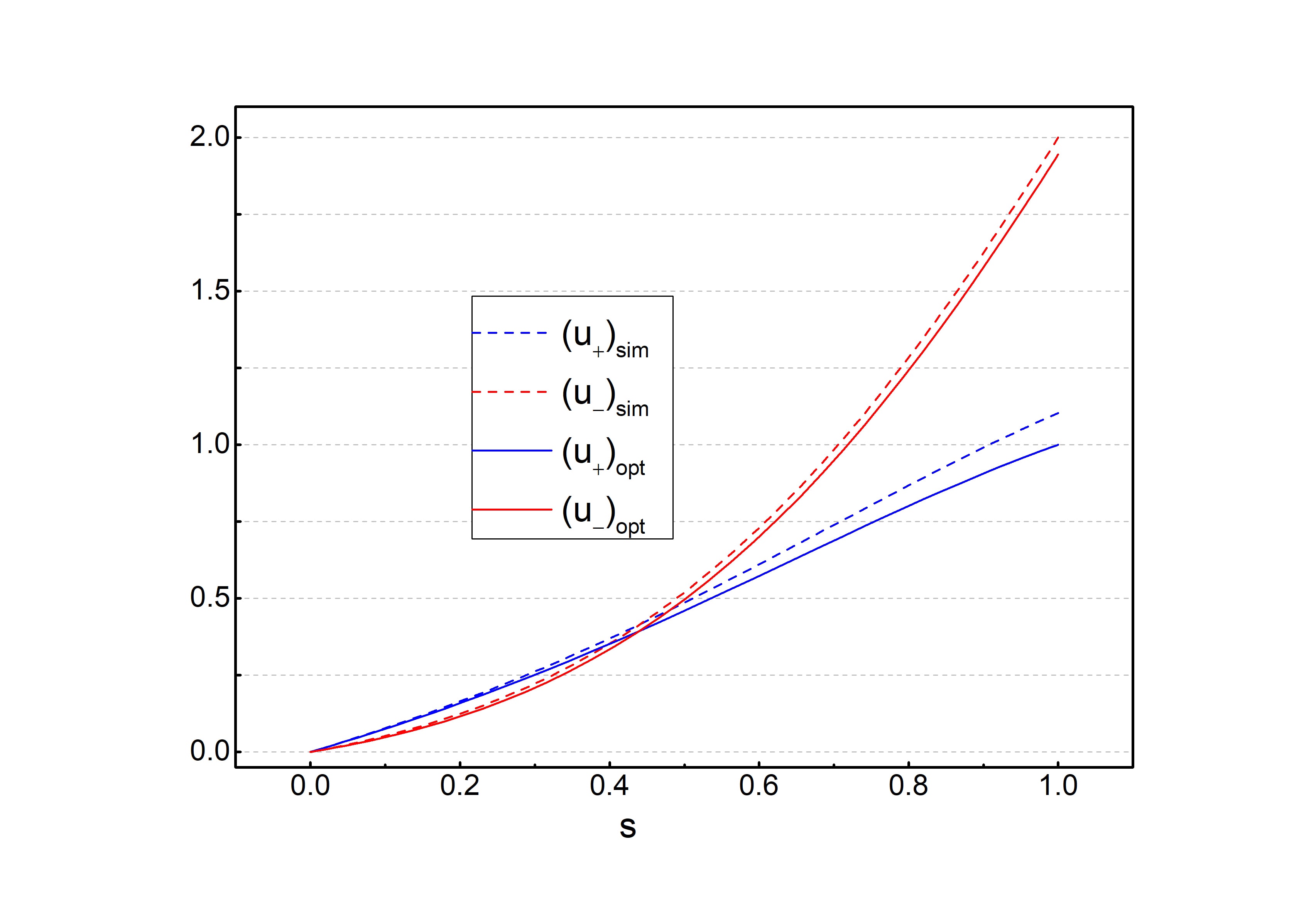}
 	 \hspace{-25pt}
 	\includegraphics[width=2.3in]{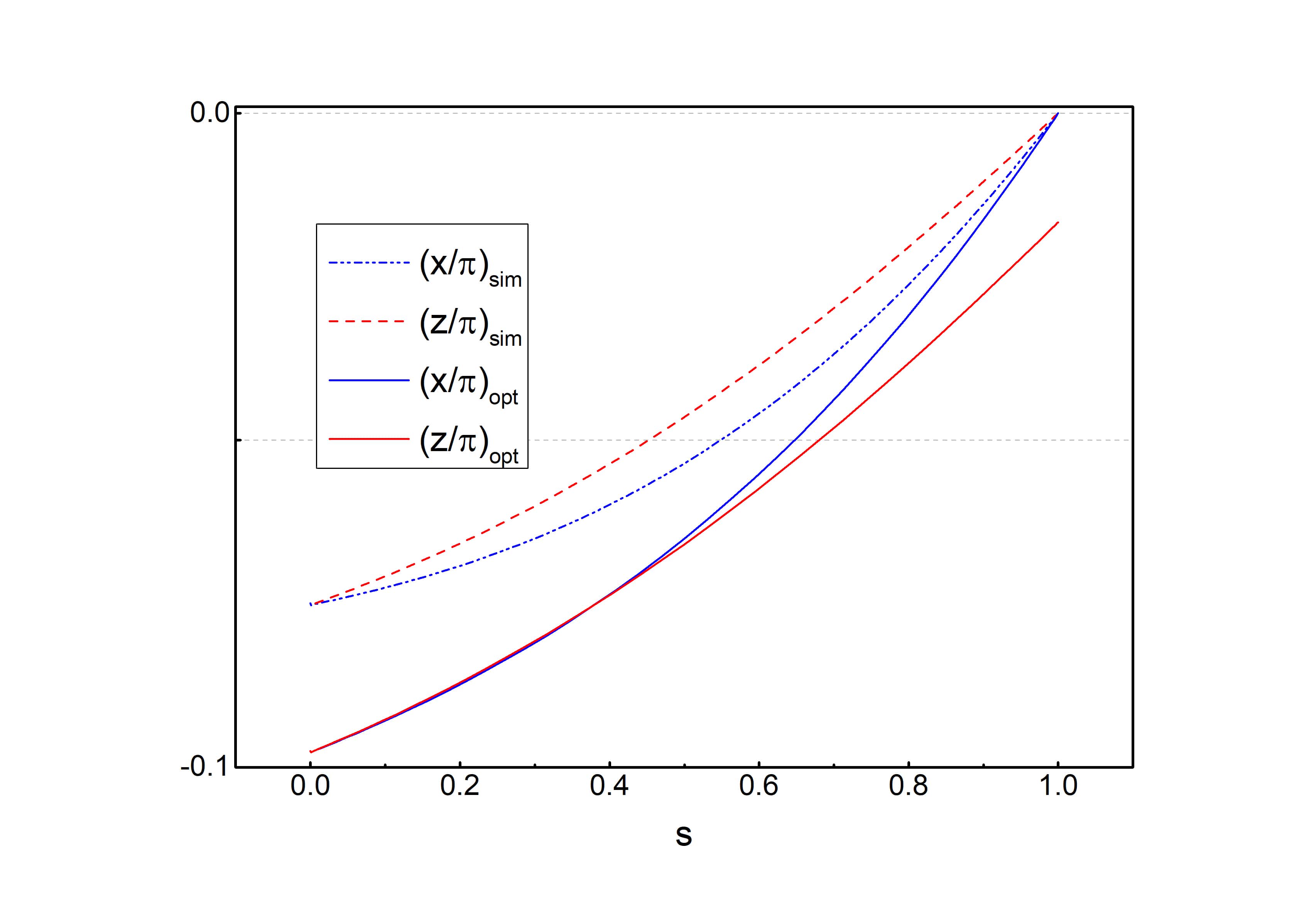}}
 	\setlength{\abovecaptionskip}{-23pt}
    \setlength{\belowcaptionskip}{0pt}
 	\caption{Example of geodesics preparing a target state with both $a_\pm$ nonvanishing. We compare the optimal geodesic and the \lq\lq{}simpler\rq\rq{} geodesic with $z_1=0$. In this example, both geodesics have the final boundary conditions: $ y_{+1}=0.1,\ y_{-1}=1.1, \ \Lambda_{+}=1.105,\  \Lambda_{-}=6.008$. For the optimal geodesic, we also have $x_0=z_0=-0.0976\pi, x_1=0, z_1=-0.0167\pi$, while for the \lq\lq{}simpler\rq\rq{} one, $x_0=z_0=-0.0749\pi, x_1=0= z_1$. Note that $y_{\pm}$ essentially coincide in both geodesics, as shown in the far left panel.}\label{gs}
 \end{figure}
 
 Figure \ref{gs} shows an example of the optimal geodesic to a target state with both $a_+$ and $a_-$ non-zero. In this situation, we do not have an analytic solution and we can see in our numerical solution that these geodesics do not take a simple form, \eg none of the coordinates follow a straight path. Similar to the previous discussion, to determine the optimal geodesic, we vary $z(s=1)$ while keeping the final state \reef{eq:A-target} fixed, evaluate the lengths of the corresponding geodesics and then choose the trajectory with the minimal length. Let us note in passing that generally these optimal geodesics pass through regions where $x(s)$ and $\rho(s)$ are nonvanishing. Therefore even though both the initial \eqref{eq:A-ref} and final \eqref{eq:A-target1} states are unentangled,  the intermediate states (all) along the optimal trajectory are entangled when preparing a target state with {both} $a_+$ and $a_-$ are nonvanishing -- see further discussion in section \ref{discuss}

\begin{figure}[h]
 	\centering
 	\subfigure{\includegraphics[width=3in]{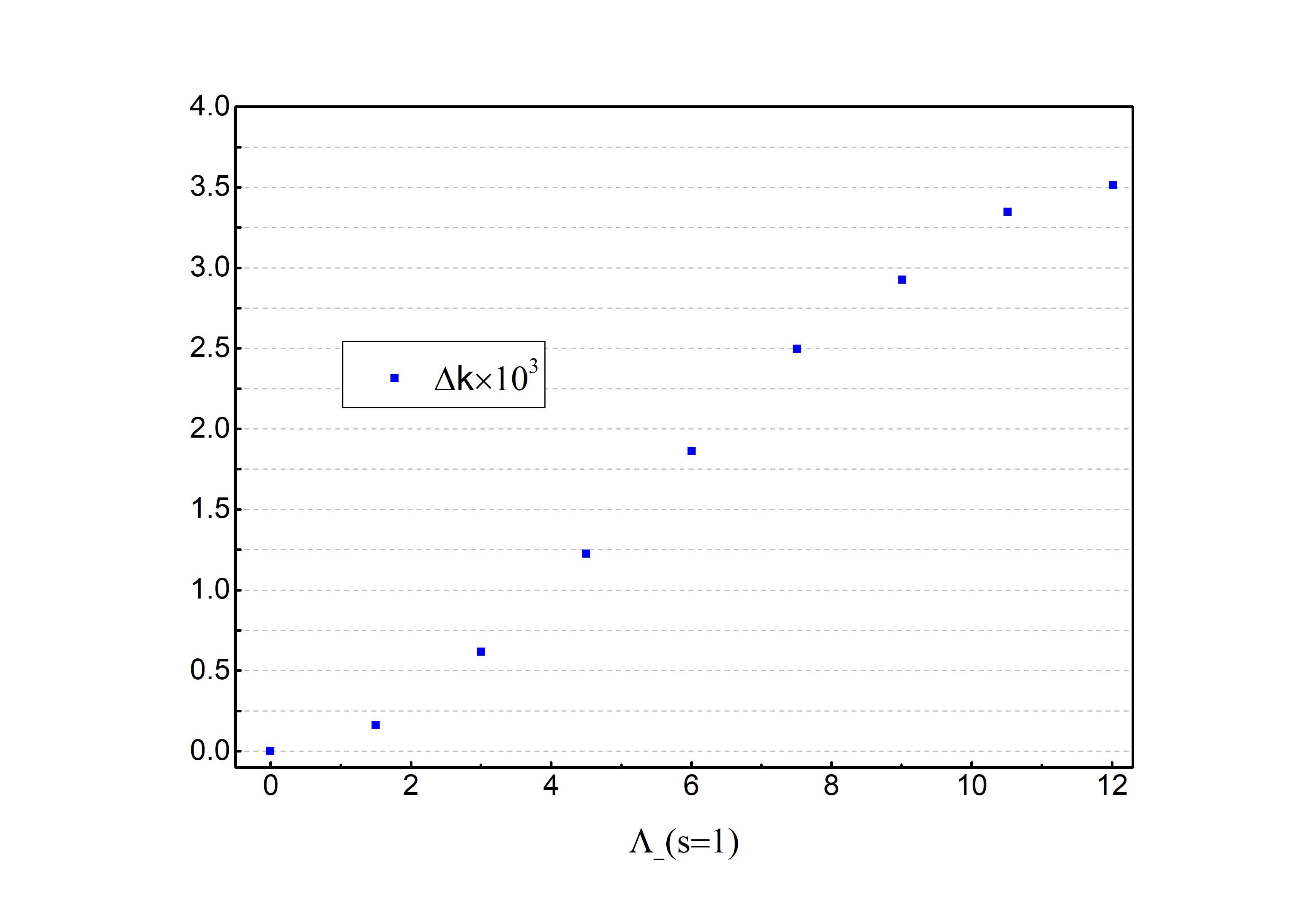}
 	\includegraphics[width=3in]{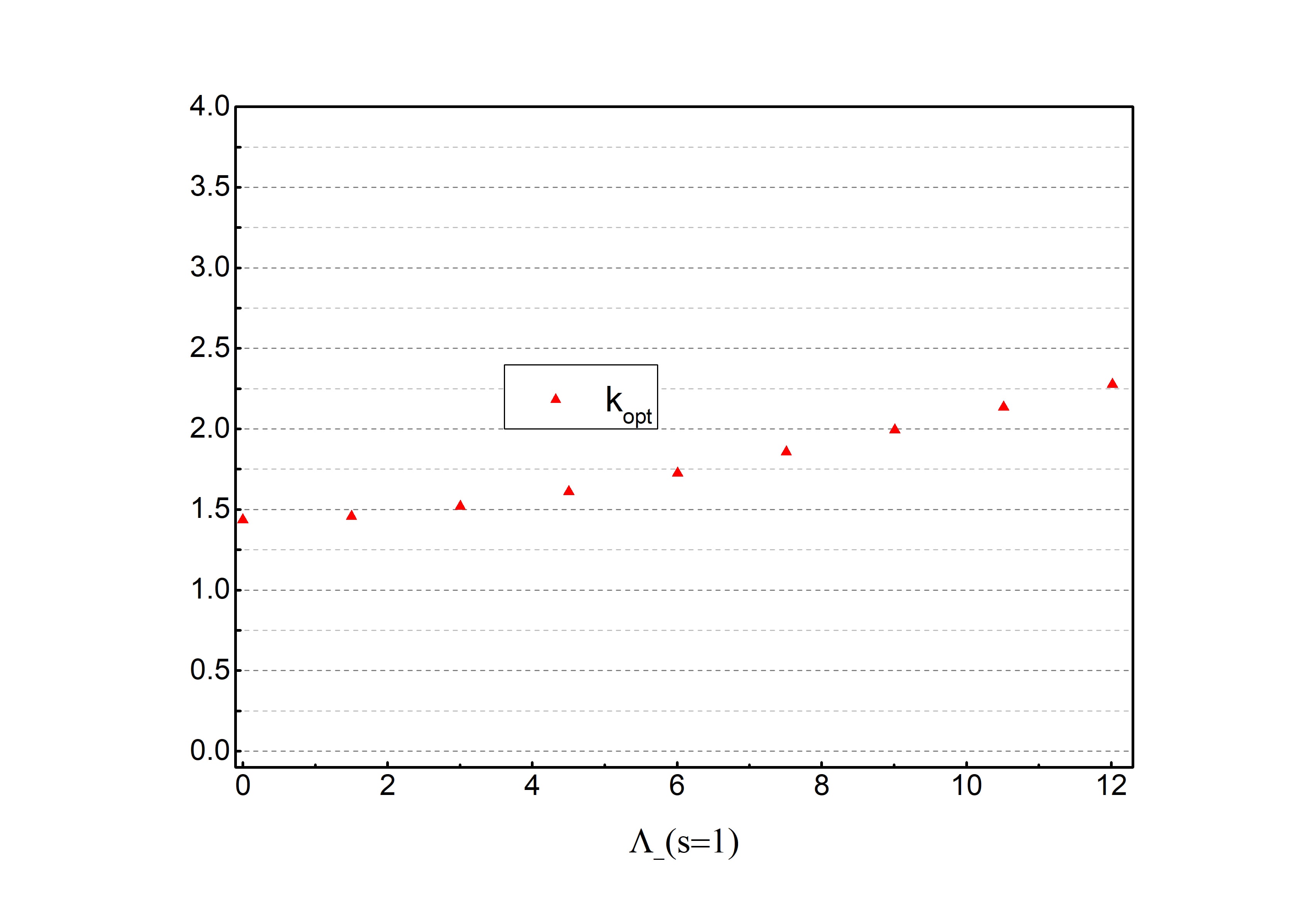}}
 	\setlength{\abovecaptionskip}{-22pt}
    \setlength{\belowcaptionskip}{0pt}
 	\caption{A comparison of the lengths of the optimal geodesic and the ``simpler'' geodesic with $z_1=0$.  $\Delta k= k_{sim} -k_{opt}$ and $k_{opt}$ are shown as functions of $\Lambda_{-}(s=1)=-\frac{m\omega_+}{x_0\wrr^2}\,a_-$. This example is characterized by the boundary conditions $ y_{+1}=0.1,\ y_{-1}=1.1$ and $\ \Lambda_{+1}=1.105$,  while $\Lambda_{-}(s=1)$ varies from 0 to 12.017. We note that while $\Delta k$ grows as $|a_-|$ increases, it represents at most an increase of $~0.2\%$ over $k_{opt}$ for the geodesics shown here.  }\label{deltak}
 \end{figure}
Recall that for the simple geodesics with only a single excitation, we found $x(s)=0=z(s)$. In contrast, when both normal modes are excited in the target state, the optimal geodesic has  nonvanishing profiles for both $x(s)$ and $z(s)$, as illustrated in figure \ref{gs}. While at the final point, $x_1=0$ in order to ensure that the target state is unentangled, as can be inferred from eq.~\reef{eq:A-target}, in general we have $z_1\ne0$ for the optimal geodesic. For comparison purposes, we can also consider the geodesic with $z_1=0$, which we will denote as the ``simpler'' geodesic, which is also shown in figure \ref{gs}. There we can see that the biggest difference between these two geodesics is in the profiles of $x(s)$ and $z(s)$. In fact, the profiles for $y_\pm(s)$ are indistinguishable in the figure. It is also interesting to compare the length of these geodesics, which we do in figure \ref{deltak}. There we introduce the new parameter $\Delta k=k_{sim}-k_{opt}$,\footnote{Of course, $\Delta k>0$ because the optimal geodesic is the shortest geodesic.} In the figure, we show the results for $\Delta k$ for geodesics where the boundary conditions are fixed as in figure \ref{gs} except that $\Lambda_{-}$ varies from 0 to 12.017. With $\Lambda_{-}=0$, $\Delta k=0$ because the two geodesics coincide with the simple geodesics found analytically in the previous section. However, we see in the figure that as $a_-$ increases, the difference in lengths increases monotonically. 
  	
\subsection{Small excitations}
\label{sec:small}

We began in section \ref{sec:simple} by considering simple geodesics for states where only one normal mode is excited. Then in section \ref{sec:numerics}, we applied numerical techniques to examine the geodesics for target states where both normal modes are excited. In particular, we noted that the resulting geodesics are driven away from the space of states with no entanglement between the two normal modes. While we cannot find the geodesics for these general target states analytically, we can at least find the leading order contributions to the length of the geodesics for small shifts, \ie when both $\a_\pm$ are nonvanishing but $|\a_\pm|\ll1$. To examine this situation, we consider small perturbations from the optimal geodesics connecting the reference state to ground state. It was already shown in~\cite{Jeff} that the optimal circuit connecting 
the reference state \reef{eq:A-ref} to the ground state,
\begin{equation}
   \label{eq:grounder}
   A_\mt{T}= m\,
   \begin{pmatrix}
   \omega_+ & 0 & 0 \\ 0 &  \omega_- & 0 \\ 0 & 0 & c_\mt{T}
   \end{pmatrix}\,, 
   \end{equation}
is the `straight line' circuit:
\beq
U(s) = {\rm exp}\left[\frac{s}2\({\rm log}\,\w_+\,M_{++}+{\rm log}\,\w_-\, M_{--}\)\right]\,.
\label{optimal0}
\eeq
In terms of the six-dimensional geometry, the corresponding geodesic is given by
\beq\label{groungeo}
y_+(s)=y_{+1}\,s\,,\quad
y_-(s)=y_{-1}\,s\,,\quad x(s)=0=z(s)=u_+(s)=u_-(s)\,,
\eeq
where $y_{+1}=\frac{1}2\log \w_+$ and $y_{-1}=\frac{1}2\log \w_-$, with $y_\pm=y\pm\rho$ as before. The $\kappa=2$ complexity \eqref{eq:distance2} is then given by the expression in eq.~\reef{eq:d-a0}.

Now we want to evaluate the leading order change to the above circuit depth \reef{eq:d-a0}
evaluated with eq.~\reef{eq:distance2} when we introduce small shifts for both normal modes, \ie $\a_+\,, \a_-\sim {\cal O}(\varepsilon)$. In particular, the final boundary conditions are then modified for $u_\pm$ but it will be true that $u_\pm(s)\,,
\dot{u}_\pm(s)\sim {\cal O}(\varepsilon)$ all along the new geodesic. This follows because the second line in~\eqref{eq:L2} is positive definite, so having $\dot{u}_\pm={\cal O}(1)$ would increase the length by an ${\cal O}(1)$ factor. The $x$ and $z$ equations of motion take the form
\beqa
	0 &=& \partial_s\left( \dot{x} - {\rm cosh}(2\rho) \dot{z} \right)\,,\\
	0 &=& \partial_s \left(2{\rm cosh}(4\rho) \dot{z} - 2 {\rm cosh}(2\rho) \dot{x} \right) + e^{-2y} {\rm sinh}(2\rho)\left(2 \cos(2z) \dot{u}_+\dot{u}_--\sin(2z)(\dot{u}_+^2 - \dot{u}_-^2) \right)\,,
	\nonumber
\eeqa
but this implies that $x,\,\dot{x},\,z,\,\dot{z} \sim {\cal O}(\varepsilon^2)$. Now if we expand the cost function, \ie determine the leading corrections to eq.~\reef{eq:L2}, we find \begin{equation}
{\cal L}_0 = \dot{y}_+^2 + \dot{y}_-^2 + e^{-2y_+} \dot{u}_+^2 + e^{-2y_-} \dot{u}_-^2 + {\cal O}(\varepsilon^4)\,.
\end{equation}
Effectively, the modified geodesics move on a four-dimensional submanifold of the full geometry \reef{metric-ds} which takes the form $\mathbb{H}^2\times \mathbb{H}^2$. Hence to leading order, the complexity becomes 
\beq
\mC_{\kappa=2} =\Delta_+^2+\Delta_-^2+{\cal O}(\varepsilon^4)
\label{nuis}
\eeq
where $\Delta_+$ is the expression in eq.~\reef{DELTA} and $\Delta_-$ is the same expression after substituting $\w_+\to\w_-$ and $\a_+\to\a_-$. The leading order change in the complexity then becomes 
\beqa
\Delta\mC_{\kappa=2}&=& \mC_{\kappa=2}-\mC_{\kappa=2,\mt{vac}}=\Delta_+^2+\Delta_-^2-\frac14\left(\log  \w_+ \right)^2-\frac14\left(\log  \w_- \right)^2+{\cal O}(\varepsilon^4)
\nonumber\\
&=&  \frac{|\log \w_+|}{|\w_+-1|} \,\w_+\a_+^2 +  \frac{|\log  \w_-|}{|\w_--1|}\,\w_- \a_-^2+{\cal O}(\varepsilon^4)\,,
\labell{complex88}
\eeqa
where we are dropping terms of the form $\a_+^4$, $\a_-^4$ and $\a_+^2\,\a_-^2$. The key result here is that the leading order corrections to complexity factorize into contributions from the individual normal modes, \ie there are no second order contributions involving the cross-term $\a_+\a_-$. 

We would like to go further and so that in fact, these geodesics on the effective $\mathbb{H}^2\times \mathbb{H}^2$ geometry are optimal, \ie that we are correctly evaluating the leading corrections to the complexity in eq.~\reef{complex88}. We can argue for a proof by contradiction of this result as follows: Imagine that we find a geodesic where the deviations of $x(s)$ and $z(s)$ from the straight-line geodesic \reef{groungeo} are the same order as $\a_\pm$, \ie $x,\,z \sim {\cal O}(\varepsilon)$. Examining the cost function, we see that the motion in $x$ and $z$ will introduce a strictly positive contribution of order $\varepsilon^2$, from the terms in the first line of eq.~\reef{eq:L2}. Similarly the second and third terms in the second line will make contributions of $\cO(\varepsilon^4)$ and $\cO(\varepsilon^3)$, respectively. There is no definite sign of these contributions but being higher order, they will never be able to make up for the $\cO(\varepsilon^2)$ increase generated by moving in the $x$ and $z$ directions. One might consider even stronger deviations, \eg $x,\,z \sim {\cal O}(1)$, but then the $\dot x,\,\dot z$ terms in eq.~\reef{eq:L2} will only increase the length of the geodesic by order one while $\dot u_\pm$ terms will still only contribute at order $\varepsilon^2$. 

\begin{figure}[t]
	\centering
	\subfigure{	 \hspace{-20pt}
	\includegraphics[width=2.1in]{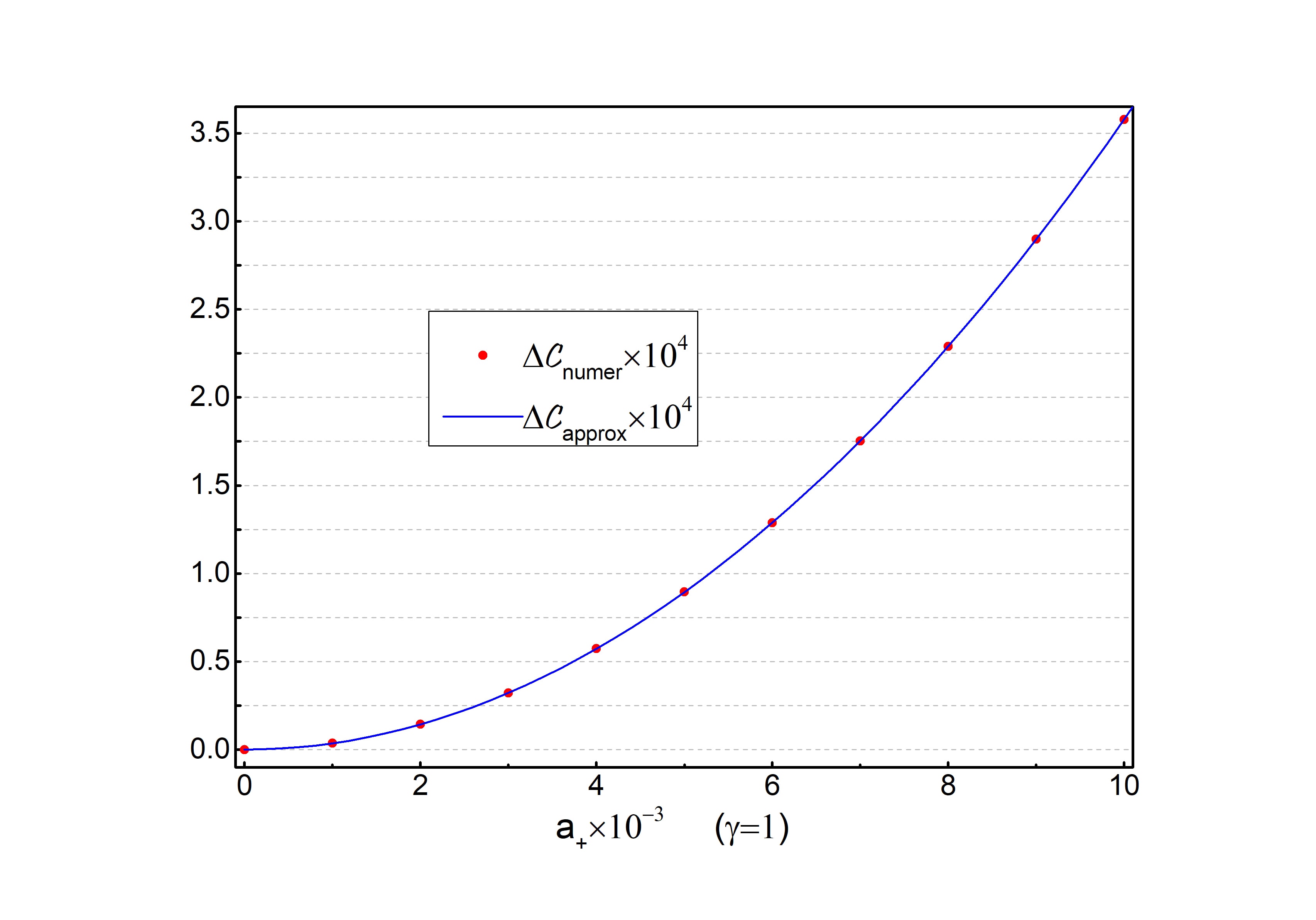}
		 \hspace{-20pt}
	\includegraphics[width=2.1in]{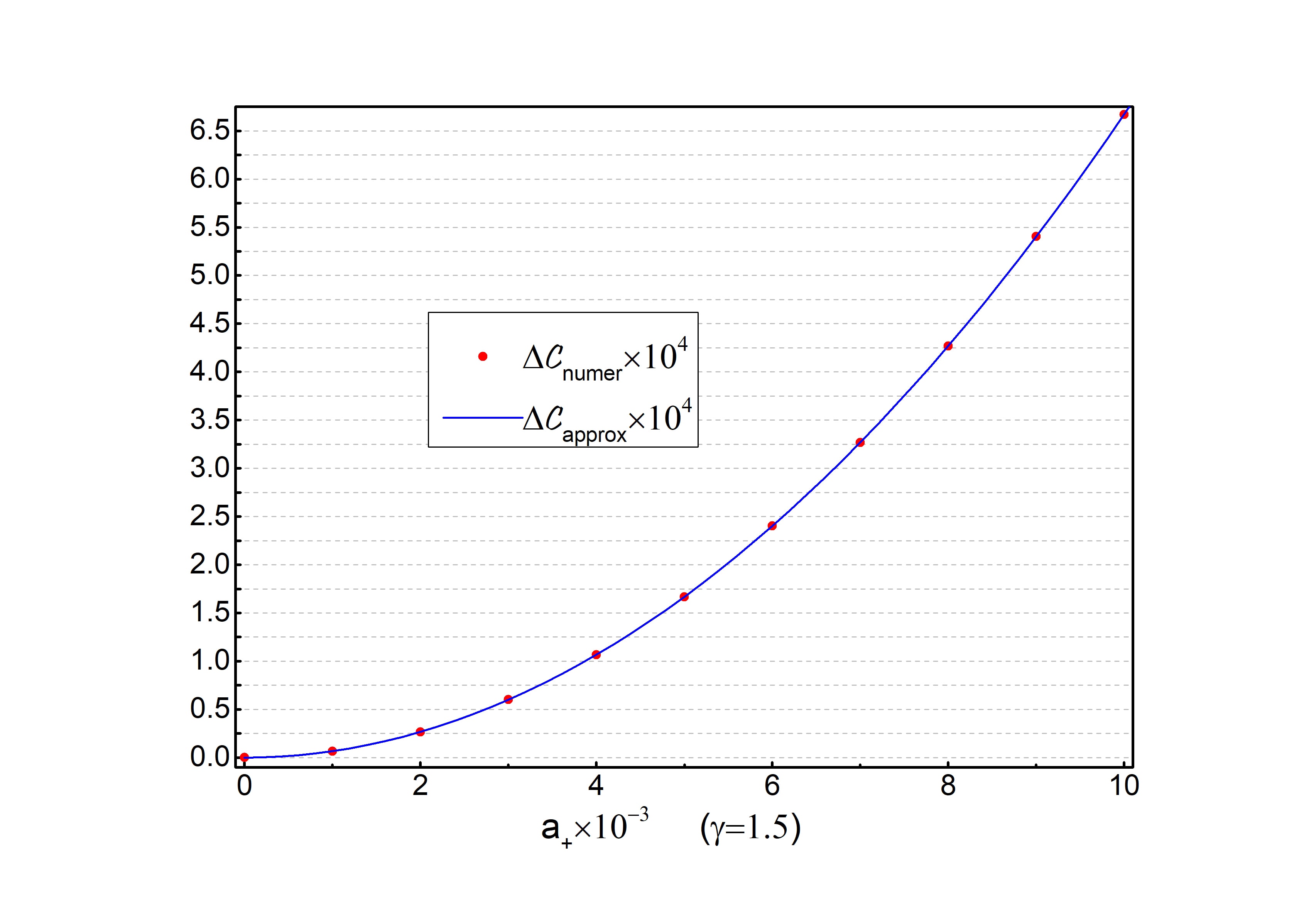}
		 \hspace{-20pt}
	\includegraphics[width=2.1in]{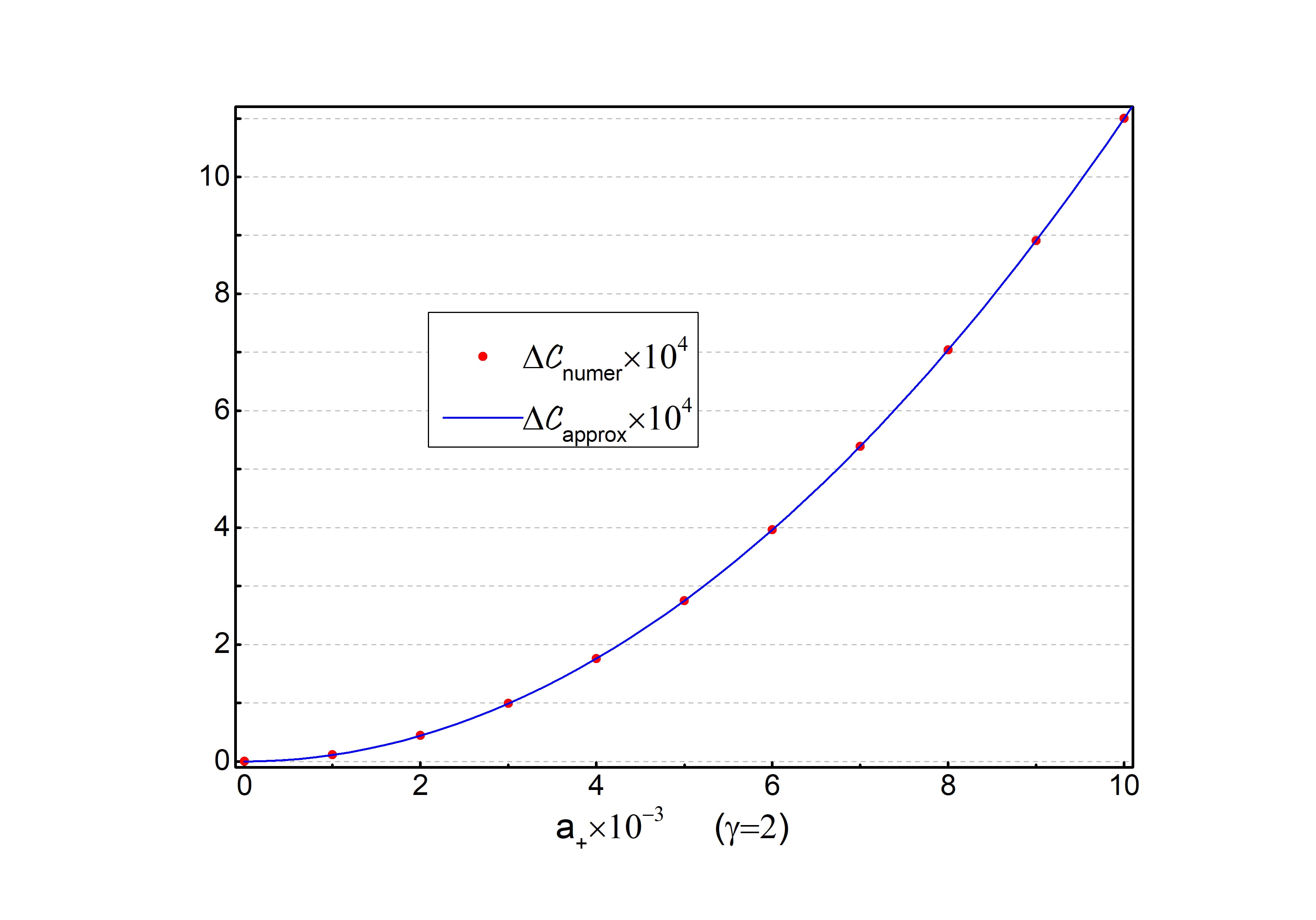}}
	\setlength{\abovecaptionskip}{-10pt}
    \setlength{\belowcaptionskip}{0pt}
	\caption{Comparison between the quadratic approximation for small excitations, \ie the second line of eq.~\eqref{complex88}, and the complexity found numerically for various target states. In the figures, we choose $\w_+=1.221$ and $\w_-=9.025$ and $a_-=\gamma a_+$. We let $a_+$ range from $0$ to $0.01$ and then compare the two results with $\gamma=1$, 1.5 and 2. The three figures show excellent agreement between the quadratic approximation and the true complexity for $\a_\pm \ll 1$. }
	\label{fig:small_a}
\end{figure}

We can also use the numerical approach developed in the previous section to find evidence that eq.~\reef{complex88} correctly provides the leading corrections to the complexity. In particular, we looked at families of states where $a_-=\gamma a_+$ with $\gamma$ being some fixed numerical constant. As $a_+$ increased from zero, we found that the numerical results matched the approximation provided in the second line of eq.~\reef{complex88} when $\a_\pm \ll 1$ in all of the cases that we examined. Figure~\ref{fig:small_a} provides an example of our numerical analysis.

\section{Complexity with alternate cost functions}
\label{sec:F1}

In \cite{Jeff},  the UV divergences in complexity of the ground state of the free scalar field were compared to those in holographic complexity (see also \cite{Chapman:2017rqy}). In particular, it was found that the $F_1$ cost function in eq.~\reef{function_F} gave the most promising comparison. In particular, the leading divergence for the $F_1$ cost function took the form $V/\delta^{d-1}\,\log(\ell/\delta)$ where $\ell$ is some undetermined length scale. This is precisely the same form as the leading divergence in holographic complexity evaluated with the CA proposal \cite{Carmi}. However, an apparent shortcoming of the $F_1$ cost function is that the complexity depends on the basis used for the gates, \eg the results will change upon rotating between the physical basis and the normal mode basis. However, in \cite{Hackl:2018ptj}, it was suggested that we  would recover the same essential results of the $F_1$ measure using the Schatten norm (with $p=1$ --- \eg see \cite{bhatia2013matrix,watrous2018theory} and further details below). The advantage of the Schatten norm is that the results are basis independent. Hence in the following, we will re-examine the complexity of the coherent states \reef{target1} for the system of two coupled harmonic oscillators introduced in section \ref{sec:gates} using these two alternatives for the cost function.  	
  	
\subsection{$F_1$ cost function} \label{costf1}

First, we turn to the task of studying the $F_1$ cost function introduced in eq.~\reef{function_F}
\be\label{Fone}
F_1(U,Y)=\sum_I \left|Y^I\right|\,,
\ee
that is, we want to study the circuits $U(s)$ that optimize the cost function
\begin{equation}\label{cost_D2}
{\cal D}_{1}(U(s))= \int^1_0 ds ~ \sum_I \left|Y^I\right|\,.  
\end{equation}
However, we re-iterate that this measure is not invariant under changes of the basis, and therefore the results depend on the choice of basis $I$ which we choose in its definition. As we saw in the previous section, the normal modes provide a natural basis to work in for the circuit optimization problem and so in the following, we simply define the $F_1$ metric in this basis. Hence for the problem of two coupled harmonic oscillators, which we focus on in the following, the index $I$ in eq.~\reef{Fone} runs over $\lbrace++,--,-+,+-,0+,0- \rbrace$.

Using the results of section \ref{sec:2ho}, we find the components $Y^I$ appearing in the cost function~\eqref{cost_D} to be
\beqa
Y^{++} &=&\dot{y}+ \dot{\rho}  \cos (2 x)-\dot{z}\sinh (2 \rho ) \sin (2 x) \,,
\nonumber\\
Y^{+-}&=&\dot{\rho}  \sin (2 x)+\dot{z} (\cosh (2 \rho )+\sinh (2 \rho ) \cos (2 x))-\dot{x}\,,
\nonumber\\
Y^{-+}&=&\dot{\rho}  \sin (2 x)- \dot{z} \left(\cosh (2 \rho )-\sinh (2 \rho ) \cos (2 x)\right)+\dot{x}\,,
\label{YI_fun}\\
Y^{--}&=&\dot{y}-\dot{\rho}  \cos (2 x)+\dot{z}\sinh (2 \rho ) \sin (2 x) \,,
\nonumber\\
Y^{0+}&=&e^{-y_+}\cos (x) (\dot{u}_+ \cos (z)+\dot{u}_- \sin (z))-e^{-y_-} \sin (x) (\dot{u}_- \cos (z)-\dot{u}_+ \sin (z))\,,
\nonumber\\
Y^{0-}&= & e^{-y_-} \cos (x) (\dot{u}_- \cos (z)-\dot{u}_+ \sin (z))+e^{-y_+}\sin (x) (\dot{u}_+ \cos (z)+\dot{u}_- \sin (z)) \,.
\nonumber
\eeqa
We will not attempt to find the extremal trajectories in complete generality here. Rather we will focus on the analog of the simple geodesics found in section \ref{sec:simple}, which prepare coherent states where only one of $\a_\pm$ is nonvanishing. We will also consider the case of small excitations, \ie $|\a_\pm|\ll 1$, to parallel the analysis in section \ref{sec:small}.

To begin let us consider constraining the trajectories with $x(s)=0=z(s)$, in which case the $F_1$ cost function \reef{cost_D2} takes a simple form 
\be
\label{eq:DF1-xz=0}
{\cal D}_{1} = \int_0^1 ds \(\left|\dot{y}_+\right|+\left|e^{y_+}\dot{u}_+\right|+\left|\dot{y}_-\right|+\left|e^{y_-}\dot{u}_-\right|\)\,,
\ee
where once again we used $y_\pm = y\pm \rho$. A key feature here is that the motions for the $+$ and $-$ coordinates have decoupled, which is reminiscent of the trajectories studied in sections \ref{sec:simple} and \ref{sec:small}. We will proceed with examining the geodesics in the $x=0=z$ subspace which extremize eq.~\reef{eq:DF1-xz=0} in a moment. However, first imagine that we have these solutions and then we wish to show that they also extremize the full cost function \reef{cost_D2} by considering perturbations away from this subspace. Let us construct a perturbative expansion with $x(s),z(s)\sim\cO(\varepsilon)$. Then keeping the leading perturbations in eq.~\reef{cost_D2} yields
\beqa
{\cal D}_1 &=& \int_0^1 ds \big[\left|\dot{y}_+\right|+\left|e^{y_+}\dot{u}_++\dot u_-\(e^{y_+}z-e^{y_-}x\)\right|
\nonumber\\
&&\qquad\quad+\left|\dot{y}_-\right|+\left|e^{y_-}\dot{u}_-+\dot u_+\(e^{y_+}x-e^{y_-}z\)\right|
\label{pert88}\\
&&\qquad\quad +\left|2\,x\,\dot{\rho} +e^{2\rho}\dot{z}-\dot{x} \right|+\left|2\,x\,\dot{\rho} -e^{-2\rho}\dot{z}+\dot{x} \right| + {\cal O}(\varepsilon^2) \big]\,.
\nonumber
\eeqa
First, let us consider the case where the excitations are small, \ie $|\a_\pm| ={\cal O}(\varepsilon)$. In this scenario, we expect $u_\pm = {\cal O}(\varepsilon)$, and therefore the $x,z$ dependent terms in the first two lines are ${\cal O}(\varepsilon^2)$. Therefore, we drop the latter and the only ${\cal O}(\varepsilon)$ terms involving these variables are in the third line and the action is minimized when they vanish, which yields
\be
\dot{x} = {\rm cosh}(2\rho)\dot{z}\,, \qquad 2x\,\dot{\rho}+{\rm sinh}(2\rho) \dot{x} = 0\,.
\label{dotdot}
\ee
These expressions in turn are solved by $x\, {\rm tanh}(\rho) = {\rm constant}$. The initial condition $\rho_0=0$ implies that the constant is zero, which in turn leads to $x(s)=0=z(s)$ as necessary conditions to minimize the action. The remainder of the cost function then takes the simple form in eq.~\eqref{eq:DF1-xz=0}, in which the $\pm$ coordinates decouple, but recall that here we assumed that both  $|\a_\pm|\ll1$.

Instead let us assume that we have a coherent state where $\a_+$ is large but that $\a_-={\cal O}(\varepsilon)$ or zero. In this situation, we again expect $u_- = {\cal O}(\varepsilon)$, which means that the $x,z$ dependent terms in the first line are ${\cal O}(\varepsilon^2)$ and can be ignored again. The ${\cal O}(\varepsilon)$ terms involving $x$ and $z$ are the second term in the second line and the two terms in the third line of eq.~\reef{pert88}. The action will be minimized if we can find a solution where they vanish. The terms in the third line vanish with $x=0=z$, with the analysis following eq.~\reef{dotdot}. Hence the action again reduces to the form given in eq.~\reef{eq:DF1-xz=0}, although we must keep in mind that the term involving $\dot{u}_-$ is ${\cal O}(\varepsilon)$.\footnote{Further let us consider states where both $|\a_\pm| \gtrsim 1$. In principle, the ${\cal O}(\varepsilon)$ terms in the first two lines could offset the contributions from the third line. However, in most situations, the relative signs between the terms in front of the $\dot{u}_+$ and $\dot{u}_-$ are different for each absolute value. Therefore we expect that for most of these states, these contributions will not be able to counter the increase in length coming from the terms in the third line. In this case, the action would still be minimized when the terms in the third line vanish, and the two modes would decouple in determining the optimal path. That is, the optimal trajectories would effectively be determined by the product geometry  $\mathbb{H}^2 \times \mathbb{H}^2$ even when $|\a_\pm|\gtrsim 1$. \label{footy3}}

\begin{figure}[htbp]
	\centering
	\hspace{-15pt}
    \includegraphics[width=0.6\textwidth]{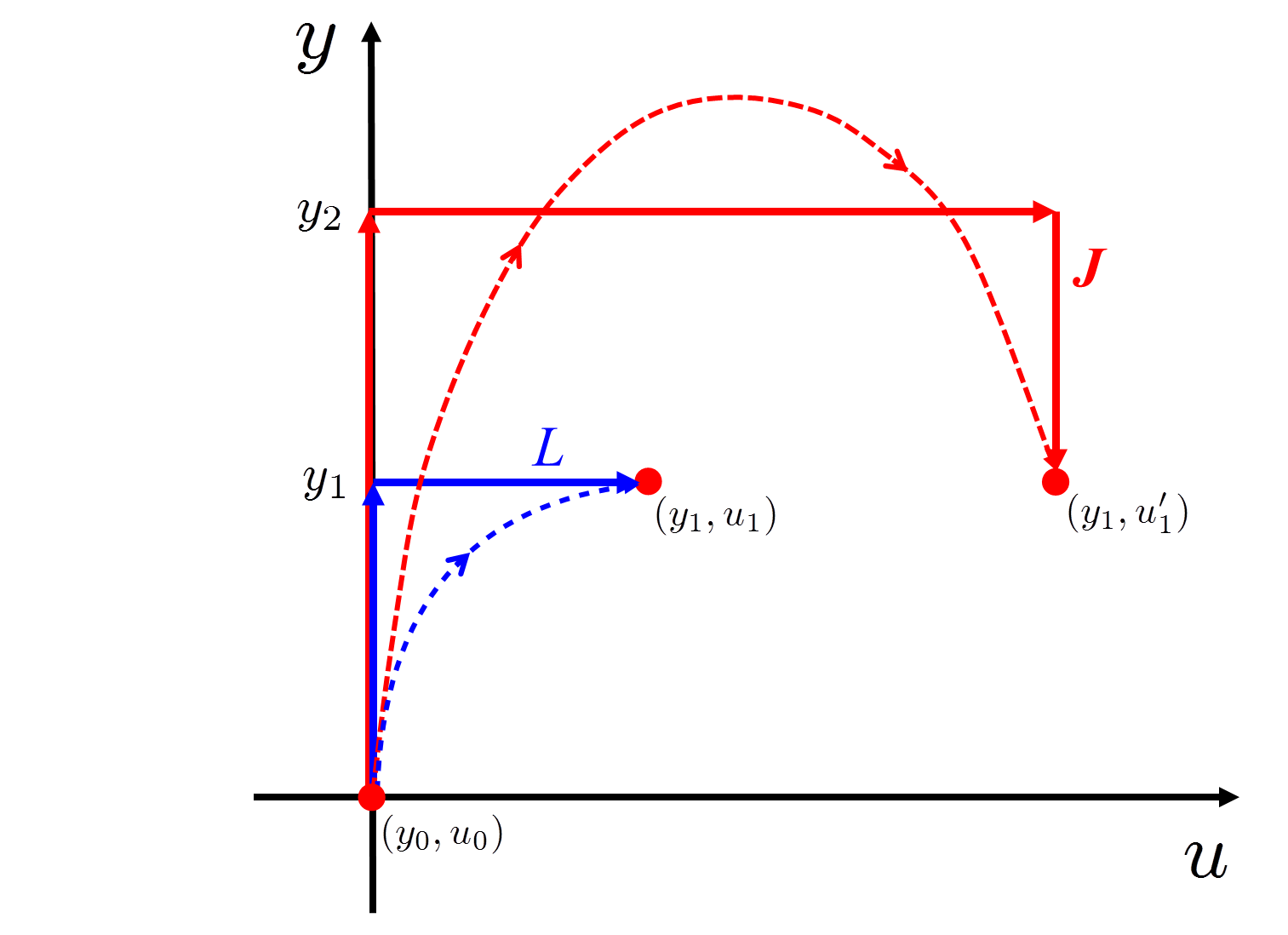}
    \vspace{-10pt}
	\caption{An illustration of the two types of geodesics arising with $F_1$ metric. The L-shaped (blue) paths move in two straight segements to the target state, while the J-shaped (red) paths have three straight segments. The first overshoots $y_{1}$ because motion in the $u$ direction is less costly at higher values of $y$. The dashed curves illustrate the corresponding simple geodesics found with the $F_2$ measure, as in section \ref{sec:simple}.}\label{fig:LJpaths}
\end{figure}

We now turn to the problem of finding the geodesics in the simple ``geometry'' appearing above in eq.~\reef{eq:DF1-xz=0}. That is, we consider
\be\label{DF1_Simple}
{\cal D}_{1} = \int_0^1 ds \(\left|\dot{y} \right|+\left|e^{-y}\, \dot{u}  \right|\)\,.
\ee
With the ``flat measure'' ${\cal D} = \int ds \(\left|\dot{y} \right| + \left|\dot{u} \right|\)$ the minimal trajectories are simply those which traverse between the initial and final endpoints without backtracking in $u$ or $y$. However, with the addition of the scaling factor $e^{-y}$ in the $\dot u$ term in eq.~\eqref{DF1_Simple}, there is a balance between backtracking in $y$ and attempting to reduce the scaling factor by going to a larger $y$ before turning back to the final value. This leads to two possible classes of paths that can minimize the distance~\eqref{DF1_Simple}, illustrated in figure~\ref{fig:LJpaths}. We call these the L and J paths.  We will assume $y_1>y_0$ and $u_1>u_0$ to simplify our discussion, but the other cases are very similar to these.

The length of the L-shaped path that is a straight line from $(y_0,u_0)$ to $(y_1,u_0)$ and then a straight line from $(y_1,u_0)$ to $(y_1,u_1)$ is 
\beq
{\cal D}_L = \Delta y + e^{-y_1} \Delta u\,,
\label{ddL}
\eeq
where $\Delta y = y_1-y_0$ and $\Delta u = u_1-u_0$.
For the J-shaped paths, there are three straight-line sections: $(y_0,u_0)$ to $(y_2,u_0)$, $(y_2,u_0)$ to $(y_2,u_1)$ and then $(y_2,u_1)$ to $(y_1,u_1)$. The length of this path is 
\beq
{\cal D}_J(\delta y) = \Delta y + 2\delta y + e^{-(y_1 +\delta y)}\Delta u\,,
\eeq
where $\delta y = y_2-y_1>0$. This cost is minimized with $\delta y_{\rm min} = {\log}\frac{\Delta u}{2} - y_1$. The optimal J path therefore goes up to $y_{2,\rm min} = \log \frac{\Delta u}{2}$, then over to $u_1$ and then back to $y_1$, and has length 
\beq
D_J(\delta y_{\rm min}) = \Delta y + 2\, \log  \frac{\Delta u}{2} - 2\,y_1 + 2 \,.
\label{ddJ}
\eeq
The L paths are shorter for $e^{-y_1} \Delta u \leqslant 2$, while the J  paths are shorter for $e^{-y_1} \Delta u \geqslant 2$.

Finally, we can express the boundary conditions in terms of parameters of the target state (and the reference state). For the reference states~\eqref{eq:refPhys} and the target states~\eqref{target1}, we find\footnote{Unlike the discussion above, $u_1<u_0$, but it is simple to show that the same argument holds with $\Delta u = -(u_0-u_1) = \sqrt{\w}\a$ in the final result.}
\be\label{walt}
y_0 = 0\,, \quad y_1 = \frac12 \log \, \w\, \quad u_0=0\,, \quad u_1= -\sqrt{\w} \a\,.
\ee
Substituting these expressions into eqs.~\reef{ddL} and \reef{ddJ}, we find
\be\label{walt1}
{\cal D}_L = \frac12 \log \, \w+ \a\,, \qquad {\cal D}_J = \frac12 \log \,\w + \log \frac{\a^2}4  +2\,.
\ee
Using eq.~\reef{walt}, we also find $e^{-y_1}\Delta u = \a$ and so we see that ${\cal D}_L$ is smaller for $\a \leqslant 2$ while ${\cal D}_J$ is smaller for $\a \geqslant 2$.

Note that since we assumed that $y_1>y_0$ above, we are implicitly considering $\w>1$. Carefully going through the argument above for the case $y_0>y_1$, we find that the following lengths for the two types of paths
\be\label{walt2}
{\cal D}_{L} = - \frac12 \log \,\w+\sqrt{\w}\,\a\,, \qquad {\cal D}_{J} = \frac12 \log \,\w+ \log \frac{ \a^2}4 +2 \,,
\ee
for $\w<1$. Here, the $J$ paths are shorter for $\sqrt{\w}\a \geqslant 2$ and otherwise the $L$ paths are shorter. Note that ${\cal D}_J$ has precisely the same form in eqs.~\reef{walt1} and \reef{walt2}. So in general, we can write the cost of these two families of extremal paths as
\be\label{duck}
{\cal D}_L= \frac12 \left|\log \,\w\right|+{\rm min}(1,\sqrt{\w})\,\left|\a\right|\,, \qquad {\cal D}_J = \frac12 \log \,\w+ \log \frac{\a^2}4 +2 \,,
\ee
where the J geodesics are defined (and shorter) for ${\rm min}(1,\sqrt{\w}) \left|\a \right|\geqslant 2$. Hence in general, the $F_1$ cost function \eqref{DF1_Simple} yields the following complexities 
\begin{equation}
\begin{split}
\label{eq:C1}
\mathcal{C}_{1}= 
\begin{cases}
{\cal D}_{L}\,, &{\rm for}\quad \w >1\,, |\a| \le 2\, ,   \quad
{\rm or}\quad \w <1\,, \sqrt{\w}|\a| \le 2 \,,\\
{\cal D}_{J}\,, & {\rm for}\quad \w >1\,, |\a| \ge 2\, ,   \quad
{\rm or}\quad \w <1\,, \sqrt{\w}|\a| \ge 2 \,.
\end{cases}
\end{split}
\end{equation}

To conclude here, let us note that if we are considering small excitations of the ground state, \ie $|\a|\ll1$, then the optimal circuit will be described by a L-shaped geodesic. Further, in this case, the change in the complexity will be linear in $|\a|$, \ie $\Delta\mC_1=\mC_1-\mC_{1,\mt{vac}}\propto |\a|$. The latter behaviour contrasts with our previous results for the $\kappa=2$ cost function, where we found $\Delta\mC_{\kappa=2}\propto\a^2$ in eq.~\reef{smalla}. Further, for large excitations with $|\a|\gg1$, the optimal circuit is descriped by a J-shaped geodesic and we find
$\Delta\mC_{1}\simeq \log\a^2$. Recall that we found in eq.~\reef{biga} that
$\Delta\mC_{\kappa=2}\simeq\(\log \a^2\)^2$ and so there is again a difference in the power of the leading contribution.

For target states where both modes are excited but with $\a_\pm\ll1$, the resulting ``geometry'' is a product space of two copies of the above geometry. The geodesics will therefore correspond to the L-shaped geodesics and following eq.~\reef{eq:C1}, the total complexity is then
\begin{equation}
{\cal C}_{1,\mt{tot}} = {\cal D}_L(\w_+,\a_+) + {\cal D}_L(\w_-,\a_-)\,,
\end{equation}
where ${\cal D}_L(\w,\a)$ is defined in~\eqref{duck}. With these small excitations, the increase in complexity above the vacuum complexity is given by
\beq\label{DelC1}
\Delta {\cal C}_{1,\mt{tot}}= 
{\rm min}(1,\sqrt{\w_+})\,\left|\a_+\right|+ {\rm min}(1,\sqrt{\w_-})\,\left|\a_-\right|
\eeq
which is linear in $|\a_\pm|$. We can also consider the complexity of states where one excitation is large, \eg $|\a_+| \gg 1$ (but the other is still small). This contribution dominates and ${\cal C}_{1,\mt{tot}} \simeq {\cal D}_J(\w_+,\a_+)$. Hence the change in the complexity becomes\footnote{As described in footnote \ref{footy3}, if both of excitation parameters $\a_\pm$ are large, we expect that two modes still decouple in the optimal preparation of most such states. In this situation,  the change in complexity would scale as $\Delta {\cal C}_1 \simeq {\rm log}\, \a_+^2 + {\rm log}\, \a_-^2$.}
\beq\label{DelC1a}
\Delta {\cal C}_1 \simeq {\rm log}\, \a_+^2\,.
\eeq
Let us note that unlike the case of the $\kappa=2$ complexity of section~\ref{sec:2ho} and of the Schatten complexity of the next section, $\Delta{\cal C}_1$ is independent of the excited frequency $\w$, unless $\w<1$ and $\sqrt{\w}|\a| \leqslant 2$, in which case it is proportional to $\sqrt{\w}$.
 
\subsection{Schatten cost function} \label{Schat}

A suggestion put forward in \cite{Hackl:2018ptj} is that we might use the $p=1$ Schatten norm (\eg see \cite{bhatia2013matrix,watrous2018theory,gil2003operator}) rather than the $F_1$ cost function. The observation was that with this new cost function that we  would recover the same leading divergence as with the $F_1$ measure for the complexity of the vacuum,\footnote{In fact, the vacuum complexity was identical for both measures, but we will see below that this does not carry over for the coherent states studied here.} however, the results are now basis independent when described in terms of the Schatten norm. This norm actually provides a family of measures based on computing the singular value decomposition of the desired transformation. Given a transformation $A$, this norm takes the form
\beq\label{Schatten}
\Vert A \Vert_p =\[
{\rm Tr}\!\(\( A^\dagger\,A\)^{p/2}\)\]^{1/p}\,.
\eeq 
Note that with $p=2$, this reduces to the standard Frobenius-Hilbert-Schmidt norm, \ie we recover the $F_2$ measure which we were studying in the previous section. As with the $F_2$ cost function, the results are basis independent for the Schatten measure for any value of $p$. Another property worth noting is that the Schatten $p$-norms are non-increasing in $p$, which means we have $\Vert A \Vert_p  \ge \Vert A \Vert_q $ for $1 \le p\le q \le \infty $. 

In the present case, the transformation of interest is the velocity tangent to the path of unitaries, namely
\begin{equation} \label{velo2}
V(\s)=\partial_\s U(\s)\, U^{-1}(\s)=  Y^I(\s)M_I=\begin{pmatrix}
Y^{++}&  Y^{+-}        & 0\\
Y^{-+} &  Y^{--}  & 0 \\
Y^{0+}  &  Y^{0-}  &   0
\end{pmatrix}\,,
\end{equation}
and the adjoint mapping is simply $V^T(\s)$. By construction, $V^T V$ (or $V V^T$) is a (non-negative) symmetric matrix with positive real eigenvalues $s_k^2$, where the $s_k\,(\ge0)$ are the singular values of $V$.\footnote{In general, we can write $V=R_1 D R_2$ where $R_1$ and $R_2$ are two independent rotation matrices while $D={\rm diag}(s_1,s_2,s_3)$ with $s_k\ge 0$. The $s_k$ are the singular values of $V$, which only agree with the eigenvalues of $V$ in special cases. For example, the two agree when $V$ is symmetric and non-negative. We note that the singular values and eigenvalues do not agree for the case of interest in eq.~\reef{velo2}.} and the Schatten norm \reef{Schatten} then becomes
\beq\label{Schatten3}
\Vert V\Vert_p=\Big[\sum_k s_k^p\Big]^{1/p}\,.
\eeq
In particular then, $\Vert V\Vert_1=\sum_k s_k$. Given eq.~\reef{velo2}, we can explicitly write out the self-adjoint operator
\beq
V^T\,V=\begin{pmatrix}
	(Y^{a+})^2&  Y^{a-} Y^{a+}        & 0\\
	Y^{a-} Y^{a+} &  (Y^{a-})^2  & 0 \\
	0  &  0  &   0
\end{pmatrix}\,,
\label{self}
\eeq
where implicitly we are summing over $a\in \lbrace +,-,0\rbrace$ in each component.
Hence we can immediately see that in the special case of interest, the third singular value is zero and we simply need to find the eigenvalues of the upper 2$\times$2 block. The latter is a simple exercise, which yields
\beq
\gamma_{1,2}=s_{1,2}^2 = \frac12 \((Y^{a+})^2+(Y^{a-})^2\pm\sqrt{((Y^{a+})^2-(Y^{a-})^2)^2  
	+4(Y^{a-} Y^{a+} )^2 } \) \,.
\label{eigenv2}
\eeq
Substituting these expressions into eq.~\reef{Schatten3} for $p=2$, we recover
\begin{equation}
\Vert V\Vert_2 = \sqrt{\gamma_1 +\gamma_{2}} = \sqrt{(Y^{a+})^2+(Y^{a-})^2}  \,,
\end{equation} 
in agreement with the $F_2$ cost function in eq.~\reef{function_F}, as expected.

Turning instead to the Schatten cost function with $p=1$, we find
\beq
\Vert V\Vert_1 = \sqrt{\gamma_1}+\sqrt{\gamma_2}\,.
\label{costS}
\eeq
It is useful to consider some simple examples, \ie some simple trajectories. First imagine that we are only scaling the two normal modes, as we would in preparing the ground state. Then from eq.~\reef{YI_fun}, we have $Y^{++}=\dot{y}+\dot{\rho}=\dot{y}_+$ and $Y^{--}=\dot{y}-\dot{\rho}=
\dot{y}_-$, and eq.~\reef{costS} reduces to 
\beq
\Vert V\Vert_1=|Y^{++}|+|Y^{--}|\,.
\eeq 
Of course, this expression has the same form as the $F_1$ cost function for these trajectories, and so both measures would yield the same complexities in situations where these simple scaling circuits were the optimal ones. But now let us consider trajectories where there is also a displacement for, say, the $+$ mode, \ie where $\dot{u}_+\ne0$. Then another component of the tangent vector is also nonvanishing, namely, $Y^{0+} =e^{-y_+}\dot{u}_+$. The cost function \reef{costS} then becomes 
\beq\label{costS2}
\Vert V\Vert_1=\sqrt{(Y^{++})^2+(Y^{0+})^2 }+|Y^{--}|=\sqrt{\dot{y}_+^2+e^{-2 y_+}\dot{u}_+^2 }+|\dot{y}_-|\,.
\eeq
Hence the `Schatten' cost of this simple trajectory is already different from the $F_1$ cost.\footnote{Further, we can anticipate that for small displacements of $u_+$, \ie small excitations of $a_+$, the total cost will have a contribution proportional to $\Delta u_+^2\sim a_+^2$.} Interestingly, because the motions associated with the $\pm$ modes are decoupled in the above cost function \eqref{costS2}, we can easily find the optimal trajectory is a geodesic in the product geometry $\mH^2 \times \mR$. The optimal trajectory which extremizes eq.~\reef{costS2} is precisely the  `simple geodesic' discussed in section \ref{sec:simple}. 

However, we have restricted the motion of the trajectories in constructing the expression in eq.~\reef{costS2} and so next we would like to show that our `simple geodesics' also extremize the full Schatten norm  \eqref{costS}.  Towards this goal, we consider a new Lagrangian (or cost function) which is the square of Schatten cost function,
\begin{equation}\label{walk}
\mathcal{L}'_0= \Vert V\Vert_1^2 =\gamma_1 + \gamma_2 +2\sqrt{\gamma_1\gamma_2}\equiv\mathcal{L}_0 +2\sqrt{\mathcal{L}_1}\,, 
\end{equation} 
and if we find trajectories which extremize $\mathcal{L}'_0$ (and yield $\Vert V\Vert_1\ne0$), then they will also extremize $\Vert V\Vert_1$, the desired cost function. Now we have divided the result in eq.~\reef{walk} into the sum of two pieces: $\mathcal{L}_0=\gamma_1+\gamma_2$, which corresponds to the $\kappa=2$ cost function,\footnote{As we saw above in evaluating the Schatten norm with $p=2$, $\gamma_1+\gamma_{2} = (Y^{a+})^2 + (Y^{a-})^2$.} and $\mathcal{L}_1=\gamma_1\gamma_2$.
Now we wish to consider the simple geodesics given by eqs.~\reef{eq:solution} and \reef{solution2}, as well as
\beq
x(s)=0= z(s)=u_-(s)\,.
\label{really}
\eeq
Now the analysis in section \ref{sec:simple} showed that these trajectories extremized the $\kappa=2$ cost function \reef{eq:L2}. Hence we already know that the simple geodesics will extremize the first part of eq.~\reef{walk}, and we need only examine the variations of the ${\mathcal L}_1$ term. These equations of motion are generally very complicated but they simplify enormously when we substitute eq.~\reef{really}. The simplied variations are
\begin{equation}\label{Schatten_EoMs}
\begin{split}
\delta_x {\mathcal L}_1&= 0 =\delta_z {\mathcal L}_1=\delta_{u_-} {\mathcal L}_1 \,,\\
\delta_{y_+} {\mathcal L}_1&= -2\dot{ y }_-\(   2\dot{y}_+\ddot{y}_-  + \dot{ y }_- \(e^{-2 y_+} \dot{ u}_+^2 +\ddot{y}_+\)\)\,, \\
\delta_{y_-} {\mathcal L}_1 &= e^{-2y_+}\dot{ u}_+ \(  -4\dot{ y }_- \ddot{u}_+  +  \dot{ u}_+ \(4\dot{y}_- \dot{ y }_+ -2\ddot{y}_-\)    \)  -2\dot{ y }_+ \( \dot{ y }_+ \ddot{y}_- +2\dot{ y }_-\ddot{y}_+ \)\,,\\
\delta_{u_+} {\mathcal L}_1 &= -2e^{-2y_+} \dot{ y}_- \(   \dot{ y}_- \ddot{u}_+ -2 \dot{ u}_+ \dot{ y }_+\dot{ y }_- +2\dot{ u}_+ \ddot{y}_-\).\\
\end{split}
\end{equation}
 However, one can easily show that the three remaining variations will vanish upon substituting the corresponding equations of motion derived from the $\kappa=2$ cost function:
\begin{equation}
\ddot{ y }_-=0\,, \qquad \ddot{u}_+-2\dot{ u}_+\dot{ y }_+=0\,,\qquad \ddot{y}_++e^{-2y_+}\dot{ u}_+^2  =0\,.
\end{equation}
Therefore we arrive at the desired conclusion that the simple geodesics in the $\mH^2 \times \mR$ geometry also describe the optimal circuits for the (full) Schatten $p=1$ cost function \eqref{costS}. 

Hence the coherent states in which a single normal mode is excited are prepared in precisely the same way as in section \ref{sec:simple}. Recall that the boundary conditions for these trajectories are given by eq.~\reef{eq:boundary-cond}. Further, using the subsequent analysis in section \ref{sec:simple}, it is then straightforward to show that the complexity measured by the Schatten cost function is then given by
\beq
\mC_\mt{Schat}=|\Delta| + |C|\,,
\label{ccc9}
\eeq
where $C$ and $\Delta$ are given in eqs.~\reef{solution2} and \reef{DELTA}, respectively. The increase in the complexity above that of the vacuum state is given by
\begin{equation}\label{complex9a}
\begin{split}
\Delta\mC_\mt{Schat}&= \mC_\mt{Schat}-\mC_{\mt{Schat},\mt{vac}}=|\Delta|-\frac12\,|\log  \w_+ |
\\
&=\left|\log\!\[\frac{(1+\a_+^2\w_++\w_+)+ \sqrt{(1+\a_+^2\w_++\w_+)^2-4\w_+}}{2\sqrt{\w_+}} \]\right|-\frac12\,|\log  \w_+ |\,.
\end{split}
\end{equation}
Expanding for small $|\a_+|$, eq.~\reef{complex9a} yields
\begin{equation}\label{smalla2}
\Delta\mC_\mt{Schat} =  \frac{\w_+ \a_+^2}{|\w_+-1|} - \frac{\w_+^2 (\w_++1)\a_+^4}{2|\w_+-1|^3} + \mathcal{O}(\a_+^6)\,,
\end{equation}
while for large $|\a_+|$, we find
\begin{equation}\label{biga2}
\Delta\mC_\mt{Schat} = \log \a_+^2 +\frac12\,\log  \w_+  - \frac12\,|\log  \w_+ | + \frac{1+\w_+}{\w_+ \a_+^2} -\frac{1+4\w_++\w_+^2}{2\w_+^2\a_+^4} + \mathcal{O}\(\frac{1}{\a_+^6}\)\,.
\end{equation}

In analogy to section \ref{sec:numerics}, one might attempt to study numerically the full equations of motion resulting from eq.~\reef{walk} to investigate the complexity of states where both of the normal modes are excited. However, we do not pursue this direction here. Instead, we turn to an analysis for such states in the regime where the excitations are small, \ie $\a_\pm\ll 1$, in analogy to section \ref{sec:small}.
We will assume that in the excited state that $a_\pm\sim\mO(\varepsilon)$ where $\varepsilon$ is a small expansion parameter in the following perturbative construction. Assuming the variation of the geodesics is smooth and starting from the geodesic with $u_\pm=0=x=z$, \ie $a_\pm=0$, the perturbed geodesic line for $a_\pm\sim\mO(\varepsilon)$ should have 
\begin{equation}
 u_\pm(s), \dot{u}_\pm(s),  x(s), \dot{x}(s), z(s), \dot{z}(s) \sim \mO(\varepsilon)\,,
\end{equation}
to leading order in our $\varepsilon$ expansion. Therefore we define the perturbative solution with $ u_\pm(s) =\varepsilon u_{\pm}^{(1)}(s) + \varepsilon^2 u_{\pm}^{(2)}(s) +\mO(\varepsilon^3)$, and similarly for $x(s)$ and $z(s)$. Substituting these expansions into the expressions in the cost function \reef{walk}, we find
\begin{equation}\label{Schatten_perturbation}
\begin{split}
\mathcal{L}_0&=\gamma_{1}+\gamma_{2} \\
&=\dot{ y }_+^2 +e^{-2y_+} \varepsilon^2(\dot{u}_+^{(1)})^2+  \dot{ y }_-^2 + e^{-2y_-} \varepsilon^2(\dot{u}_-^{(1)})^2\\
&\quad+ 2\varepsilon^2\((\dot{x}^{(1)})^2 -2\cosh (2\rho) \dot{x}^{(1)}\dot{z}^{(1)} +\cosh (4\rho) (\dot{z}^{(1)})^2 \) +\mO(\varepsilon^3)\,,\\
\mathcal{L}_1&=\gamma_{1}\gamma_{2} =\( \dot{ y }_+^2 +e^{-2y_+} \varepsilon^2(\dot{u}_+^{(1)})^2\)\(  \dot{ y }_-^2 + e^{-2y_-} \varepsilon^2(\dot{u}_-^{(1)})^2\) \\
&\qquad\qquad + \dot{ y }_+\dot{ y }_- \varepsilon^2\( (\dot{x}^{(1)})^2 -2\cosh (2\rho) \dot{x}^{(1)}\dot{z}^{(1)} + (\dot{z}^{(1)})^2 \)+\mO(\varepsilon^3)\,.
\end{split}
\end{equation}
So let us consider the solutions extremizing $\mL_0$ first: It is consistent to solve with $x^{(1)}(s)=0=z^{(1)}(s)$.\footnote{We stress that at higher orders, we expect that $x(s)$ and $z(s)$ will be nonvanishing one sees in the full numerical solutions for finite values of $a_\pm$.} With this choice, the $\pm$ modes decouple at this order and the solutions correspond to geodesics on $\mH^2\times\mH^2$. Further note that for either mode on these geodesics, $\dot{y}^2 + e^{-2y} \varepsilon^2(\dot{u}^{(1)})^2=\Delta $ is a constant of the motion. That is, this is like a conserved energy, which corresponds to $\Delta^2$ in eq.~\reef{DELTA} for the simple geodesics.
Now we move to consider variations of $\mL_1$. Again it is straightforward to show that $x^{(1)}(s)=0=z^{(1)}(s)$ is a consistent solution. To facilitate the discussion, we can then write 
$$
\mL_1=\(\dot{ y }_+^2 +e^{-2y_+} \varepsilon^2(\dot{u}_+^{(1)})^2 \)\( \dot{ y }_-^2 + e^{-2y_-} \varepsilon^2(\dot{u}_-^{(1)})^2 \) + \mO(\varepsilon^3)\,,
$$
where we should only really pay attention to the terms to $\mO(\varepsilon^2)$. But given this form, the variations with respect to $y_\pm$ and $u_\pm$ are all proportional either to equations of motion from $\mL_0$ or to $\partial_s(\dot{y}^2 + e^{-2y} \varepsilon^2\dot{u}^2)$, both of which vanish for the perturbative solutions of the equations of motion from $\mL_0$.
Therefore to leading order, the two modes decouple and we can just consider geodesics on $\mH^2\times\mH^2$. From eq.~\reef{smalla2}, the resulting change in complexity is then just
\begin{equation}
\Delta\mC_\mt{Schat} =  \frac{\w_+ \a_+^2}{|\w_+-1|}+ \frac{\w_- \a_-^2}{|\w_--1|}+ \mathcal{O}(\varepsilon^4)\,.
\end{equation}

\section{Complexity of coherent states in QFT}
\label{sec:QFT}

In the previous section, we examined the complexity of coherent states in a system of two coupled harmonic oscillators. In this section, we extend the results to the quantum field theory describing a free scalar. In particular, we consider a free scalar theory in $d$ spacetime dimensions with the Hamiltonian 
\begin{equation}\label{Ha_scalarQFT}
H=\frac{1}{2}\int d^{d-1}x\left[\pi(x)^2+(\vec\nabla\phi(x))^2+\mu^2\,\phi(x)^2\right]~.
\end{equation}
Following \cite{Jeff}, we regulate the theory by putting it on a lattice with lattice spacing $\delta$, in which case the regulated theory becomes a family of coupled harmonic oscillators. The lattice Hamiltonian can be written as\footnote{We approximate the spatial derivatives as $\partial_i \phi(x)  \simeq \frac1\delta( \phi(x )- \phi(x-\delta\, \hat{x}^i))$ and we designate the lattice sites with $\vec{n}=n_i\,\hat{x}^i$, where $\hat{x}^i$ are unit normals along the spatial axes.}  
\begin{equation}\label{ham88}
\begin{split}
H&=\frac{1}{2}\sum_{\vec{n}}\delta^{d-1}\left[ \pi(\vec{n})^2+\frac{1}{\delta^2}\sum_i\(\phi(\vec{n})-\phi(\vec{n}-\hat{x}_i)\)^2+\mu^2\,\phi(\vec{n})^2\right]\\
&=\sum_{\vec{n}}\left\{\frac{P(\vec{n})^2}{2m}+\frac12 m \left[\omega^2 X(\vec{n})^2+\Omega^2\sum_i\( X(\vec{n})-X(\vec{n}-\hat{x}_i)\)^2\right]\right\},
\end{split}
\end{equation}
where in the second line, we have defined $X(\vec{n})=\delta^{d/2}\phi(\vec{n})$, $P(\vec{n})=\delta^{(d-2)/2}\pi(\vec{n})$, $m=1/\delta$, $\omega=\mu$ and $\Omega=1/\delta$.
Hence as noted above, the lattice Hamiltonian describes a system of the coupled harmonic oscillators on an ($d$--1)-dimensional lattice. 
For simplicity in the following, let us consider the case of $d=2$. That is, we will consider $N$ oscillators on a one-dimensional lattice with the Hamiltonian
\begin{equation}\label{lattice_SHO}
 \begin{aligned}
H=\frac{1}{2m}\sum^{N}_{a=1}[\bar{p}_a^2+m^2\omega^2\bar{x}_a^2+m^2\Omega^2(\bar{x}_a-\bar{x}_{a+1})^2]\,,
 \end{aligned}
 \end{equation}
and periodic boundary conditions $\bar{x}_{a+N}=\bar{x}_a$. The Hamiltonian is straightforwardly rewritten in terms of normal modes,
 \begin{equation}\label{green}
H=\frac{1}{2m}\sum^{N}_{k=1}\(|{p}_k|^2+m^2{\omega}_k^2\,|x_k|^2\)\,, 
 \end{equation}
where the normal modes and the corresponding frequencies are given by
  \begin{equation}\label{foury}
x_k\equiv\frac{1}{\sqrt{N}}\sum^{N}_{a=1}\exp\!\left(-\frac{2\pi i }{N}\,k\,a\right)\bar{x}_a \qquad
{\rm and}\qquad \omega_k^2=\omega^2+4\Omega^2
\sin^2\frac{\pi k}{N}
 \end{equation}
with $k\in\{1,...,N\}$. The conjugates are $x_k^\dagger = x_{-k}=x_{N-k}$ and similarly for $p_k$. In the normal mode basis, the ground state wave function becomes 
 \begin{equation}\label{ground_Lattice}
 \psi_0(x_k) = \prod^{N}_{k=1} \,\(\frac{m\omega_k}{{\pi}}\)^{1/4}\mathrm{exp}\!\left( -\frac{1}{2}\,m\omega_k\,  |x_k|^2 \right)\,.
 \end{equation}
The complexity associated with preparing the ground state from an unentangled reference state\footnote{As in eq.~\reef{eq:refPhys}, this wave function is unentangled both in the normal mode basis and in the physical position basis.}
\begin{equation}
\label{ref-state}
 \psi_\mt{R}(x_k) =\(\frac{\wrr^2}{{\pi}}\)^{N/4} \prod^{N}_{k=1} \,\mathrm{exp}\!\left( -\frac{1}{2}\,\wrr^2\,  |x_k|^2 \right)\,.
\end{equation}
was analyzed in \cite{Jeff} and, \eg with the $\kappa=2$ cost function \reef{function_Fkappa}, the complexity is given by 
\beq
\mC_{\kappa=2,\mt{vac}}=\frac14\,{\sum_{k=1}^N [\log(m\omega_k/\wrr^2)]^2}
=\frac14\,{\sum_{k=1}^N [\log\w_k]^2}\,.
\label{vacC}
\eeq
where in the latter expression, we substituted the notation introduced in eq.~\reef{dimless}.

We now consider the complexity of coherent states in the (regulated) scalar field theory of the form
 \begin{equation}\label{coherent}
 \psi_\mt{T}(x_k) = \prod^{N}_{k=1} \(\frac{m\omega_k}{{\pi}}\)^{1/4}\,\mathrm{exp}\!\left[-\frac{1}{2}\,m\omega_k |x_k-a_k|^2\right]\,.
 \end{equation}
Of course, this question is a simple  extension to  $N$ modes of that examined in the previous section for two coupled harmonic oscillators. Hence the construction of the circuits preparing $\psi_\mt{T}(x_k)$ from the reference state $\psi_\mt{R}(x_k)$ also only requires a straightforward extension of the previous discussion. For example, to define the gates, we need only extend the range of the indices in eq.~\reef{eq:6gates}, \ie $i\in\lbrace 1,\ldots,N\rbrace$ and $a\in\lbrace 0,\ldots,N\rbrace$.\footnote{Note that we reserve $x_0$ to denote the c-number scale appearing in the shift gates. The zero modes are accounted for in $x_N$.} With this set of  gates, the group structure in eq.~\reef{groupX} is generalized to $\mR^{N} \rtimes GL(N,\mR)$, and eq.~\reef{oblique} is replaced by a representation of $(N+1)\!\times\!(N+1)$ matrices taking the form
\begin{equation}\label{lattice_U}
U=
\begin{pmatrix}
U_N  & \mathbf{0} \\
       \mathbf{u}^T       &   1 
\end{pmatrix} \,,
\end{equation}
where $\mathbf{u}^T = (u_1,\ldots,u_N)\in \mR^N$ and $U_N \in GL(N,\mR)$. In principle, we can then construct a metric on the
corresponding $N(N+1)$-dimensional space of unitaries, analogous to eq.~\eqref{metric-ds}, and the geodesics would be solutions extremizing the following particle action, analogous to eq.~\reef{eq:L2},
\beq
{\cal L}_0= \delta_{IJ}\, \text{tr}(\dot{U}\,U^{-1}\,M^T_I)\, \text{tr}(\dot{U}\,U^{-1}\,M^T_J)\,.
\label{action_def}
\eeq 
However, parametrizing the transformations in eq.~\reef{lattice_U} (and in particular, the $GL(N,R)$ transformation $U_N$) is far more complicated. In any event, given our experience in the previous section, we do not expect that we will be able to find analytical solutions for geodesics preparing general states of the form in eq.~\eqref{coherent}.

Instead then, let us focus on the special case where only a single mode in the lattice model is shifted. In particular, we focus on target states of the form
 \begin{equation}\label{lattice_ref}
 \psi'_\mt{T}(x_k ) = \prod^{N}_{k=1,k \ne i} \(\frac{m\omega_k}{{\pi}}\)^{1/4}\,\mathrm{exp}\!\left(-\frac{1}{2}\,m\omega_k  |x_k|^2\right)\ \times \(\frac{m\omega_i}{{\pi}}\)^{1/4}\,\mathrm{exp}\!\left(-\frac{1}{2}\,m\omega_i |x_i-a_i|^2\right)\,,
 \end{equation}
where only the $i$'th mode is excited by shifting to $\langle x_i\rangle=a_i$. Motivated by the results in last section and also in \cite{Jeff}, we are led to conjecture that the optimal circuit preparing this state corresponds to a geodesic in the geometry $\mathbb{H}^2 \times \mathbb{R}^{N-1}$. That is, for the optimal transformation preparing the above state \reef{lattice_ref},  eq.~\reef{lattice_U} reduces to
\begin{equation}\label{lattice_U2}
U(s)=
\begin{pmatrix}
D_N  & \mathbf{0} \\
       \mathbf{d}^T       &   1 
\end{pmatrix} \qquad{\rm where}\ \ 
\begin{cases}
	 \quad D_N={\rm diag}(e^{y_1(s)}, \cdots,e^{y_N(s)}) \,,\\ 
	 \quad \mathbf{d}^T=(0,\cdots,0,u_i(s),0,\cdots,0) \,. \\ 
\end{cases}
\end{equation}
The $y_j(s)$ with $j\ne i$ would simply grow linearly with $s$, while $y_i(s)$ and $u_i(s)$ satisfy the geodesic equations on the hyperbolic space $\mathbb{H}^2$. This suggestion generalizes the geodesics on $\mathbb{H}^2 \times \mathbb{R}$ found for two coupled harmonic oscillators in the previous section, and if we set $a_i=0$ (and hence $u_i(s)=0$), the motion would be restricted to the $\mathbb{R}^N$ parametrized by $y_i$, which was dubbed the normal mode subspace in \cite{Jeff}.  In fact, we can prove that eq.~\reef{lattice_U2} do indeed yield a family of simple geodesics in the full $N(N+1)$-dimensional manifold described by eq.~\reef{lattice_U}. We save the proof for the next subsection where we consider the more general geodesics necessary to prepare coherent states where more than one of the $a_i$ are nonvanishing.
 
 Given these simple geodesics describing coherent states \eqref{lattice_ref} with a single excited mode, we can easily find their complexity as in section \ref{sec:simple}.

 \subsection{Perturbations of simple geodesics}
 \label{sec:perturabtion}
Here we would like to examine the effect of exciting some subset of the normal modes with a shift producing $\langle x_i\rangle=a_i$. To motivate the conjecture which we will prove in the following, let us review: First, with no such excitations at all, it was found in \cite{Jeff} that the optimal geodesics preparing the ground state were confined to an $\mathbb{R}^N$ submanifold of the full $GL(N,\mathbb{R})$ geometry. In the previous section, it was found that for two coupled oscillators that the geodesics preparing a coherent state in which a single normal mode was excited were confined to a $\mathbb{H}^2\times\mathbb{R}$ submanifold of the full $\mathbb{R}^2\rtimes GL(2,\mathbb{R})$ geometry. That is, the motion of the geodesic was still confined to the normal mode subspace for the second unexcited oscillator. However, when both normal modes were excited, we had to consider the geodesic motion in the full six-dimensional geometry, as described in section \ref{sec:numerics}. Now the geodesics describing optimal circuits to prepare coherent states in our (regulated) scalar field theory are governed by the $N(N+1)$-dimensional geometry $\mR^{N} \rtimes GL(N,\mR)$. However, given the previous observations, it is natural to conjecture that if we are considering coherent states \reef{coherent} where only $K$ of the normal modes are excited, then the motion is confined to the normal mode subspace $\mR^{N-K}$ for the unexcited modes, while the geodesics explore the full $\mR^K\rtimes GL(K,\mR)$ subspace describing all of the gates acting on the remaining normal modes. That is, for these states, the optimal geodesics are confined to a $(K^2+N)$-dimensional submanifold of the full geometry, described by
\beq\label{simpU}
U=\begin{pmatrix}
   U_K\ \     &  0\ \  \   & 0          \\
   0\ \ \   &  D\ \  \  & 0        \\
   \mathbf{d}^T\ \     &  0\ \  \    &   1 
   \end{pmatrix} \,,
   \end{equation}
where $U_K\in GL(K,\mR)$, $\mathbf{d}^T$ is a $K$-dimensional vector (with entries $u_i(s)$), and $D$ is an $(N-K)\!\times\!(N-K)$ diagonal matrix (with entries $e^{y_a(s)}$). For convenience, we have arranged the basis of normal modes so that the first $K$ modes are being excited with $a_i\ne0$. 
 
Given the ansatz \reef{simpU}, we can use eq.~\reef{action_def} to describe geodesics restricted to move on this $(K^2+N)$-dimensional submanifold. However, we would like to show that geodesics lying within this subpace are in fact geodesics of the full $\mR^{N} \rtimes GL(N,\mR)$ geometry. Hence we consider perturbing the above trajectories as follows
   \begin{equation}
   \label{eq:Upert}
   \hat{U}=U+\veps\,\delta U\qquad
   {\rm with}\ \ \ \delta U=
   \begin{pmatrix}
   0\ \    &  X\ \   & 0          \\
   Y\ \   &  Z\ \   & 0         \\
   0\ \   &  V\ \    &   0
   \end{pmatrix} \,,
   \end{equation}
where $V$, $X$, $Y$ and $Z$ represent small first-order excursions  away from the submanifold described by eq.~\reef{simpU}. Here, $V$, $X$ and $Y$ fill out the three `zero' blocks on the left-hand side of $U$ and $Z$ comprises the off-diagonal components of the central $(N-K)\!\times\!(N-k)$ block. We have also introduced a small expansion parameter $\veps$ here and so that if we substitute $\hat U$ into eq.~\reef{action_def}, the particle action can be expanded as
\beq
{\cal L}_0(\hat U)={\cal L}_0(U)+\veps\, {\cal L}'_0+\frac12\,\veps^2\, {\cal L}^{''}_0+\cdots\,.
\label{action3}
\eeq
If we set $\delta U=0$, the variation of ${\cal L}_0(U)$ yields the geodesic equations on the submanifold of interest, \ie $\mR^K\rtimes (GL(K,\mR)\times \mR^{N-K})$. The order $\veps$ and higher order terms will contribute to determine the geodesics in the full geometry as they move away from the submanifold. However, the terms of order $\veps^2$ and higher will vanish in the equations of motion if we simply set the components of $\delta U$ to zero. The dangerous terms are those linear in $\veps$ since they may yield nonvanishing terms which do not vanish in the equations derived from variations of the components of $\delta U$, \ie these terms may produce source terms which drive the geodesics away from the submanifold. Therefore our goal is to verify that in fact ${\cal L}'_0$ vanishes.

Towards the latter goal, let us begin by writing the inverse of $\hat{U}$ to first order in perturbations: $\hat{U}^{-1} =U^{-1}-\veps \,U^{-1}\delta U U^{-1}+\cdots$ where
\beqa
   U^{-1}&=&\begin{pmatrix}
   U_K^{-1}\ \     &  0\ \  \   & 0          \\
   0\ \ \   &  D^{-1}\ \  \  & 0        \\
   -\mathbf{d}^T\,U_K^{-1}\ \     &  0\ \  \    &   1
   \end{pmatrix} 
\labell{invers4}\\
   U^{-1}\delta U U^{-1}
   &=& \begin{pmatrix}
   0 & U_K^{-1} X D^{-1}  & 0 \\
     D^{-1} Y  U_K^{-1}\ \  & {D}^{-1}Z {D}^{-1}\ \   & 0\\
   0&  V D^{-1}-\mathbf{d}^TU_K^{-1}XD^{-1} & 0
   \end{pmatrix} \,.
   \nonumber
\eeqa
We then examine the expansion of the tangent vector
\beq
Y^I(s) M_I=\partial_s\hat{U}\, \hat{U}^{-1}=\partial_s{U}\, {U}^{-1}+\veps \(\partial_s\delta{U}\, {U}^{-1}-\partial_s{U}\, U^{-1}\delta U U^{-1}\)+\cdots\,,
\label{tan4}
\eeq
and let us explicitly write out the zero'th order term
\beq\label{tan0}
\partial_s{U}\, {U}^{-1}=\begin{pmatrix}
   \partial_s U_K\,U_K^{-1}\ \     &  0\ \  \   & 0          \\
   0\ \ \   &  \partial_s D\,D^{-1}\ \  \  & 0        \\
   \partial_s\mathbf{d}^T\,U_K^{-1}\ \     &  0\ \  \    &   0
   \end{pmatrix} \,.
   \end{equation}
Given this last expression and the form of the action \reef{action_def}, we can conclude that a nonvanishing order $\veps$ term will arise in eq.~\reef{action3} if and only if 
$\partial_s\delta{U}\, {U}^{-1}-\partial_s{U}\, U^{-1}\delta U U^{-1}$ has contributions proportional to the same matrix generators as appear in $\partial_s{U}\, {U}^{-1}$, \ie
the ${\cal O}(\veps)$ term in eq.~\reef{tan4} has nonvanishing components in the same entries as eq.~\reef{tan0}. However, given our explict expressions above, it is straightforward to show that all of these entries vanish. For example, 
\beq
\partial_s\delta{U}\, {U}^{-1}=\begin{pmatrix}
   0\ \    &  \partial_sXD^{-1}\ \   & 0          \\
   \partial_sYU_K^{-1}\ \   &  \partial_sZD^{-1}\ \   & 0         \\
   0\ \   &  \partial_sVD^{-1}\ \    &   0
   \end{pmatrix} \,,
\label{tan55}
\eeq
where the only potential overlap with eq.~\reef{tan0} is in the central block. However, since $D$ is a diagonal matrix while $Z$ has only off-diagonal components, these contributions are orthogonal in the sense of the inner product \reef{inner} on the matrix generators.

As we argued above, since we were able to show that ${\cal L}'_0$ vanishes in eq.~\reef{action3} above, we can conclude that the geodesics determined with ${\cal L}_0(U)$ on the $\mR^K\rtimes (GL(K,\mR)\times \mR^{N-K})$ submanifold are in fact geodesics in the full geometry $\mR^N\rtimes GL(N,\mR)$. 
In particular, notice that if we choose $K=1$, \ie our target state is only excited in one normal mode, this proof shows that  there is a simple geodesic in an $(N+1)$-dimensional slice of the full geometry which takes the form $\mR\rtimes (GL(1,\mR)\times \mR^{N-1})  = \mH^2 \times \mR^{N-1}$, discussed in the previous section. In this case, the present argument is a generalization of that presented in section~\ref{sec:simple}, in which we showed this geometry plays a role in determining simple geodesics for $N=2$. 

We will not examine here the geodesics in the more general case where $K\ge2$, as it seems that this will demand rather intensive numerical work. For example, the numerical results in section \ref{sec:numerics} are easily extended to the case of $K=2$ for the present discussion with larger values of $N$. However, we would remark that if we excite $K$ normal modes but all with small amplitudes, it is straightforward to show that to leading order the optimal geodesics can be evaluated using an $(N+K)$-dimensional submanifold of the form $(\mH^2)^K\times \mR^{N-K}$ -- see discussion in the next section. Hence, for example, eq.~\reef{complex88} would be easily extended here to give the change in the $\kappa=2$ complexity for a QFT state of this form, as we discuss in the next section.

Recall that it seems that the simple geodesics found in section~\ref{sec:simple} (\ie $K=1$ and $N=2$) actually seem to provide the optimal geodesics for the corresponding family of target states. Strong evidence for this claim came from our numerical studies in section \ref{sec:numerics}. An interesting open question is whether the generalization of these simple geodesics  found here for larger values of $N$ and $K$ will actually provide the optimal geodesics.

 \subsection{Complexity for simple target states}
 \label{sec:complexity}

Now we would like to evaluate the complexity of coherent states in the free scalar field theory using various cost functions. We will focus on two situations: a) where a single mode is excited and b) where many modes are excited but all with small amplitudes. Recall that the complexity of the ground state \reef{ground_Lattice} is  divergent because the complexity is dominated by contributions of the UV modes \cite{Jeff,Chapman:2017rqy}. In particular, with the $F_2$ cost function \reef{function_F}, this leading divergence takes the form
\beq\label{lead2}
\mC_{2,\mt{vac}}\sim \(\frac{V}{\delta^{d-1}}\)^{1/2}\,|\log(\delta\,\wrr)|\,,
\eeq
where $V$ is the spatial volume, $\delta$ is the short-distance cutoff (\ie the lattice spacing) and $d$ is the spacetime dimension of the scalar field theory. Further we have introduced $m=1/\delta$ (as in eq.~\reef{ham88}) and $\omega_k\sim 1/\delta$ for a typical UV mode. The form of this divergence did not match the leading divergence \reef{leaderH} found for holographic complexity \cite{Carmi} and hence the $\kappa$ measures \reef{function_Fkappa} were introduced in \cite{Jeff} to ameliorate this problem. With these cost functions, the leading divergence becomes
\beq\label{leadk}
\mC_{\kappa,\mt{vac}}\sim \frac{V}{\delta^{d-1}}\ |\log(\delta\,\wrr)|^\kappa\,.
\eeq
Let us add that following the reasoning presented in \cite{Hackl:2018ptj}, one can show  that the Schatten $p=1$ cost function  yields the same leading divergence in the vacuum complexity as for the $\kappa=1$ complexity (or the $F_1$ cost function).\\

\noindent{\bf a) Single excitation:} In the previous section, we have argued that the simple geodesics found in section \ref{sec:simple} also describe the optimal circuit preparing QFT coherent states \eqref{lattice_ref} with a single excited mode, for the $F_2$ and $\kappa=2$ cost functions. Hence we can apply our earlier results to evaluate the complexity of these states. For example with the $\kappa=2$ cost function \reef{function_Fkappa}, we would have
 \beq \label{lattice_distance}
\mC_{\kappa=2} =  \Delta_i^2 + \sum_{k=1,k\ne i}^{N}  C_k^2
\eeq
where, in analogy to eqs.~\reef{solution2} and \reef{DELTA}, we have
\beq\label{light8}
\Delta_i = \log\!\[\frac{(1+\a_i^2\w_i+\w_i)+ \sqrt{(1+\a_i^2\w_i+\w_i)^2-4\w_i}}{2\sqrt{\w_i}} \]\,,
\qquad  C_k= \frac 12 \log \w_k \,,
\eeq
with $\a_i = a/x_0$ and $\w_j= m \omega_j/\wrr^2$. In particular, we can evaluate the difference between the complexity of this coherent state and the complexity \reef{vacC} of the ground state, which yields precisely the same result as for two coupled harmonic oscillators in eq.~\reef{complex8} with the substitution $\a_+,\w_+\to\a_i,\w_i$,
\beq
\Delta\mC_{\kappa=2}\(\a_i\)=\(\log\!\[\frac{(1+\a_i^2\w_i+\w_i)+ \sqrt{(1+\a_i^2\w_i+\w_i)^2-4\w_i}}{2\sqrt{\w_i}} \]\)^2-\frac14\left(\log  \w_i \right)^2\,.
\label{complex89}
\eeq
Further, we can consider various limits of this result in analogy to those presented at the end of section \ref{sec:simple}. For example, to leading order for $\a_i\ll1$, we have $\Delta\mC_{\kappa=2}\propto \a_i^2$ as in eq.~\reef{smalla}.

The arguments in the previous section can also be extended to the $F_1$ metric \reef{Fone} and $p=1$ Schatten cost function \reef{Schatten}. This would combine the reasoning given in section \ref{sec:perturabtion} and extending the perturbative arguments given below for the case of small amplitudes, but we do not present the details here. Hence the results for the simple geodesics can be extended to give the Schatten complexity for QFT coherent states with a single excited mode, with
 \beq \label{distant9}
\mC_\mt{Schatten} =  |\Delta_i| + \sum_{k=1,k\ne i}^{N}  |C_k|\,,
\eeq
where again $\Delta_i$ and $C_k$ are given in eq.~\reef{light8}. Similarly, for this class of states, the $F_1$ cost function would be extremized by the $L$- or $J$-shaped paths described in section \ref{costf1}. Hence from eq.~\reef{eq:C1}, the $F_1$ complexity becomes
 \beq \label{distant9}
\mC_1 =  {\cal D}_{L,J}(\w_i,\a_i) + \sum_{k=1,k\ne i}^{N}  |C_k|\,,
\eeq
where ${\cal D}_L$ and ${\cal D}_J$ are the costs given in eq.~\reef{duck}. Of course, as described for eq.~\reef{eq:C1},
the $L$ cost applies for $\w >1, \,|\a| \le 2$ or $\w <1,\, \sqrt{\w}|\a| \le 2$, and the $J$ cost applies otherwise.

One may also be tempted to extend the previous analysis to the higher $\kappa$ cost functions \reef{function_Fkappa}. In this case, we would have
\beq\label{ld99}
\Delta\mC_{\kappa} =  \int_0^1ds\[(\dot y_i)^\kappa + (e^{-y_i}
\dot u_i)^\kappa\] -  C_i^\kappa\,,
\eeq
where $C_i$ are given in eq.~\reef{light8}. This expression can be evaluated numerically for the simple geodesics given the expressions in eqs.~\reef{eq:solution} and \reef{solution2}.
In the limit $\a_i\to0$, the simple geodesic reduces to a straight-line geodesic with $u_i(s)=0$ and of course, $\Delta\mC_{\kappa} \to 0$. It was shown in \cite{Jeff} that these straight-line geodesics were still optimal geodesics preparing the vacuum  state for general $\kappa$ measures, not just for $\kappa=2$. Hence the vacuum complexity is correctly given by $\mC_{\kappa,\mt{vac}} =  \sum_{k=1}^{N}  (C_k)^\kappa$, but for the coherent states with $a_i\ne0$, eq.~\reef{ld99} only provides an upper bound on the complexity. Hence the above expression 
provides an upper bound on the increase of the complexity for these special states.
 
An interesting feature of the result in eq.~\reef{complex89} is that $\Delta\mC_{\kappa=2}$ is finite, \ie there is a single contribution from the excited mode. In particular, the sum over the contributions from the UV modes causes  $\mC_{\kappa=2}$ to diverge for both the vacuum and the coherent state (in the limit $\delta\to0$), but these UV divergences cancel in the difference.  In contrast, we might carry out the analogous calculations with the $F_2$ cost function but in this case, the leading contribution takes the form
\beq\label{delta22}
\Delta\mC_{2}\(\a_i\)=\frac12\,\frac{\Delta\mC_{\kappa=2}\(\a_i\)}{\mC_{2,\mt{vac}}}\,,
\eeq
where $\mC_{2,\mt{vac}}$ and $\Delta\mC_{\kappa=2}$ are given by eqs.~\reef{lead2} and \reef{complex89}, respectively. Hence combining these expressions, we find that $\Delta\mC_{2}$ vanishes as $\delta^{\frac{d-1}2}/V^{\frac12}$ as $\delta\to0$.\\

\noindent{\bf b) Small amplitudes:} 
Another interesting case which is relatively easy to analyze is that of excited states where a number of modes are excited but all with small amplitudes, \ie with $\a_k \ll 1$ for every mode. As alluded to in the previous section, the final result will be that, to leading order, the modes decouple and $\Delta\mC$ is simply given by the sum of the leading results found when simply exciting a single mode. In the following, we will demonstrate that this result applies for the $F_1$, $\kappa=2$ and $p=1$ Schatten measures, using the techniques developed in section \ref{sec:perturabtion}. Let us summarize the (leading) result for the increase in the complexity for each of these cost functions here,
\begin{equation}\label{collections0}
\begin{split}
&\Delta \mC_1 \simeq  \sum_{\w_k \le  1}\sqrt{\w_k}\,|\a_k| + \sum_{\w_k \ge  1} |\a_k|\\
\Delta\mC_{\kappa=2}  \simeq \sum_k & \frac{\log\w_k}{\w_k-1}\,\w_k \,\a_k^2 \,,\qquad
\Delta\mC_\mt{Schat} \simeq \sum_k \frac{\w_k\,\a_k^2}{|\w_k-1|} \,,
\end{split}
\end{equation}
where the sums run over the excited modes. 

Let us begin with the $\kappa=2$ cost function, where we are generalizing the arguments made for two coupled harmonic oscillators in section \ref{sec:small}. Imagine that we have a number of modes excited but that $\a_k\ll1$.  We wish to construct a perturbation expansion in which we designate the excitations as first order, \ie $\a_+ \sim \mO(\veps)$. Then using the formalism of section \ref{sec:perturabtion}, we assume that $\mathbf{u}^T = \veps \mathbf{u}^{(1)T}  +\mO(\veps^2)$ and 
\begin{equation}
\hat{U} = \begin{pmatrix}
U_N & 0 \\ 0 & 1
\end{pmatrix} + \varepsilon \begin{pmatrix}
0 & 0 \\ \mathbf{u}^{(1)T} & 0
\end{pmatrix} +\mO(\veps^2)\,.
\end{equation}
Expanding the tangent then yields 
\begin{equation}
V=\partial_s \hat{U} \hat{U}^{-1} = \begin{pmatrix}
\dot{U}_N U_N^{-1} & 0 \\ 0 & 0
\end{pmatrix} + \varepsilon \begin{pmatrix}
0 & 0 \\ \dot{\mathbf{u}}^{(1)T}  U_N^{-1} & 0
\end{pmatrix}+\mO(\veps^2)\,,
\end{equation}
and the particle action \reef{action_def} in the $\kappa=2$ cost function becomes
\begin{equation}\label{full3}
{\cal L}_0(\hat{U} ) = {\rm tr}\left(\dot{U}_N U_N^{-1} U_N^{-1T} \dot{U}_N^T \right) + \varepsilon^2 \dot{\mathbf{u}}^{(1)T} U_N^{-1} U_N^{-1T} \dot{\mathbf{u}}^{(1)} +\mO(\veps^3)\,.
\end{equation}

The ${\cal O}(1)$ part of the Lagrangian is the precisely same as for considered in~\cite{Jeff}, and the solution which prepares Gaussian states (with $\a_k=0$) is a diagonal matrix and $\mathbf{u}^T=0$. Therefore in our perturbative expansion for small excitations,  we may assume 
\begin{equation}\label{U_N_expansion}
U_N = D + \varepsilon Z^{(1)}+\veps^2 Z^{(2)} +\mO(\veps^3)\,,
\end{equation}
where $D$ is a diagonal matrix and $Z^{(i)}$ are completely off-diagonal. Furthermore, since the diagonal is a local minimum of the zeroth order Lagrangian, substituting this into eq.~\reef{full3} gives an expression of the form
\begin{equation}
{\cal L}_0 = {\rm tr}\left(\dot{D}^2 D^{-2}\right) + \varepsilon^2 \left( \dot{\mathbf{u}}^{(1)T} D^{-2}\, \dot{\mathbf{u}}^{(1)} + F(Z^{(1)},\dot{Z}^{(1)};D,\dot{D})\right)+{\cal O}(\varepsilon^4)\,,
\end{equation}
where $F(Z^{(1)},\dot{Z}^{(1)};D,\dot{D})$ is quadratic in $Z^{(1)}$, positive semidefinite and vanishes if and only if $Z^{(1)}=\dot{Z}^{(1)}=0$. Because there is no term linear in $\veps$, the optimal solution for $Z^{(1)}$ is just zero since any nonvanishing $Z^{(1)}$ would only increase the cost at second order of $\varepsilon$. With this choice, the separate modes simply decouple and the resulting cost function indeed describes motion on $(\mathbb{H}^2)^N$. That is, the motion in the full $\mR^N\rtimes GL(N,\mathbb{R})$ geometry is restricted to a  $(\mathbb{H}^2)^N$ submanifold to leading order when the excitations are small. According to the previous result \reef{smalla} for the simple geodesics in the hyperbolic geometry, we find the leading order change of complexity $\Delta\mC_{\kappa=2}$ is given by the expression in eq.~\reef{collections0} above. 

Of course, extremizing the $\kappa=2$ cost function \reef{function_Fkappa} also extremizes the $F_2$ cost function \reef{function_F}. Hence using eq.~\reef{delta22}, we can evaluate
$\Delta\mC_{2}$  for only small excitations as
\begin{equation}
\Delta\mC_{2} =\frac{1}{2\mC_{2,\mt{vac}}}\,\sum_{k} \frac{|\log \w_k|}{|\w_k-1|} \,\w_k\a_k^2  +\mathcal{O}(\varepsilon^4)\,.
\end{equation}\\

Now we turn to the $p=1$ Schatten norm, where at the end of section \ref{Schat}, we already argued that for two coupled harmonic oscillators, the leading order result for $\Delta\mC$ is simply the sum of those for the individual modes. It is straightforward to extend the above perturbative argument to the Schatten norm for  the free scalar field theory, \ie for $N$ coupled modes. In order to proceed, we need to consider the eigenvalues of the square matrix as 
\begin{equation}
V^TV= (\partial_s \hat{U} \hat{U}^{-1})^T(\partial_s\hat{U} \hat{U}^{-1})  = \begin{pmatrix}
M\ \     &  0            \\
0\ \     &  0
\end{pmatrix} \,,
\end{equation}	
where $\hat{U}$ is defined as in eq.~\eqref{lattice_U} and then the matrix $M$ is given by
\begin{equation}\label{M_matrix}
M=	(\partial_s U_N\,U_N^{-1})^T(\partial_s U_N\,U_N^{-1}) +(\partial_s\mathbf{u}^T\,U_N^{-1})^T(\partial_s\mathbf{u}^T \,U_N^{-1})\,.
\end{equation}
The eigenvalues of $M$ are labeled as $\gamma_{i}$ with $i=1,\cdots N$. The general $p=1$ Schatten cost function is then defined as 
\begin{equation}
\Vert V\Vert_1 = \sum_{i=1}^N \sqrt{\gamma_i}\,.
\end{equation}

Our perturbative construction again begins with the small excitations where $\a_k \sim \mO(\varepsilon) $. It is straightforward to show that
the zeroth order solution is then given by $\mathbf{u}(s)=0$ and
\begin{equation}
\gamma_{i}^{(0)} = (Y^{ii})^2= \( \dot{ y }_i^{(0)}\)^2= C_i^2,
\end{equation}
which means that here the straight-line geodesics also provide the optimal circuit for the Schatten cost function. Perturbing around these solutions as above, we have
\begin{equation}\label{lap9}
u_i= \varepsilon u_i^{(1)} +\mO(\veps^2)\,,\quad	U_N= D + \varepsilon Z^{(1)}+\mO(\veps^2)\,,
\end{equation}
where $D$ is a diagonal matrix with $D_{ii}=e^{2y_i}$ and $Z$ is a completely off-diagonal perturbation. The leading perturbations of the eigenvalues are now given by
\begin{equation}\label{lap8}
\delta \gamma_i =  \vec{v_i}^T \cdot  \Delta M \cdot \vec{ v_i}, \quad \text{with}\quad M^{(0)}\cdot \vec{ v_i}=\gamma_i \,\vec{v_i}\,,
\end{equation}
where substituting eq.~\reef{lap9} into eq.~\reef{M_matrix} has produced an expansion $M=M^{(0)}+\veps M^{(1)}+\veps^2 M^{(2)}+\mO(\veps^3)$ and we combine all of the higher order terms as $\Delta M=M-M^{(0)}$. Further, since $U_N$ is diagonal to leading order, $M^{(0)}$ is a diagonal matrix as well and so the eigenvectors take the special form: $(v_i)_a=\delta_{ia}$. As a result, the perturbations $\delta \gamma_i$ all come from the diagonal components of $\Delta M$, \ie eq.~\reef{lap8} yields $\delta \gamma_i=(\Delta M)_{ii}$. However, it is straightforward to show that $M^{(1)}$ has only off-diagonal components and so there is no $\mO(\veps)$ contribution to $\delta \gamma_i$. Hence we may focus on $M^{(2)}$ to find the leading perturbations of the eigenvalues. First of all, we can easily see the second term in \eqref{M_matrix} provides us with the term like $\varepsilon^2 e^{-2y_i} \dot{u}_i^{(1) }\dot{u}_i^{(1)} $ in  $\delta \gamma_{i}$, which implies the leading  corrections contain the  hyperbolic geometry $\mH^2$ for every mode. Secondly, the corrections on $\delta \gamma_{i}$ from the first term in \eqref{M_matrix} provide  terms which are quadratic in $Z^{(1)}$ and $\dot{Z}^{(1)}$, and hence we solve the corresponding equations of motion with $Z^{(1)} =0$. Finally, as eq.~\reef{walk}, we consider the square of Schatten norm 
\begin{equation}
\Vert V\Vert_1^2 = \(\sum_{i=1}^N \sqrt{\gamma_i}\)^2 = \sum_{i=1}^N \gamma_i  \, + 2\, \sum_{i> j} \sqrt{\gamma_{i}\gamma_{j}}\,.
\end{equation}
Now the first term is precisely the $\kappa=2$ cost function, which in our perturbative expansion describes motion in the restricted subspace $(\mH^2)^N$ of the full geometry, as above. Combining these observations with the arguments at the end of section \ref{Schat}, it is straightforward to show that to leading order, the simple geodesics for each of the individual modes extremize  the above squared cost function and then the $p=1$ Schatten cost function. Hence we may simply sum the leading order results for the change in complexity given in eq.~\reef{smalla2} for each of the decoupled modes to find the expression for $\Delta\mC_\mt{Schat} $ given in eq.~\reef{collections0}.\footnote{Let us note that this discussion can be easily adapted to show that for states in which a single mode is excited, the full result of the simple geodesics can be applied. That is, the increase in the complexity is given by eq.~\reef{complex9a} with the substitution $\w_+,\a_+\to\w_i,\a_i$, where the subscript $i$ indicates the mode which is excited. The discussion is almost the same. We only need to replace the original solutions by $\mathbf{u}(s)=(0,\cdots u_i \cdots,0)$ and notice the eigenvector for this mode is also $(0,\cdots1\cdots,0)$. While the simple geodesic is clearly a solution of the restricted cost function analogous to eq.~\reef{costS2}, we can perturb around this trajectory to find that it is also extremizes the full cost function.}\\

Lastly, to close this section, we consider the case of small excitations with $F_1$ cost function. Firstly, we re-iterate that the $F_1$ cost function depends on the choice of the basis of generators $M_I$. Here we work with the normal mode basis where the $M$ take the simple form given in eq.~\eqref{matrix}, \ie $\left[M_{ai}\right]{}_{cd} = \delta_{ac}\,\delta_{id}$. Again we construct a perturbative expansion with the $\a_i\sim\mO(\veps)$ and at zeroth order, we begin with the simple straight-line solution (without any excitations). We then consider perturbations of the $F_1$ cost function,
\begin{equation}\label{F1_def}
F_1(U,Y)=\sum_I \left|Y^I\right|\,,
\end{equation}
where the index $I\in\{ij,0i\}$ with $i,j=1,\cdots N$, and given the simple form of the generators, the components $Y^I$ are read off from the entries of $V=\partial_s \hat{U} \hat{U}^{-1}$. As above, we
assume $\mathbf{u}^T = \veps \mathbf{u}^{(1)T}  +\mO(\veps^2)$ and expand $U_N$ as in eq.~\eqref{U_N_expansion}, which yields
\begin{equation}
V=\partial_s \hat{U}\, \hat{U}^{-1} 
=\begin{pmatrix}
\dot{D} D^{-1} & 0 \\ 0 & 0
\end{pmatrix} + \varepsilon \begin{pmatrix}
\dot{Z}^{(1)} D^{-1}-\dot{D}D^{-1}Z^{(1)}D^{-1} & 0 \\ 
\dot{\mathbf{u}}^{(1)T}  D^{-1} & 0
\end{pmatrix}+\mO(\veps^2)\,.
\end{equation}  
Now the leading perturbation of eq.~\reef{F1_def} comes from the second term above which produces $\mO(\veps)$ contributions with $|Y^{ij}|$ with $i\ne j$ and $|Y^{0i}|$. Here we are using the fact that the original simple solution, \ie the first term, only contains $Y^{ii}$ components. Now because of the absolute value for all of the terms in \eqref{F1_def} and the boundary conditions $Z^{(1)}(s=0)=0=Z^{(1)}(s=1)$, we minimize the $|Y^{ij}|$ (with $i\ne j$) contribution by setting $Z^{(1)}(s)=0$. Finally the measure of the optimal path should have $N$ copies of the analogous structure in eq.~\eqref{DF1_Simple}, which are extremized by the $L$-shaped paths (for small $\a_i$). Hence to leading order in our expansion, the $F_1$ complexity becomes the sum of the  ${\cal D}_L$ costs in eq.~\reef{duck} for the individual modes and then $\Delta\mC_1$ is given by the expression in eq.~\reef{collections0}. 

\section{Fubini-Study approach for circuit complexity}
  \label{app:info-metric}

In this section, we apply  the Fubini-Study approach proposed in \cite{Chapman:2017rqy} to examine the complexity of coherent states \reef{target1} for a pair of coupled harmonic oscillators. In contrast to the Nielsen approach, which defines a geometry on the space of unitaries \reef{unitaries}, this method makes use of the Fubini-Study metric to define a geometry on the space of states.

First, to introduce the basic definitions, let us imagine that the space of states of interest is covered by some convenient set of coordinates $\lambda^\mu$ -- we will be explicit about the coordinates in our calculations but for the time being one might think of the coordinates in eq.~\reef{eq:A-target}. In the following, we focus on a family of pure states $|\psi(\lambda)\rangle$ and then we can consider the quantum fidelity as the inner product between two such states, \eg \cite{nielsen2002quantum,Zgeometry},
\beq\label{fidel}
F(\lambda,\lambda')  = |\langle\psi(\lambda)|\psi(\lambda')\rangle|\,.
\eeq
The quantum information metric then measures the distance between nearby states as
\beq\label{fidel2}
F(\lambda,\lambda+d\lambda)=1-\frac12 g_{\mu\nu}
\,d\lambda^\mu\,d\lambda^\nu +\cO(d\lambda^3)
\eeq
with 
\beq\label{FSmetric}
 g_{\mu\nu}=\frac{1}{2}\( \langle\partial_\mu \psi|\partial_\nu\psi\rangle +  \langle\partial_\nu \psi|\partial_\mu\psi\rangle  \) - \langle\partial_\mu \psi|\psi\rangle  \langle\psi|\partial_\nu\psi\rangle\,.
\eeq
The quantum information metric is also known as the fidelity susceptibility since it encodes the response of the fidelity to small changes in one of the states.\footnote{We might add that in the context of the AdS/CFT correspondence, the information metric or fidelity susceptibility for boundary states deformed by a marginal operator was proposed to be described by the volume of maximal time slice in AdS spacetime in  \cite{distance}. Of course, the latter is also the conjectured dual of complexity according to the CV proposal \cite{Susskind:2014rva,Stanford:2014jda}. Different proposals for the holographic dual of information metric are also discussed in \cite{canonical-energy,fidelity2,fidelity3,fidelity4}.} In the present case of pure states, eq.~\reef{FSmetric} also corresponds to the desired Fubini-Study metric. This metric may also be evaluated with the following expression 
  \begin{equation}\label{metric_limit}
  g_{\mu\nu} = -\left. \frac{\partial^2 F(\lambda,\lambda') }{\partial{\lambda^{\mu}} \ \partial{\lambda^\nu}}\right|_{\lambda'=\lambda}  \,.
  \end{equation}
Then following \cite{Chapman:2017rqy}, we consider curves $\lambda^\mu(\s)$ on the space of states parameterized by $\s \in [0,1]$ which take us from the reference state to the desired target state, \ie
\beq\label{FSbc}
|\psi(\s=0)\rangle=|\psi_\mt{R}\rangle\,,\qquad
|\psi(\s=1)\rangle=|\psi_\mt{T}\rangle\,.
\eeq
We then assign a cost to each of these trajectories as the distance as measured by the Fubini-Study metric \reef{FSmetric},
  \begin{equation}\label{Def_FScomplexity}
  \mathcal{D}_\mt{FS}= \int_0^1\!\! ds\   \sqrt{g_{\mu\nu}\, \dot{\lambda}^\mu\, \dot{\lambda}^\nu}\,,
  \end{equation}
where  $\dot{\lambda}{}^\mu(s) = \frac{d\lambda^\mu(s)}{ds}$ specifies the tangent vector to the trajectory. The complexity assigned to the target state is then the minimal distance according to this measure, \ie the complexity is the length of the geodesic in the state space equipped with the Fubini-Study metric.

Before proceeding with our calculation of the Fubini-Study complexity for coherent states, it is interesting to express this approach in a way that is closer to the circuit construction introduced in eq.~\reef{unitaries}. In particular, given a trajectory described by a particular choice of $\lambda^\mu(\s)$, we may express the corresponding states as
  \begin{equation}\label{Pa_evolution2}
  \ket{\psi(\s)} = \cev{\mathcal{P}} \exp \[ -i \int^\s_0\!\!\! d s\, {\cal H}( s)\]\,\ket{\psi_\mr}  \qquad \rm{where}\ \ 
  {\cal H}(s)= \sum_\mu \dot{\lambda}{}^\mu(s)\,\mathcal{O}_\mu(\lambda)
  \end{equation} 
where $\mathcal{O}_\mu(\lambda)$ is the set of Hermitian operators which generate the evolution of state $\ket{\psi(\lambda)}$ in the $\lambda^\mu$ direction, \ie
  \begin{equation}\label{PDevolution}
  i \partial_\mu \ket{\psi(\lambda)} = {\mathcal{O}}_\mu(\lambda)\, \ket{\psi(\lambda)}\,.
  \end{equation}
Note that we may think of the operators $\mathcal{O}_\mu(\lambda)$ as being linear combinations of the $\cO_I$ appearing in eq.~\reef{unitaries}. We show a $\lambda$ dependence to indicate that these linear combinations vary as we move through the space of states. However, this leaves the definition of the $\mathcal{O}_\mu(\lambda)$ ambiguous since, at any particular point, there will be degenerate operations which leave the state unchanged, \ie $\cO_0(\lambda)\ket{\psi(\lambda)}=0$. Therefore, in general, one finds that the space of states has a smaller dimension than the space of unitaries, as will be illustrated by the example discussed below.  Given eq.~\eqref{PDevolution}, we can also rewrite the Fubini-Study metric as connected correlation functions of the operators ${\cal O_\mu}$,
  \begin{equation}\label{metric_operator}
  \begin{split}
  g_{\mu\nu}(\lambda) 
  &= \frac12 \bra{\psi(\lambda)} \{ {\mathcal{O}}_\mu\,,{\mathcal{O}}_\nu\} \ket{\psi(\lambda)}   -
  \bra{\psi(\lambda)} {\mathcal{O}}_\mu\ket{\psi(\lambda)} \bra{\psi(\lambda)} {\mathcal{O}}_\nu\ket{\psi(\lambda)},\\
  &=\frac12 \langle \{{\mathcal{O}}_\mu -  \langle {\mathcal{O}}_\mu  \rangle_\lambda\,,   {\mathcal{O}}_\nu  -  \langle {\mathcal{O}}_\nu\rangle_{\lambda}\} \rangle_{\lambda}\,.
  \end{split}
  \end{equation}

Let us also add that ref.~\cite{Chapman:2017rqy} also proposed an alternative formulation where only one gate acts at any given point in the circuit. In preparing the vacuum state of the scalar field theory, this formulation gave a result similar to that of the $F_1$ measure. However, this formulation was developed in \cite{Chapman:2017rqy} with a limited gate set and so it would be interesting to extend it to the more general setting discussed here.\footnote{We thank Shira Chapman for a discussion on this point.}

\subsection{FS complexity of two harmonic oscillators} \label{FStwo}
  
We would like to describe the coherent states discussed in section \ref{sec:2ho} as 
\begin{equation}\label{family}
\psi(x_+,x_-) = \frac{\left({\rm det} A_2\right)^{1/4}}{\sqrt{\pi}} \exp \left[-\frac12 (x_i-a_i)\, [A_2]^{ij} (x_j-a_j)\right]\,,
\end{equation} 
where $i,j\in\lbrace +,\,-\rbrace$. The 2$\times$2 coefficient matrix $A_2$ is given by
\beq\label{mator2}
A_2
= U_2\,A_\mr\, U_2^T \qquad{\rm where}\ \ A_\mr= \wrr^2\, \id_2
\eeq
and $U_2$ is the $GL(2,\mR)$ matrix given in eq.~\reef{GL2_matrix}. The explicit form of $A_2$ is given by the upper left 2$\times$2 block found in eq.~\reef{eq:A-target}, and as noted there, $A_2$ is independent of $z$.\footnote{Therefore det$A_2=e^y\,\wrr$ in the normalization factor in eq.~\reef{family}.}  Let us parametrize the displacements of the coherent states in terms of the dimensionless coordinates $v_\pm$ with $a_\pm\equiv \tilde{x}_0\,v_\pm$, where we have introduced a convenient length scale $\tilde{x}_0$ in this definition.\footnote{This scale appears in a similar role to $x_0$ in section \ref{sec:2ho} but we use the notation $\tilde{x}_0$ here to distinguish the two. We also emphasize that $\tilde{x}_0$ was introduced here for the convenience of producing the dimensionless coordinates $v_\pm$ but in the end, this scale will not appear in the results for the complexity.} Then our family \reef{family} of coherent states is described by five dimensionless coordinates $\lambda^\mu=\lbrace y,\, \rho,\, x,\, v_\pm\rbrace$, and by construction, the origin of this coordinate system corresponds to the reference state \reef{eq:refPhys}. Of course, this is one less coordinate than described the unitary transformations in section \ref{sec:2ho}.
 
Now by the methods introduced above, we can define the Fubini-Study metric for the space of states $\ket{\psi(y,\rho, x, v_\pm)}$. The metric can be constructed with eq.~\reef{FSmetric} by evaluating the integrals
\beq
  g_{\mu\nu}  =\frac 12\int \!\!dx_+ dx_-  \(\partial_\mu \bar{\psi}\,\partial_\nu \psi +\partial_\nu \bar{\psi}\,\partial_\mu \psi \) -  \int \!\!dx_+dx_-\ \psi \, \partial_\mu\bar{\psi} \times \int \!\!dx_+dx_-\  \bar{\psi}\,\partial_\nu \psi \,,
\eeq
where the wave function $\psi(x_+,x_-;y,\rho,x,v_\pm)$ is defined in eq.~\eqref{family}. Alternatively, we can calculate the fidelity \reef{fidel}
  \begin{equation}
  F(\lambda,\lambda')=\int \!\!dx_+dx_-\,\bar{\psi}(x_+,x_-;y,\rho,x,v_\pm)\, \psi(x_+,x_-;y',\rho',x',v'_\pm) \,,
  \end{equation}
and then evaluate the metric with eq.~\eqref{metric_limit}.
 
Using either method, we find the Fubini-Study metric is given by
\beqa
ds^2_\mt{FS} &=& dy^2 +d\rho^2 +\sinh^2(2\rho)\,dx^2+ \frac{\tilde{\kappa}^2}2\, e^{2y}\,\Big[
2\sin(2x)\sinh(2\rho)\,dv_+ dv_-
\labell{FS-5D}\\
&&\quad+ \(\cosh(2\rho)  +\cos(2x)\sinh(2\rho) \) dv_+^2+ \(\cosh(2\rho) - \cos(2x)\sinh(2\rho) \) dv_-^2\Big]\,,
\nonumber
\eeqa
where $\tilde{\kappa}=\wrr\, \tilde{x}_0$. 

If we begin by focusing on Gaussian states with $a_\pm=0$, we expect that the optimal trajectories will not involve motion in the $v_\pm$ directions and hence we focus on the first three terms in eq.~\reef{FS-5D}. This three-dimensional subspace has the geometry $\mR \times \mH^2$. As noted above, the reference state corresponds to the origin, \ie $y=0=\rho$ (while the angle $x$ is unspecified). Hence the geodesics are simply lines moving along the $\mR$ and radially outward in the hyperbolic space, \ie $y=y_1\,s,\, \rho=\rho_1\,s$ and $x=x_1$ where $(y_1,\,\rho_1,\,x_1)$ is the position specifying the target state \cite{Chapman:2017rqy}. However, the complexity or the length of the geodesic is precisely the same as found using the Nielsen approach \cite{Jeff}, except for the overall constant factor.\footnote{Note that our conventions were such that the metric \reef{metric-ds} for the Nielsen geometry had an extra overall factor of 2 compared to the Fubini-Study metric \reef{FS-5D}, \ie $ds_\mt{Nielsen}^2=2dy^2 +2d\rho^2+\cdots$ while $ds^2_\mt{FS} = dy^2 +d\rho^2 +\cdots$.}

The full Fubini-Study metric \reef{FS-5D} has a form similar to the Nielsen metric in eq.~\eqref{metric-ds} defined on the space of unitaries, although the dimension of the geometry differs by one as we already noted.  In order to define complexity with this metric \eqref{FS-5D}, we would need to solve the corresponding geodesic equations, but generally the only tractable approach is to find numerical solutions, as we did in section \ref{sec:numerics} for the Nielsen geometry. However, as in section \ref{sec:simple}, we can find a simple analytic solution here for states with a single excitation, \eg $a_+\ne0$ and $a_-=0$. Examining the full geodesic equations, we find it is consistent to set
$x=0$ and $v_-=0$ in this case. Hence we are simply solving for the geodesic equations in the reduced Fubini-Study metric 
\begin{equation}\label{FS_3D2}
\begin{split}
ds^2_\mt{FS} &=\frac12\( dy_+^2 +dy_-^2 +  e^{2y_+}\, dv_+^2\)\,,
\end{split}
\end{equation}
where as before $y_\pm=y\pm\rho$ and for convenience, we have set $\tilde\kappa=1$. We note that this geometry again has the familiar form $\mH^2 \times \mR$ but comparing to the corresponding geometry in section \ref{sec:simple}, we see that to identify this metric with eq.~\reef{eq:metric3d2} and the corresponding geodesic equations, we must set 
\beq
(y_+,y_-, v_+)_\mt{FS} = (-y_+,y_-, u_+)_\mt{Nielsen}\,.
\label{match9}
\eeq
The initial boundary conditions are simply $y_{+0}=0=y_{-0}=v_{+0}$ and to match the final target state \reef{target1} with $a_-=0$, the final boundary conditions are 
\begin{equation}
\label{eq:boundary-condX}
y_{+1} = \frac{1}{2} \log  \w_+\,,\qquad y_{-1} = \frac{1}{2} \log  \w_-\,, \qquad v_{+1} =\ha_+\,,
\end{equation}
where $\w_\pm$ are the same dimensionless ratios as in eq.~\reef{dimless}, while $\ha_\pm\equiv\wrr\,a_\pm$.\footnote{Recall that in eq.~\reef{FS_3D2}, we set $\tilde\kappa=1$ and hence $\tilde{x}_0=1/\wrr$.} We note that, of course, the boundary conditions for $y_\pm$ are the same here as in eq.~\reef{eq:boundary-cond}, but $u_+$ and $v_+$ are different coordinates and so their boundary conditions do not match.

Using the above observations, we can use the solution found in section \ref{sec:simple} given by eqs.~\reef{eq:solution} and \reef{solution2} to produce the simple geodesics for the Fubini-Study metric \eqref{FS_3D2}, which take the form 
\beq
y_+(s)=-\frac{1}{2}\log(\frac{\Delta_\mt{FS}^2}{B_\mt{FS}^2}\text{sech}^{2}(\alpha_\mt{FS}(s)))\,,\quad
v_+(s)=\frac{\Delta_\mt{FS}}{B_\mt{FS}}\tanh(\alpha_\mt{FS}(s))+\frac{A_\mt{FS}}{B_\mt{FS}}\,,\quad
y_-(s)= C s\,,
\eeq
where $\Delta_\mt{FS} = \sqrt{A_\mt{FS}^2+B_\mt{FS}^2}$ and $\alpha_\mt{FS}(s)=s\Delta_\mt{FS} -\text{arctanh}(\frac{A_\mt{FS}}{\Delta_\mt{FS}})$.
The final boundary conditions~\eqref{eq:boundary-cond} fixes the integration constants as
\begin{equation}\label{solution_FS}
\begin{split}
A_\mt{FS} &= \frac{\tilde{\a}_+^2\w_+-\w_+ +1 }{\sqrt{(\tilde{\a}_+^2\w_+- \w_+ +1)^2+4\tilde{\a}_+^2 \w_+^2}}\, \text{arccosh} \( \frac{ \tilde{\a}_+^2\w_++\w_++1}{2\sqrt{\w_+}}\)   \,,\\
B_\mt{FS} &= \pm 2 \sqrt{\frac{\tilde{\a}_+^2\w_+^2}{(\tilde{\a}_+^2\w_+-\w_+  +1)^2+4\tilde{\a}_+^2 \w_+^2}}\, \text{arccosh} \( \frac{\w_+ \tilde{\a}_+^2+\w_++1}{2\sqrt{\w_+}}\)  \,,\\
C &= \frac 12 \log \w_- \,,
\end{split}
\end{equation}
where the sign of $B_\mt{FS}$ is chosen to match that of $\tilde{\a}_+$. The coefficients above can also be derived from eq.~\eqref{solution2} by replacing $\a_+^2 \w_+ \rightarrow \tilde{\a}_+^2$ and $\w_+ \rightarrow 1/\w_+$ at the same time. Further, one can verify the above solutions satisfy
\begin{equation}
\dot{y}_+^2 +e^{2y_+} \dot{v}_+^2 = \Delta_\mt{FS}^2\,,
\end{equation}
and as expected, this combination of the velocities is constant along the geodesic.

\begin{figure}[t]
	\centering
	\subfigure{	 \hspace{-15pt}
	\includegraphics[width=2.2in]{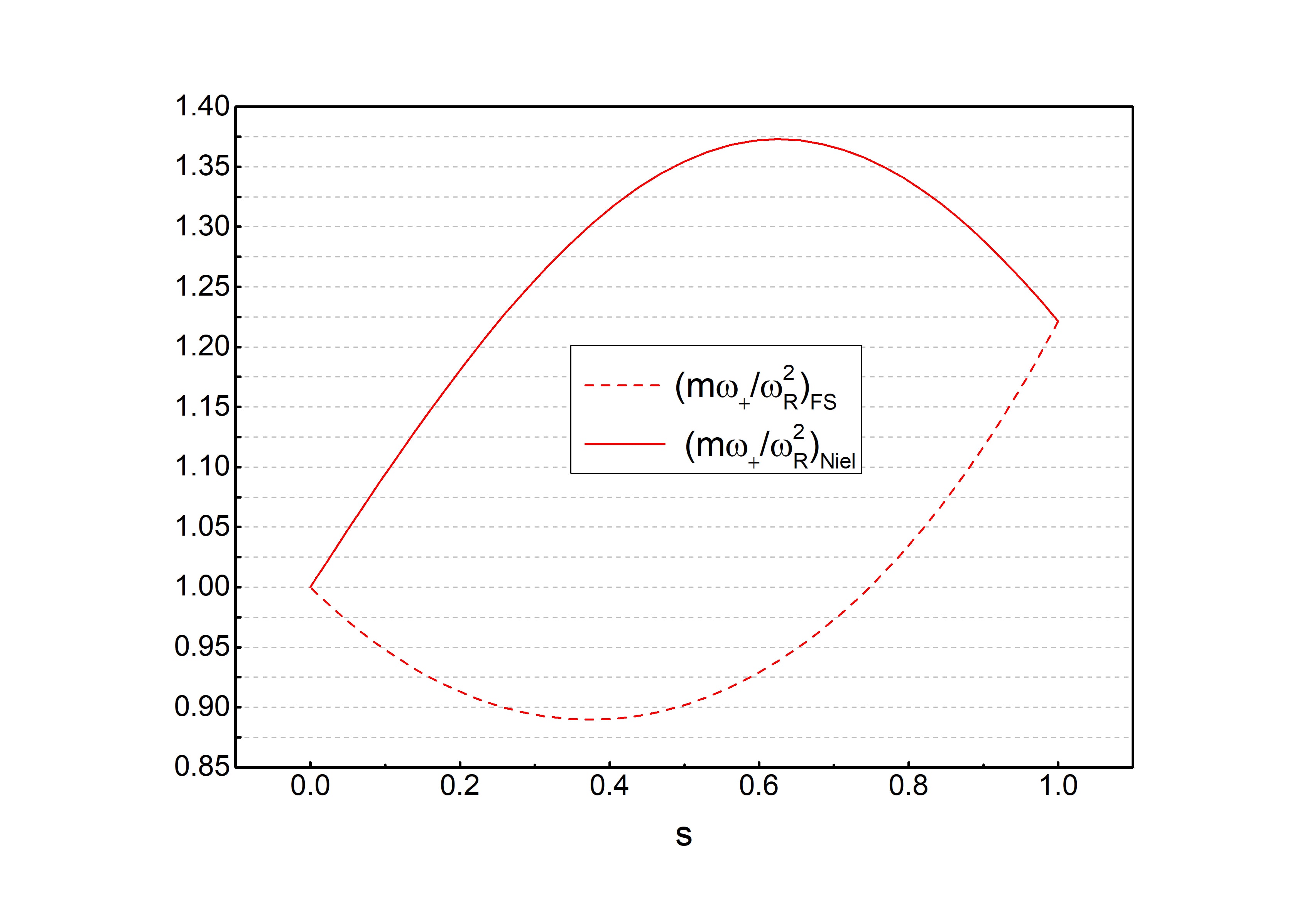}
		 \hspace{-15pt}
	\includegraphics[width=2.2in]{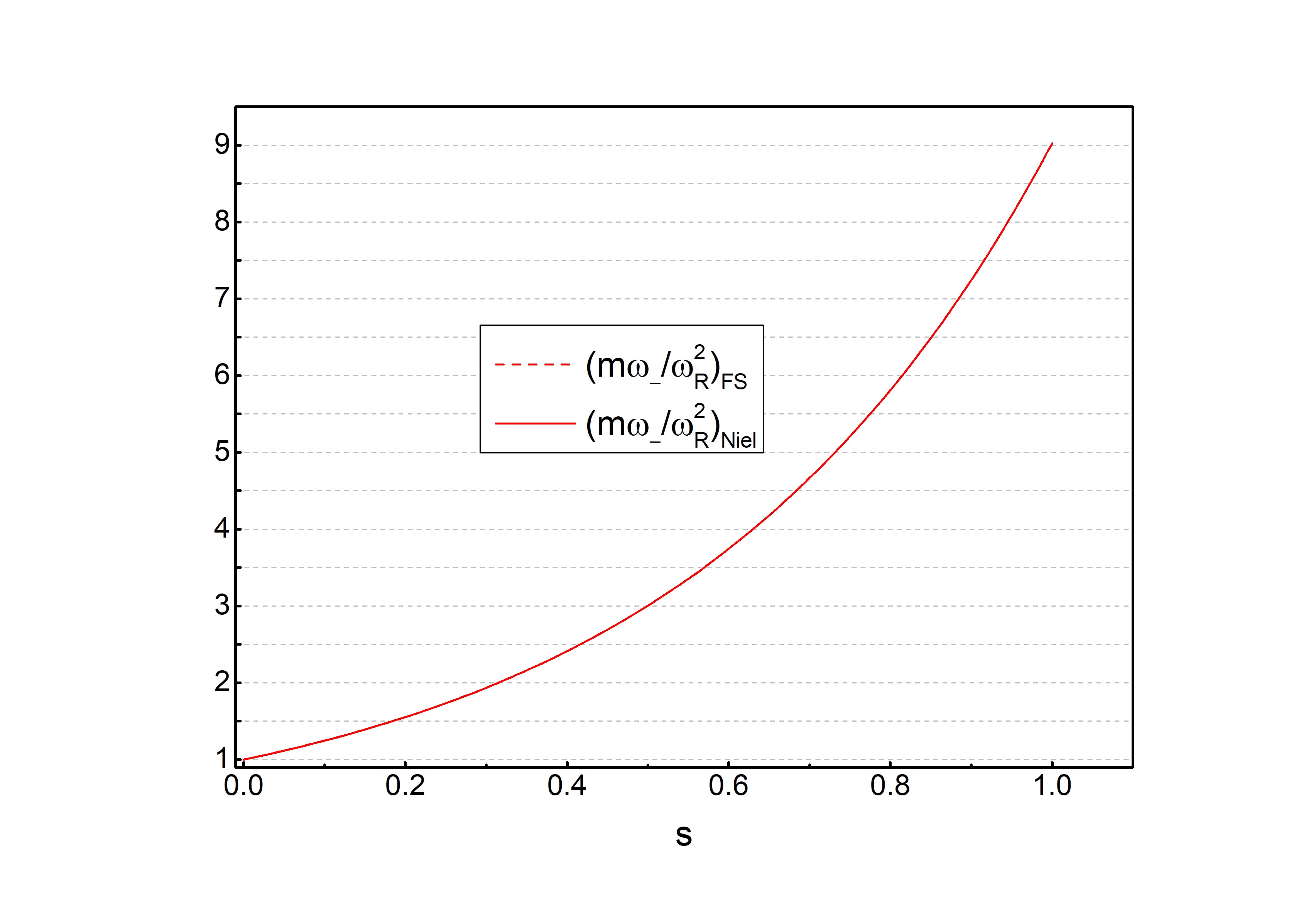}
		 \hspace{-15pt}
	\includegraphics[width=2.2in]{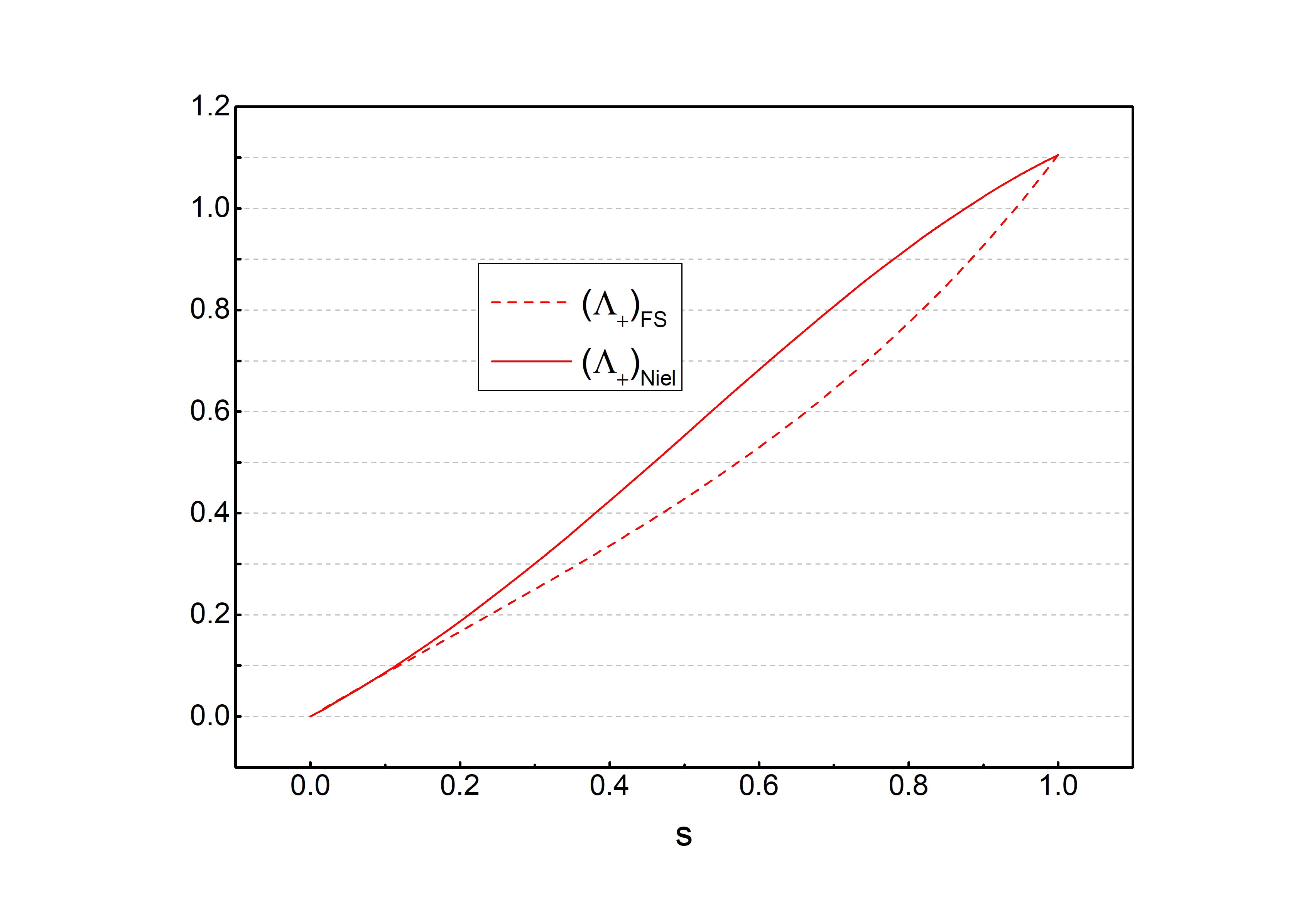}}
	\setlength{\abovecaptionskip}{-15pt}
    \setlength{\belowcaptionskip}{0pt}
	\caption{Example of  simple geodesics connecting the reference state to a same target state with $a_+$ nonvanishing, using Fubini-Study metric and Nielson approaches. In this particular example, we chose $\frac{m}{\wrr^2}\omega_{+}(s=1)=1.221,\ \frac{m}{\wrr^2}\omega_{-}(s=1)=9.025,\  \Lambda_+(s=1)=1.105$. (Recall from eq.~\reef{eq:A-target} that $\Lambda_\pm=[A]^{0\pm}/\wrr^2$, and we defined $m\omega_\pm=[A]^{\pm\pm}$.) We note that $[A]^{+-}(s)=0= \Lambda_-(s)$ throughout the preparation of the target state. This figure shows that the optimal circuits from the two approaches follow different trajectories even though they begin and end at the same states. This difference appears most dramatically for $m\omega_+(s)/\wrr^2$ in the first plot.}
	\label{compare_FS1}
\end{figure}

Certainly the new trajectories in the Fubini-Study geometry \reef{FS_3D2} should be different from those in the Nielsen geometry \reef{eq:metric3d2} because of the differences in $A_\mt{FS}, B_\mt{FS}$ compared to $A,B$ in eq.~\reef{solution2}. Of course, the $y_-$ part of the trajectory is identical in both cases. However, to make clear that the simple geodesics describe distinct circuits in the Nielsen and Fubini-Study geometries, we compare the evolution of the states as described by the 3$\times$3 coefficient matrix in eq.~\reef{eq:A-target}, which for the simple geodesics reduces to
\begin{equation}
A_\mt{Niel}(s)=
\wrr^2\,\left(
\begin{array}{ccc}
e^{2 y_{+}} & 0 &  e^{y_+} u_+\\
0 & e^{2 y_{-}}  & 0 \\
e^{y_+} u_+ & 0 & c_\mt{T} \\
\end{array}
\right)\,,\quad
A_\mt{FS}(s) =\wrr^2\,\left(
\begin{array}{ccc}
e^{2 y_{+}} & 0 &  -\frac{e^{2y_+} v_+}{\wrr x_0}\\
0 & e^{2 y_{-}}  & 0 \\
-\frac{e^{2y_+} v_+}{\wrr x_0} & 0 & c_\mt{T} \\
\end{array}
\right)
\,.\label{horse2}
\end{equation}
The scale $x_0$ appears in $A_\mt{FS}$ because by definition this 3$\times$3 matrix is contracted with $x_a=(x_+,x_-,x_0)$ to construct the wave function. In general, the comparison depends on the combination $\wrr x_0$ appearing in $[A_\mt{FS}]^{0+}$,\footnote{We might note that the coefficient in $[A_\mt{FS}]^{0+}$ is actually $\tilde{x}_0/x_0$, but we set $\tilde{x}_0=1/\wrr$ above to simplify the metric \reef{FS_3D2} and the subsequent analysis.} however, to simplify the comparison we might simply set $\wrr x_0=1$. With this choice, figure \ref{compare_FS1} illustrates an example comparing the components of $A_\mt{Niel}$ and $A_\mt{FS}$ for a fixed target state (and reference state).   As expected, the evolution of $[A]^{--}\propto \omega_-$ is identical in both approaches because this component is controlled entirely by $y_-(s)$, which we already noted is the same in the two cases. The evolution of $[A]^{0+}\propto \Lambda_+$ distinguishes the two trajectories, but in both cases, this component is monotonically increasing from zero to the final value in both cases. The difference between the two circuits is shown most dramatically in $[A]^{++}\propto \omega_+$. In the Nielsen approach, the evolution of this component is concave down, \ie it begins by increasing but it overshoots the final value and so it must decrease again towards the end of the trajectory. In contrast, for the Fubini-Study approach the evolution is concave up, \ie this component begins by decreasing but this direction is reversed in the latter part of the geodesic so that it can reach the final positive value. This reversal of the concavity might be expected from the fact that the metrics look identical under the identification~\eqref{match9}, \ie where the sign of $y_+$ is reversed.

We should also examine the length of the simple geodesics in the Fubini-Study geometry. Towards this end, we first evaluate $\Delta_\mt{FS} = \sqrt{A_\mt{FS}^2+B_\mt{FS}^2}$ using the expression in eq.~\reef{solution_FS} to find
\begin{equation} \label{horse8}
\Delta_\mt{FS} = \text{arccosh} \( \frac{\w_+ \tilde{\a}_+^2+\w_++1}{2\sqrt{\w_+}}\)=\text{arccosh} \( \frac{ \wrr^2x_0^2\,\w_+ \a_+^2+\w_++1}{2\sqrt{\w_+}}\) \,,
\end{equation}
where we are again using $\tilde{x}_0=1/\wrr$ in expressing the result in terms of $\a_+$. Comparing this result to $\Delta$ in eq.~\reef{DELTA} for the Nielsen construction, we see that the two expressions generally differ because of the factor of  $\wrr^2x_0^2$ in the final expression above. However, quite remarkably if we set $\wrr x_0=1$ (as above), we find that $\Delta_\mt{FS}=\Delta$. Now the Fubini-Study complexity is given by
\begin{equation}\label{FS_complexity}
\mC_\mt{FS}(\omega_\pm,a_+)= \sqrt{\Delta_\mt{FS}^2+C^2}\,.
\end{equation}
This result is naturally compared with the $F_2$ complexity in the Nielsen approach, and while the two complexities differ in general, there is a remarkable agreement between the two complexities if we simplify the analysis by choosing $\wrr x_0=1$. We emphasize that with this choice, the Fubini-Study and Nielsen complexities agree even though we have shown above that the two approaches are constructing different  optimal circuits.

We can highlight the difference between the Nielsen and Fubini-Study geometries by introducing the same coordinate systems for both. In fact, the coordinates $(y,\rho,x,v_\pm)$ match those introduced in footnote \ref{footy44}. If we focus on the subspace $x=0=v_-$, we can compare the coefficient matrices in eq.~\reef{horse2} to find $ u_+ =-e^{y_+}v_+$ (where we also set $\wrr x_0=1$ as before).
Then the Nielsen geometry \reef{eq:metric3d2} becomes
\beq
\label{N3geo}
ds^2_\mt{Niel} = (1+v_+^2)dy_+^2+2 v_+\,dv_+\, dy_+ +dv_+^2  +dy_-^2\,.
\eeq
Comparing this expression to the Fubini-Study metric \reef{FS_3D2} we see that even though both describe an $\mH^2\times\mR$ geometry, the physical states are assigned to the geometries in different ways. That is, if we choose a particular state described by particular values of the coordinates $(y_+,y_-,v_+)$ in eq.~\reef{horse2}, then the distances to nearby states $(y_++\delta y_+,y_-+\delta y_-,v_++\delta v_+)$ are very different in the two metrics in eqs.~\reef{FS_3D2} and \reef{N3geo}.  Of course, if we fix our attention on the plane $v_+=0$, \eg to evaluate the ground state complexity, the metrics are the same (except for an overall factor of 1/2).
However, when we move away from this  `normal mode subspace' \cite{Jeff}, we  should not expect that the optimal circuits between two states (or the corresponding complexities) to be the same in the two geometries.
\\
\\
\begin{figure}[h]
	\centering
	 \vspace{-30pt}
	\subfigure{\includegraphics[width=2.8in]{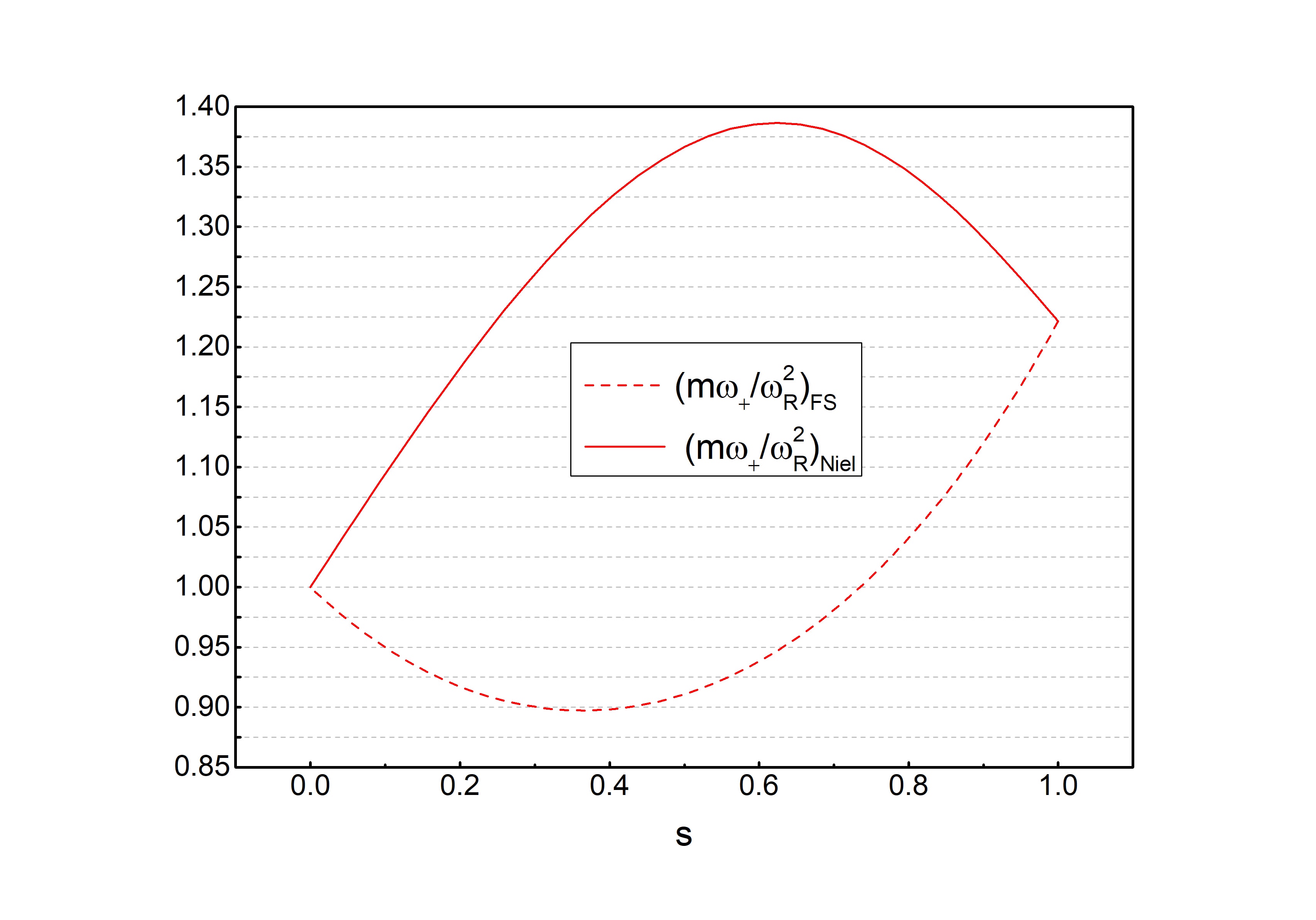}
		 \hspace{-30pt}
	\includegraphics[width=2.8in]{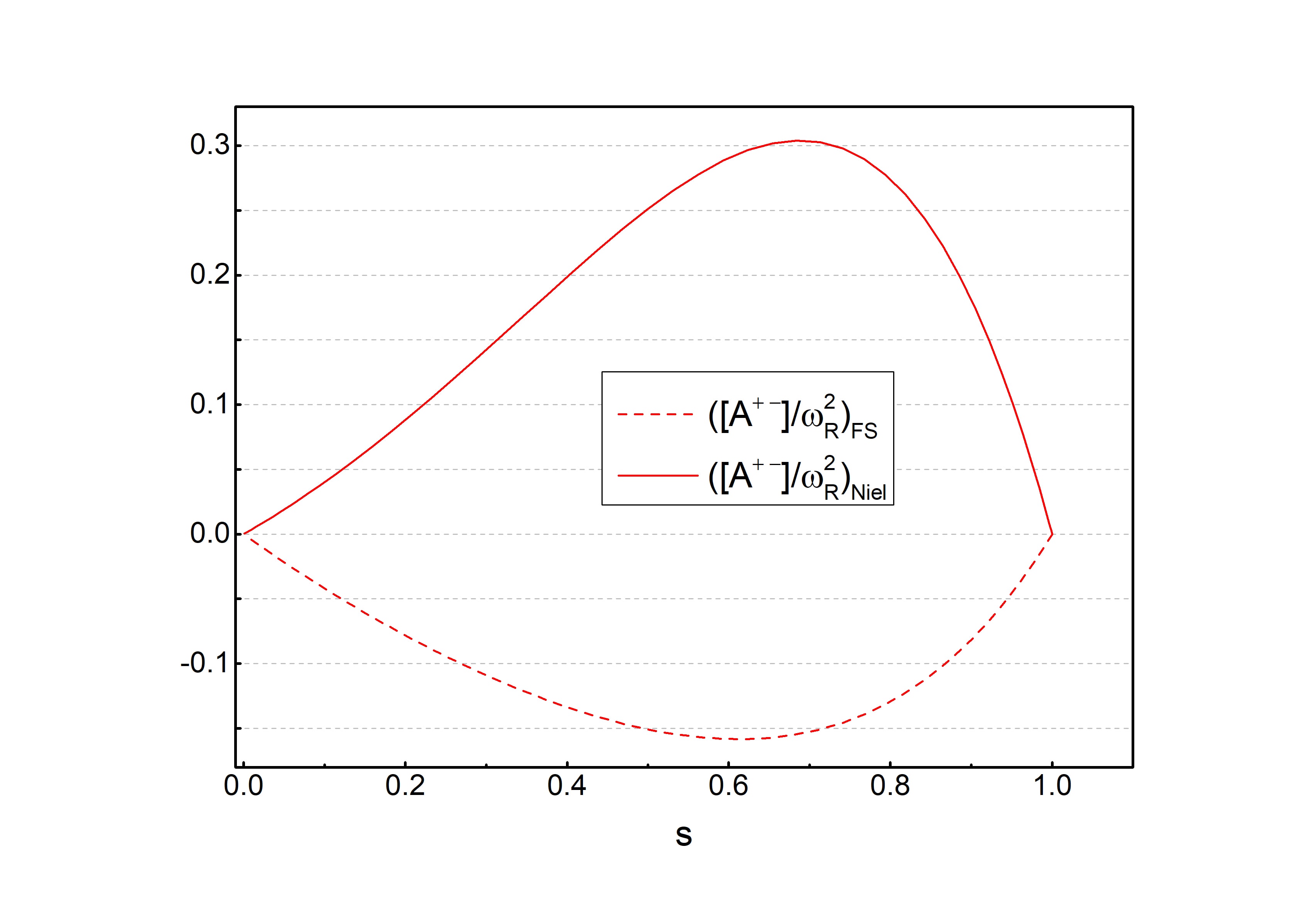}}
    \vspace{-10pt}
	\subfigure{\includegraphics[width=2in]{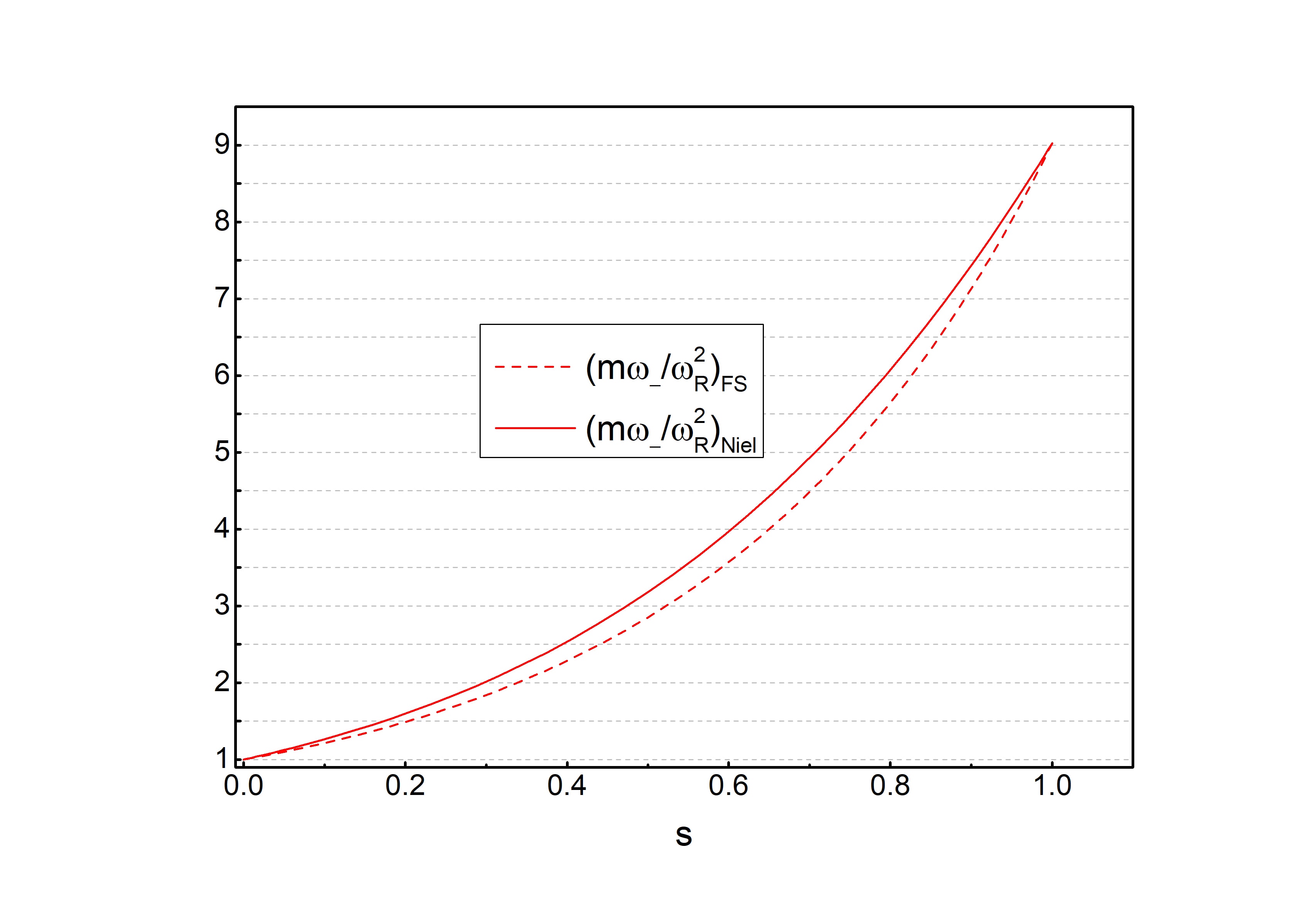}
	\includegraphics[width=2in]{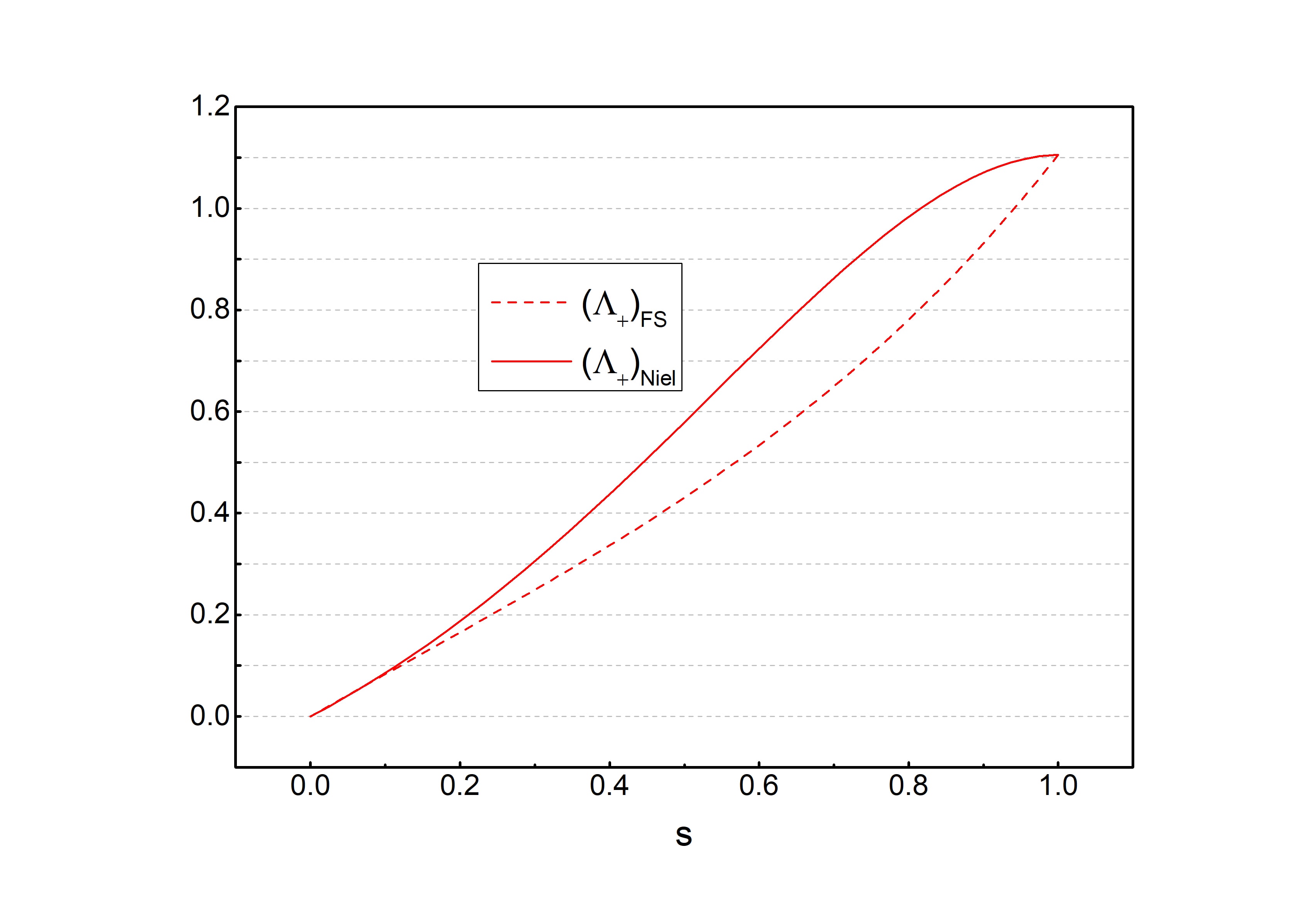}
	\includegraphics[width=2in]{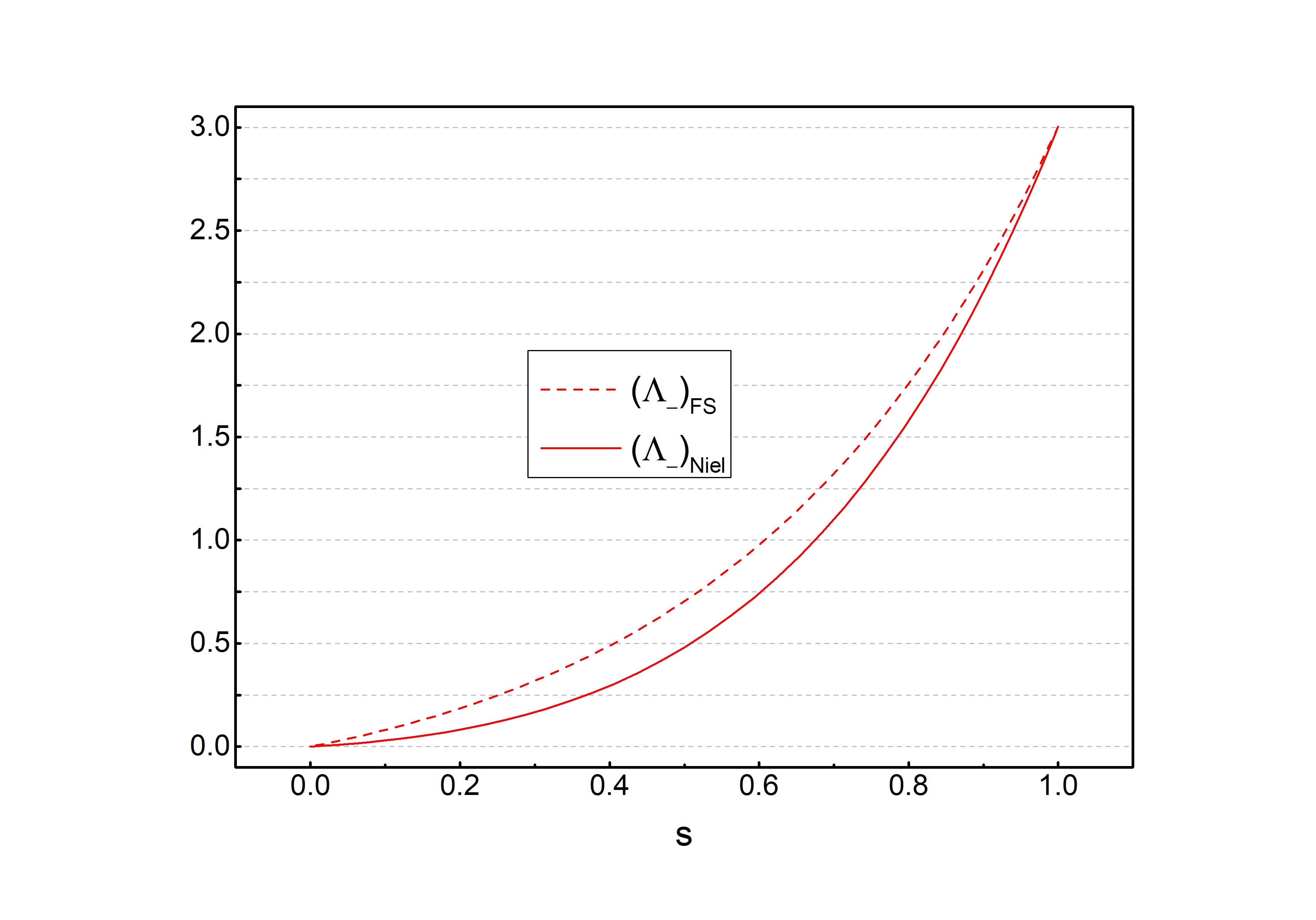}}
	\setlength{\abovecaptionskip}{-15pt}
    \setlength{\belowcaptionskip}{0pt}
	\caption{Example of  geodesics connecting the reference state to a same target state with both $a_\pm$ nonvanishing using Fubini-Study metric and Nielson approaches. In this particular example, we chose $\frac{m}{\wrr^2}\omega_{+}(s=1)=1.221,\ \frac{m}{\wrr^2}\omega_{-}(s=1)=9.025,\  \Lambda_+(s=1)=1.105$ and $\Lambda_-(s=1)=3.004$. (Recall from eq.~\reef{eq:A-target} that $\Lambda_\pm=[A]^{0\pm}/\wrr^2$, and we defined $m\omega_\pm=[A]^{\pm\pm}$.) From this figure, we see that the optimal circuits from the two approaches follow different trajectories even though they begin and end at the same states. This difference appears most dramatically for $m\omega_+(s)/\wrr^2$ and $[A]^{+-}(s)/\wrr^2$ in the upper two plots.
	\label{compare_FS}}
\end{figure}
\\
The comparison above highlights that when considering coherent states, the Nielsen and Fubini-Study approaches to complexity are really different systems. That is, for a given reference state and target state, the optimal preparation chosen by these two approaches moves through different families of intermediate states. This stands in contrast to the preparation of simple Gaussian states, where the optimal trajectories were the same for both approaches. We have seen that these differences already arise for the simple geodesics preparing states where only one of $a_\pm$ is nonvanishing. We provide another example in figure \ref{compare_FS}  for a general target state in which both $a_\pm$ are nonzero (where we again set $\wrr x_0=1$). We solved for the optimal trajectory for the Fubini-Study and the Nielsen approaches numerically and then translated to trajectories into the corresponding (physical) components of the $A$ matrix in eq.~\reef{waveA}. As shown in the plots, the two approaches prepare the same target state through different families of intermediate states. Further for general states like this in which both $a_\pm$ are nonvanishing, the complexity derived by the two methods also differs. In the example shown in the figure, $\mC_\mt{FS}=1.518$ while $\mC_1=1.511$.\footnote{The difference is small but significant, \ie we are confident that the accuracy of our numerical calculations goes well beyond the fourth significant digit here.} For comparison purposes, let us note that the ground state complexity is $\mC_\mt{FS,vac}=\mC_{1,\mt{vac}}=1.221$ for this example.

 \section{Discussion}\label{discuss}
 
Refs.~\cite{Jeff,Chapman:2017rqy} provided the first calculations of complexity in quantum field theory. In the present paper, we extended this analysis, which examined the ground state of a free scalar field theory, to evaluate the complexity of excited states in the same theory. In particular, we considered coherent states with a nonvanishing expectation value of the scalar field (but for which the expectation value of the conjugate momentum was vanishing). Following the analysis of \cite{Jeff}, we began by examining in detail the complexity of the analogous coherent states in a pair of coupled harmonic oscillators in sections \ref{sec:2ho} and \ref{sec:F1}, and then extending the results to the free scalar in section \ref{sec:QFT}. The generators of the gates preparing our coherent state naturally gave rise to a group structure $\mR^{N} \rtimes GL(N,\mR)$, which is a simple extension of the $GL(N,\mR)$ structure found in \cite{Jeff}.\footnote{We reiterate that for the most general coherent states (\eg for which the expectation values of the momenta are also nonvanishing), this group structure is enlarged to $\mR^{2N} \rtimes Sp(2N,\mR)$, which extends the $Sp(2N,\mR)$ for general bosonic Gaussian states (with vanishing expectation values) \cite{Hackl:2018ptj,therm0}.} While this analysis focused on Nielsen's geometric approach \cite{Nielsen:2006,nielsen2006quantum,nielsen2008} for evaluating circuit complexity, we also considered the Fubini-Study approach proposed by \cite{Chapman:2017rqy} in section \ref{app:info-metric}. 

Before proceeding, let us remind the reader that a brief discussion of the complexity of coherent states appeared in \cite{Yang:2017nfn}. This recent work was one of the first investigations of the application of Nielsen's geometric approach to evaluate state complexity in a quantum field theory, and as an example in a free scalar field theory, they consider coherent states  where both $\langle x\rangle$ and $\langle p\rangle$ can be nonvanishing for a single mode. However, their analysis differs from ours in a number of essential ways: First of all, rather than considering an unentangled reference state, \cite{Yang:2017nfn} considers preparing their coherent states beginning with the vacuum state of the field theory. Further, the gate scale introduced for the shift gates in eq.~\reef{4gates} is implicitly set by the frequency of the excited mode in \cite{Yang:2017nfn}. In particular, $x_0^2=2/(m\omega_k)$ is chosen there. Finally, we would add that the complexity is evaluated there by optimizing a somewhat unconventional cost function and the circuits considered are generally not unitary. Hence there is no substantive overlap between our work and the discussion in \cite{Yang:2017nfn}.

\subsection*{Optimal Trajectories/Circuits:}

\begin{figure}[htbp]
	\centering
	\subfigure{\includegraphics[width=4in]{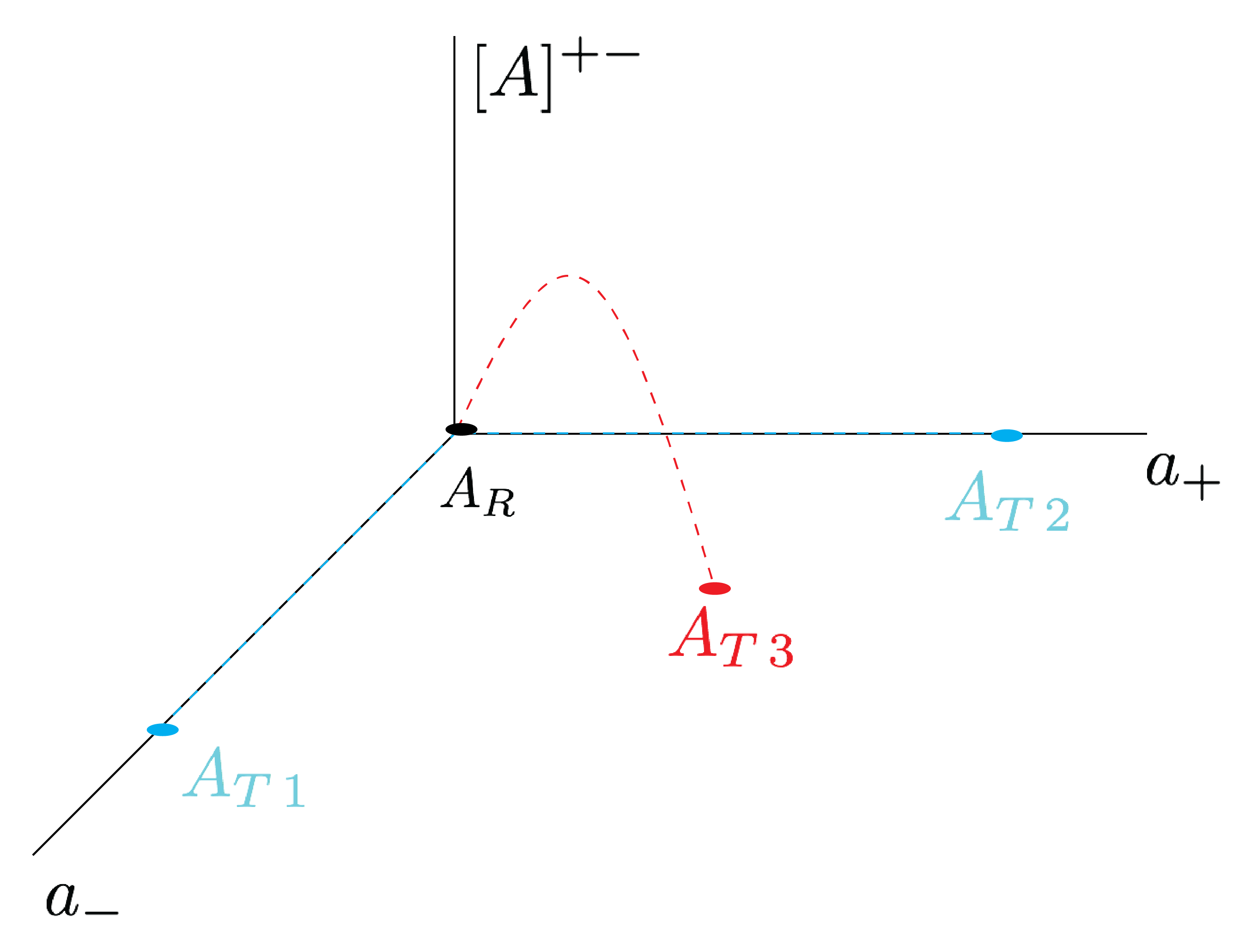}}
	\caption{Schematic diagram of the paths followed by the optimal circuits connecting the unentangled product state $A_\mt{R}$ to three different target states $A_{\mt{T}i}$. The states $A_{\mt{T}1}$ and $A_{\mt{T}2}$ have $a_+=0$ and $a_-=0$, respectively. The optimal circuits preparing such states in which only one of the normal modes is shifted remain in the $x = 0$ plane, \ie $[A]^{+-}= 0$. Therefore the normal modes are unentangled for all of the states along these trajectories. In contrast, the trajectory preparing $A_{\mt{T}3}$ begins and ends with $[A]^{+-}= 0$ but this component is nonvanishing everywhere away from these endpoints. That is, both the reference state and target state are unentangled but the optimal circuit introduces entanglement in the intermediate states when both $a_\pm\ne0$ in the final state. }\label{fig:paths}
\end{figure}

When applying the Nielsen or the Fubini-Study approach to  coherent states for the system of two coupled harmonic oscillators, we could only find the desired geodesics numerically for states in which both normal modes were excited, \ie both $a_\pm\ne0$. 
One of the interesting features of these geodesics was that generally they pass through nonvanishing values of $x$. The physical significance of this feature appears in eq.~\reef{eq:A-target}, where we see that $[A]^{+-}\ne 0\, (\ne [A]^{-+})$ with $x\ne0$ (and also $\rho\ne0$). Therefore, even though the two normal modes are unentangled in both the reference state~\eqref{eq:A-ref} and target state~\eqref{eq:A-target1}, they become entangled in the intermediate states that appear in the optimal circuit joining these states. This behaviour is illustrated schematically in figure~\ref{fig:paths}. It is also exhibited by the explicit examples shown in figures \ref{gs} and \ref{compare_FS}.
We emphasize again that this behaviour is common to both the Nielsen and Fubini-Study approaches.

In section \ref{sec:QFT}, we showed that the `complexity' of determining the optimal trajectories grew with a larger number of excitations. In particular, determining the optimal circuit for states with $K$ normal modes excited, required studying the geodesic equations on a $\mR^K\rtimes GL(K,\mR)$ manifold.  The remaining unexcited modes decouple and they are simply prepared with the linear application of the corresponding scaling gates. It maybe interesting to use numerical methods to investigate the general properties of optimal circuits and corresponding complexity for states where $K\ge3$.

However, a particularly simple case is $K=1$, \ie only one normal mode was excited. In this case, we found analytic solutions for a class of simple geodesics for the Nielsen approach in section \ref{sec:simple}. These geodesics moved in a $\mathbb{H}^2$ slice of the full geometry, involving the coordinates corresponding to scaling and shift gates for the excited mode, \eg $y_+$ and $u_+$ in eq.~\reef{eq:metric3d2}. However, we still had to rely on numerical tests to support the claim that these simple geodesics were the optimal geodesics connecting the reference state~\eqref{eq:A-ref} to a target state~\eqref{eq:A-target1} with only one normal mode excited. These simple geodesics played a role not just for the $F_2$ and $\kappa=2$ cost functions as discussed in section \ref{sec:simple}, but also for the Schatten $p=1$ cost function as described in section \ref{Schat} and also with the Fubini-Study approach in section \ref{app:info-metric}. The analysis of these states showed a similar behaviour for the $F_1$ cost function, in that the optimal circuit only involved the scaling and shift gates for the excited mode, while the other modes decoupled.

\subsection*{Schatten measures:}

In section \ref{Schat}, we investigated the complexity of our coherent states for a cost function constructed from the $p=1$ Schatten norm \reef{Schatten}. This cost function was first suggested in \cite{Hackl:2018ptj} as a replacement for the $F_1$ cost function. There it was observed that in preparing the ground state, this Schatten cost function would produce the same optimal circuit and complexity as the $F_1$ measure constructed in the normal mode basis. However, these results are basis independent when described in terms of the Schatten norm. 
In comparing the results in sections \ref{costf1} and \ref{Schat}, one of the most striking results is that the $F_1$ and Schatten measures no longer give the same circuits or complexities when considering coherent states. For example, the increase above the complexity produced by a small amplitude excitation produced by the $F_1$ measure gave $\Delta\mC_1\propto|\a|$ (see eq.~\reef{DelC1}) while the Schatten norm gave $\Delta\mC_\mt{Schat}\propto\a^2$ (see eq.~\reef{smalla}). We return to these different behaviours for these two cost functions below.

The examination of the Schatten cost function in section \ref{Schat} focused on the complexity of coherent states for two coupled harmonic oscillators. However, this is easily extended to the (regulated) scalar field theory where the circuits act in the group
$\mR^{N} \rtimes GL(N,\mathbb{R})$. For example, with $p=1$, eq.~\reef{costS} is replaced by
	\beq
	\Vert V\Vert_1 = \sum_{i=1}^N \sqrt{\gamma_i}\,,
	\label{costS_Gen}
	\eeq
where the $\gamma_{i}$ are the eigenvalues of $V^TV$. Note that the range of $i$ implicitly indicates that $\gamma_{N+1}=0$, \ie $V^TV$ is represented by a square ($N$+1)$\times$($N$+1) matrix but one of the eigenvalues automatically vanishes (because the last column of $V$ is filled by zeros, as in eq.~\reef{velo2}). We note that the number of eigenvalues matches the number of types of gates that are applied to prepare the ground state, \ie the optimal circuit only uses the scaling gates for each of the $N$ normal modes. This match is why the $p=1$ Schatten complexity agrees with the $F_1$ complexity for the ground state. However,  the $N$ eigenvalues encode information about the shift gates, as well as the scaling gates, when preparing the coherent states, and so as noted above, this agreement does not extend to these states.

Generally the $p=1$ Schatten cost function also involves a complicated coupling between the different modes, \eg as is implicit in the singular values given in eq.~\reef{eigenv2}. However, the modes seem to decouple when evaluating the complexity of coherent states where a single mode is excited, and the optimal circuit follow the same simple geodesics described above for the $F_2$ or $\kappa=2$ cost function. We were able to prove these geodesics extremized the full Schatten norm \reef{costS} by considering a new cost function $\mL_0'=\Vert V\Vert_1^2$, see eq.~\reef{walk}. This could be decomposed into two parts: $\mL_0=\gamma_1+\gamma_2$ and $\mL_1=\gamma_1 \gamma_2$. The first coincides with the $\kappa=2$ cost function and so the simple geodesics extremized this term. It was then straightforward to show that they also extremized $\mL_1$.

If we recall that there is a family of Schatten norms \reef{Schatten} labeled by a positive integer $p$, it is interesting that the previous reasoning can be extended to the higher $p$ norms. That is, we can argue that the simple geodesics extremize the Schatten cost functions for general $p$ as follows: First it is straightforward to show the recursion relation
\begin{equation}\label{mystery}
	\(	\Vert V\Vert_{p+1}\)^{p+1}=\(	\Vert V\Vert_p\)^p \, \Vert V\Vert_1- \sqrt{\mL_1} \,\(	\Vert V\Vert_{p-1}\)^{p-1}\,.
	\end{equation}
Now we have shown that the simple geodesics extremize $\Vert V\Vert_1$ and $\mL_1$ and therefore if they also extremize  $\Vert V\Vert_p$ and $\Vert V\Vert_{p-1}$, then the same geodesics will
extremize $\Vert V\Vert_{p+1}$.\footnote{Further, evaluated on the simple geodesics, we have $\gamma_1=\Delta^2$ and $\gamma_2=C^2$, with $C$ and $\Delta$ given in eqs.~\reef{solution2} and \reef{DELTA}, respectively. It is important that the singular values are both constants along the simple geodesic because this  eliminates potential contributions arising from integration by parts in the following argument.}
Since the $p=2$ norm corresponds to the $F_2$ norm, it is also extremized by these simple geodesics. Hence beginning with $p=1,2$, we can work iteratively to show that our simple geodesics are in fact also geodesics for the general $p$ Schatten cost functions. 

Given the previous result, we can apply the interesting property of Schatten norms that $\Vert A \Vert_p  \ge \Vert A \Vert_q $ for $1 \le p\le q \le \infty$ \cite{gil2003operator}. This leads us to conclude that given a particular simple geodesic describing the optimal circuit for a particular state, the complexity of the same circuit increases if we increase the index of the Schatten norm with which the complexity is evaluated, \ie 
 \begin{equation}\label{Schatten:compare}
 \mC_{\mt{Schat},p} (\w_+,\a_+)  \ge  \mC_{\mt{Schat},q} (\w_+,\a_+) \,, \qquad{\rm for}\ \ \  1 \le p\le q \le \infty \,.
 \end{equation}

 We stress that our discussion above focused on the simple geodesics describing the optimal circuits for states with a single excitation (and this discussion easily generalizes to the case of $N$ normal modes but only a single excitation). General geodesics of the $F_2$ or $\kappa=2$ measures, \ie for states with multiple excitations, will not extremize the auxillary functional $\mL_1$ and so they will not be optimal trajectories for any of the Schatten norms except $p=2$. However, we did argue at the end of section \ref{sec:complexity} that it is possible to consider multiple excitations as long as the amplitudes are small, \ie $\a_i\ll1$. In this case, the different normal modes can be decoupled at least to first order in a perturbative expansion.

To close here, we would like to point out that we can use a modified Schatten cost function of the form,
\beq\label{Schatten2}
\(\Vert A \Vert_p\)^p ={\rm Tr}\!\[\( A^\dagger\,A\)^{p/2}\]\,,
\eeq
\ie we eliminate the overall $p$'th root in eq.~\reef{Schatten}.
These cost functions are rather analogous to the $\kappa$ cost functions \reef{function_Fkappa} with $\kappa=p$, \ie optimizing these new cost functions would yield the same optimal circuit and complexity as the $\kappa=p$ cost function constructed in the normal mode basis when considering Gaussian states. Therefore, the divergence structure of the ground state complexity would match for these two sets of cost functions. Of course, the advantage of using eq.~\reef{Schatten2} would be that the results are basis independent. However, as with the case of $\kappa=p=1$, this agreement would not extend to the coherent states considered here. We should also note that like the $\kappa$ measures, these modified Schatten cost functions are not homogeneous, \ie the total cost associated with a path is generally not invariant under reparametrizations of $s$. 

\subsection*{Fubini-Study approach:}

In Section \ref{app:info-metric}, we examined  the Fubini-Study approach developed in \cite{Chapman:2017rqy} in some detail. In particular, we applied this approach to examine the complexity of coherent states 
for a pair of coupled harmonic oscillators, the same problem that we studied using the Nielsen approach in section \ref{sec:2ho}. Both the Nielsen and the Fubini-Study approaches identify the complexity of a state as the distance from a simple reference state in some geometry.  Nielsen's method \cite{nielsen2006quantum,nielsen2008,Nielsen:2006} is motivated by the definition of complexity as the number of elementary gates in the optimal circuit, and so in this case, a metric is defined on the space of quantum circuits or unitary transformations, \eg as in eq.~\eqref{metric-ds}. Optimizing the trajectory in this space then has a direct interpretation as minimizing the number gates
used in the circuit preparing the desired target state (or at least, optimizing this number according to some cost function). The Fubini-Study approach instead accounts for the complexity by keeping track of the changes of the state throughout the preparation of the target state. As its title indicates, this method makes use of the Fubini-Study metric, which defines a geometry directly on the space of states. An important difference is then that the latter geometry assigns a variable cost to specific gates, \ie the cost depends on the details of the state on which they act, whereas the gates are assigned fixed costs in the Nielsen approach. Further, at any point in the space of states, there will be degenerate operations which leave the state unchanged, \ie $\ket{\psi}= U_0\ket{\psi}$. Therefore, in general, one finds that the space of unitaries has a larger dimension than the space of states, as illustrated by comparing the geometries in sections \ref{sec:2ho} and \ref{app:info-metric}.\footnote{At a pragmatic level, this proves to be an advantage for the Fubini-Study approach since in many cases, one will find a single geodesic connecting the reference state and the target state. In contrast, as discussed in section \ref{sec:2ho}, the Nielsen
approach yields a family of geodesics connecting these states and the complexity is determined by the length of the shortest geodesic in this family.} For a more detailed discussion comparing these two approaches, the interested reader is referred to \cite{comparison-paper}.

However, we want to stress that the definition of Fubini-Study metric only depends on the physical parameters which characterize the states. This is clear from the definitions in terms of the fidelity in eqs.~\reef{fidel2} and \reef{metric_limit}. For example, even though the coordinates $\lambda^\mu$ may be dimensionful, producing a dimensionful metric, the cost is dimensionless due to the appearance of the compensating factors of $\dot{\lambda}$ in eq.~\eqref{Def_FScomplexity}.  Hence, the parameter $\tilde{x}_0$, which was introduced to define the dimensionless coordinates $v_\pm=a_\pm/ \tilde{x}_0$ and which appears in the metric \reef{FS-5D}, will never appear in the complexity or in the distance along any trajectories. Instead it will be absorbed by the boundary conditions which would be defined in terms of the dimensionful displacements $a_\pm$. In contrast, the parameter $x_0$ is an essential ingredient in the definition of the shift gates \reef{4gates}, which must have dimensionless generators.\footnote{A similar gate scale appears in defining gates for the full $Sp(2N,\mR)$ group of Bogoliubov transformations acting on bosonic Gaussian states, \eg see \cite{therm0}.} This parameter reflects a true freedom in the choice of the fundamental gates and it will affect the final complexity evaluated using the Nielsen approach. For example, it implicitly appears in eqs.~\reef{smalla} and \reef{biga} through the definition of $\a_\pm=a_\pm/x_0$

Hence we see that the Fubini-Study and Nielsen approaches must define different complexities for the optimal circuit with the same target and reference state. However, we remind the reader that the ground state complexities, and in fact the optimal circuits, were found to agree with these two different approaches \cite{Jeff,Chapman:2017rqy}. In this case, the optimal circuits only involved $GL(N,\mR)$ gates and so no additional scale was needed to define the corresponding generators. In fact, in this case, the Fubini-Study geometry can be embedded in the corresponding Nielsen geometry. However, in the case of coherent states,  we saw in section \ref{FStwo} that the Nielsen and Fubini-Study approaches produced different optimal circuits for a fixed pair of reference and target states. We were able to show this analytically for the simple geodesics where only one of $a_\pm$ is nonvanishing. However, even though the optimal circuits are clearly different (see figure \ref{compare_FS1}), a somewhat surprising result was that the Fubini-Study complexity still matched the Nielsen complexity (measured with the $F_2$ cost function) if we make the choice $x_0=1/\wrr$. It would be interesting to better understand this agreement. Nevertheless, when we explored the geodesics for coherent states with both $a_\pm$ nonvanishing, we found that the optimal circuits produced by the Nielsen and Fubini-Study approaches were again different (see figure \ref{compare_FS}) and that the corresponding complexities were also distinct.

\subsection*{Complexity for free scalar field:}

As we described in section \ref{sec:complexity},  the complexity of coherent states (or any state) in the free scalar field theory is UV divergent. However, considering the difference $\Delta\mC=\mC_\mt{coh}-\mC_\mt{vac}$ yields an interesting UV finite quantity. Hence in the following, we focus on discussing this difference, \ie the increase of the complexity of the coherent state over the complexity of the vacuum state. However, we must add that as explained with eq.~\reef{delta22}, this difference vanishes for the $F_2$ complexity. This same reasoning would apply for the complexity evaluated with the Schatten cost functions \reef{Schatten} with $p\ge2$. Further, this difference would also vanish for the Fubini-Study complexities if we were to extend the result of section \ref{app:info-metric} to the quantum field theory. However, we can still consider this difference when evaluating the complexity with $F_1$ cost function, $\kappa=2$ cost function and the $p=1$ Schatten norm, and as we will discuss below the QFT complexities produced with these cost functions are  most closely aligned with the  result of holographic complexity.

If we only excite a single mode of the field theory, we can use the analytic results for the simple geodesics found for the $\kappa=2$ cost function  or the $p=1$ Schatten norm. That is, eqs.~\reef{complex8} and \reef{complex9a} would produce $\Delta\mC_{\kappa=2}$ and $\Delta\mC_{\mt{Schat}}$ for the full field theory with $\w_+,\a_+$ corresponding to the frequency and shift of the excited mode. Similarly, eq.~\reef{eq:C1} could be used to evaluate $\Delta\mC_{1}$ for a field theory state with a single excitation. In principle, one could use numerical methods, \eg as in section \ref{sec:numerics}, to study the increase in complexity for coherent states in which more than one mode is excited. 

However, a simpler and more interesting situation is one where many modes are excited in the coherent state but with small shifts, \ie $\a_k\ll1$ for all of the modes. As we argued in section \ref{sec:complexity} for these three cost functions, to leading order, the shift in the complexities for each of the individual modes can be added together to produce
\begin{equation}\label{collections}
\begin{split}
&\Delta \mC_1 \simeq  \sum_{\w_k \le  1}\sqrt{\w_k}\,|\a_k| + \sum_{\w_k \ge  1} |\a_k|\\
\Delta\mC_{\kappa=2}  \simeq \sum_k & \frac{\log\w_k}{|\w_k-1|}\,\w_k \,\a_k^2 \,,\qquad
\Delta\mC_\mt{Schat} \simeq \sum_k \frac{\w_k\,\a_k^2}{|\w_k-1|} \,,
\end{split}
\end{equation}
where the sums run over the excited modes. We would like to stress that verifying these results required a nontrivial analysis and relied on the special form of the simple trajectories for the individual modes. Here we might recall the definitions of the dimensionless ratios from eq.~\reef{dimless}
\beq\label{dimless2}
\w_k=\frac{\omega_k}{\delta\,\wrr^2} \qquad{\rm and}\qquad \a_k =\frac{\delta^{d/2}\,\langle \phi_k\rangle}{x_0}
\eeq
where we have also substituted $m=1/\delta$ (\ie the inverse of the lattice spacing) and $a_k=\delta^{d/2}\,\langle \phi_k\rangle$ from the discussion of the lattice regularization of the scalar field theory at the beginning of section \ref{sec:QFT}.  While the full dispersion relation for arbitrary modes is given in eq.~\reef{foury}, we would typically only be interested in exciting low energy modes, \ie with $\omega_k\ll 1/\delta$, and so the dispersion relation would be well approximated by $\omega_k^2=|\vec k|^2+\mu^2$ (where $\mu$ is the mass of the scalar in eq.~\reef{Ha_scalarQFT}). One interesting difference here is that the leading contribution for the $F_1$ complexity scales as $\Delta \mC_1\propto |\a_k|$ while in the other two cases, we have $\Delta \mC \propto \a_k^2$. We return to this point below.

Another observation is that, at least with the $\kappa=2$ and $p=1$ Schatten metrics, the appropriate expansion parameter is actually the combination
\beq
\w_k\, \a_k^2= \frac{\delta^{\frac{d-2}2}}{\wrr^2\, x_0^2}\ \omega_k\, \langle \phi_k\rangle^2\,.
\label{energy}
\eeq
This is immediately obvious from examining eqs.~\reef{complex8} and \reef{complex9a} and seeing that the shift only appears in this combination $\w_k \a_k^2$ for the full nonlinear results for the cost of the simple geodesics. A further comment is that if we make the choice $x_0=1/\wrr$, then the above expression simplifies to $\w_k\, \a_k^2= \delta^{\frac{d-2}2}\,\omega_k\, \langle \phi_k\rangle^2$, which is now only dependent on physical parameters defining the state (and with $\delta$, defining the quantum field theory). This choice of identifying $x_0$, the scale appearing in the shift gates, with $\wrr$, the frequency defining the reference state simplifies our complexity model in that with this choice, there is a single (dimensionful) free parameter appearing in the definition of the complexity -- of course, there is still also the freedom in choosing the cost function. Recall that $\wrr x_0=1$ also appeared in section \ref{app:info-metric} where this choice ensured that the $F_2$ complexity of the simple geodesics matched the Fubini-Study complexity. 

Examining eq.~\reef{collections}, we can see that generally $\Delta\mC$ increases as $\omega_k$ increases (when we begin with a small $\w_k$). However, we cannot rely on these expressions for very large energies because we explained above the correct expansion parameter is the combination of the frequency and amplitude given in eq.~\reef{energy}. Hence let us focus on coherent states with a single excited mode, for which our full nonlinear results for the simple geodesics apply, and consider the limit when $\w_k$ becomes large with a fixed value of $\a_k$.  Then using eqs.~\reef{complex8}, \reef{eq:C1} and \reef{complex9a}, we find that this limit yields
\beqa
\Delta \mC_1 &=& |\a_k|\ \ {\rm for}\ |\a_k|<2\,,\quad  {\rm or}\quad \log\frac{\a_k^2}4 + 2 \ \ {\rm for}\ |\a_k|>2\,,
\label{bigw}\\
\Delta\mC_{\kappa=2}  &=&\log \left(1+\a_k^2\right) \log \left(\left(1+\a_k^2\right) \w_k \right)+ \frac{\a_k^2  \log\( (1+\a_k^2)^2\w_k \)}{\left(1+\a_k^2\right)^2 \w_k } +{\cal O}\!\(\frac{\log \w_k}{\w_k^2}\)\,,
\nonumber\\
\Delta\mC_\mt{Schat} &=&  \log \( 1 +\a_k^2 \) +\frac{\a_k^2}{(1+\a_k^2)^2 \w_k} + \frac{\a_k^2 (2-\a_k^2)}{2(1+\a_k^2)^4 \w_k^2}+{\cal O}\!\(\frac{1}{\w_k^3}\) \,.
\nonumber
\eeqa
Hence we see that in fact with the $F_1$ and the Schatten metrics, the increase in the complexity saturates at some fixed value determined by $\a_k$ at large energies. In contrast, the $\kappa=2$ complexity continues to grow logarithmically at very large energies. 

It is interesting to compare this behaviour to that of the complexity for excited states in free fermion theories found in \cite{Hackl:2018ptj}. As discussed there, a broad class of states with particle and antiparticle excitations remain Gaussian states and so their complexity is easily computed using the same methods (\ie the same gates) as were used to evaluate the complexity of the vacuum state. In particular, the space of Gaussian fermionic states has two disconnected components, \ie states with odd and even particle number, where the component with even particle number contains the vacuum. It is the excited states in this component whose complexity was evaluated in \cite{Hackl:2018ptj}. The precise increase in the complexity depends on the details of the excited state, but generally $\Delta\mC$ is finite and larger for lower energy modes. For example, considering the class of states with $n$ particle excitations and $n$ antiparticle excitations, but where the momenta of all of these excitations are different,
\begin{align}
\Delta\mathcal{C}_{\kappa=2}&=n\,\pi^2-\sum_i \left[ \tan^{-1}\!\left({|\vec{k}_i|}/ {\mu}\right)\right]^2\,,
\label{hotpot}\\
\Delta\mathcal{C}_\mt{Schat}&=n\,\pi-\sum_i  \tan^{-1}\!\left({|\vec{k}_i|}/{\mu}\right)\,,
\nonumber
\end{align}
where $\mu$ is the fermion mass. For these states, we see that with the $\kappa=2$ cost function, $\Delta\mathcal{C}_{\kappa=2} \simeq n\,\pi^2$ if all of the excitations have low energy (\ie $|\vec{k}_i|\ll \mu$) whereas $\Delta\mathcal{C}_{\kappa=2} \simeq\frac12\, n\,\pi^2$ with all high energy excitations (\ie $|\vec{k}_i|\gg \mu$). Even more dramatically, $p=1$ Schatten cost function yields $\Delta\mathcal{C}_\mt{Schat} \simeq n\,\pi$ if all  $|\vec{k}_i|\ll \mu$ and $\Delta\mathcal{C}_\mt{Schat} \simeq 0$ if all  $|\vec{k}_i|\gg \mu$. Hence the behaviour of the fermionic states (with even particle number) contrasts with the bosonic coherent states above since for the latter, excitations in the higher momentum modes generally produces a larger $\Delta\mC$.

Let us conclude with a few comments on possible future extensions. One obvious extension would be to consider more general coherent states with expectation values for both the field modes and their conjugate momenta. As we commented before, this would require extending the $\mR^{N} \rtimes GL(N,\mR)$ group structure found here to $\mR^{2N} \rtimes Sp(2N,\mR)$. In particular, this would allow us to follow the time evolution of the coherent states. An obvious question would be to then to examine if the complexity increases, decreases or remains constant as a coherent state evolves. Coherent states also provide an interesting forum to compare to the QFT complexity with holographic complexity. Recall that the leading divergences appearing in the QFT calculations of complexity compared well with those appearing in holographic complexity \reef{leaderH} with an appropriate choice for the cost function \cite{Jeff,Chapman:2017rqy}. The holographic analog of our coherent states would be a bulk configuration where a bulk scalar has excited in the vacuum AdS spacetime. Here we observe that to leading order, modification of the bulk geometry will be proportional to the square of the scalar amplitude since the bulk scalar backreacts on the geometry through the stress tensor in Einstein's equations, which is quadratic in scalar field. Hence we expect that the change in the holographic complexity must also be quadratic in the scalar amplitude, which is in agreement with our results in eq.~\reef{collections} for the $\kappa=2$ and $p=1$ Schatten cost functions. However, the $F_1$ cost function does not exhibit this behaviour. We plan to return to this topic and make a detailed comparison between our results for the complexity of QFT coherent states and holographic complexity in \cite{prep0}.


\section*{Acknowledgments}
It is a pleasure to thank Alice Bernamonti, Shira Chapman, Federico Galli, Lucas Hackl, Markus Hauru, Hugo Marrochio and Joan Simon for useful conversations.  Research at Perimeter Institute is supported by the Government of Canada through the Department of Innovation, Science and Economic Development and by the Province of Ontario through the Ministry of Research \& Innovation. RCM, JH and SMR were  supported in part by research funding from the Simons Foundation through the ``It from Qubit'' Collaboration. RCM is also supported by an NSERC Discovery grant.  RCM also thanks the Galileo Galilei Institute for Theoretical Physics for hospitality and the INFN for partial support during part of this work. MG is supported in part by NNSFC Grants No.~11775022 and No.~11375026 and also by the China Scholarship Council.  MG also gratefully acknowledges the support of the Perimeter Institute Visiting Graduate Fellows program. JH is also supported by the Government of Ontario's Ministry of Advanced Education and Skills Development through a Queen Elizabeth II Graduate Scholarship in Science and Technology. 
 
\begin{appendix}

\end{appendix}

\bibliographystyle{JHEP}
\bibliography{biography_circuit_complexity}

\end{document}